\documentclass[]{pasj01} 
\Received{$\langle$Octorber--5--2020$\rangle$}
\Accepted{$\langle$May--19--2021$\rangle$}
\Published{$\langle$publication date$\rangle$}

\usepackage{longtable}
\usepackage{graphicx}
\usepackage{txfonts}
\usepackage{natbib}
\usepackage{subfloat}

\begin{document}

\title{Hot Molecular Core Candidates in the Galactic Center 50~km\,s$^{-1}$  Molecular Cloud}

\author{Ryosuke Miyawaki$^1$, Masato Tsuboi$^{2, 3}$, Kenta Uehara$^3$, and Atsushi Miyazaki$^4$}%
%
\altaffiltext{1}{College of Arts and Sciences, J.F. Oberlin University, Machida, Tokyo 194-0294, Japan}
\email{miyawaki@obirin.ac.jp}
\altaffiltext{2}{Institute of Space and Astronautical Science, Japan Aerospace Exploration Agency,\\
3-1-1 Yoshinodai, Chuo-ku, Sagamihara, Kanagawa 252-5210, Japan }
\altaffiltext{3}{Department of Astronomy, The University of Tokyo, Bunkyo, Tokyo 113-0033, Japan}
\altaffiltext{4}{Japan Space Forum, Kanda-surugadai, Chiyoda-ku,Tokyo,101-0062, Japan}


\maketitle

\begin{abstract}

We present the results based on the 2\,\farcs\,5-resolution observations using Atacama Large Millimeter/Submillmeter Array (ALMA) of the Galactic Center Molecular Cloud G$-$0.02$-$0.07, or the 50 \,km\,s$^{-1}$ Molecular Cloud (50~MC), in the SO ({\it N}$_J$=2$_2$-1$_1)$  line and 86-GHz continuum emission, the combination of which is considered to trace ``hot molecular core candidates'' (HMCCs) appearing in the early stage of massive-star formation.
In the 86-GHz continuum image,  we identified nine dust cores in the central part of the 50~MC, in which four famous compact HII regions are located. 
No new ultra-compact HII regions were found.
We also identified 28 HMCCs in the 50~MC with the SO line. 
The overall SO distribution had no clear positional correlation with  the identified HII regions. 
The HMCCs in the 50~MC  showed a variety of association and non-association with dust and Class-I CH$_3$OH maser emissions. 
The variety suggests that they are not in a single evolutionary stage or environment.
 Nevertheless, the masses of the identified HMCCs  were found to be well approximated  by a single power law of their radii,  {\it M$_{\rm LTE}$}/(M$_\odot$)=5.44 $\times $10$^5$ (r/(pc))$^{2.17}$ at  {\it T}$_{\mathrm{ex}}$  = 50--100 K.
The derived HMCC masses were larger than those of the molecular cores with the same radii in the 50~MC and also than those of the molecular clumps in the Galactic disk.
Additional observations are needed to confirm the nature of these HMCCs in the 50 MC.

\end{abstract}

\section{INTRODUCTION}

The Galactic Center 50 \,km\,s$^{-1}$ Molecular Cloud (50~MC) is located only 3\' \  from Sagittarius A$^\ast$ (Sgr A$^\ast$) in the Central Molecular Zone (CMZ) \citep{Morris1996}.
The hot and turbulent medium of the 50~MC  is believed to have been generated by a strong tidal field, cloud-cloud collisions (CCCs), 
stellar winds, supernova shocks \citep{Morris1996}. 

In the CMZ, a very steep power-law linewidth-size relation of N$_2$H$^+$ molecule has been observed down to 0.1 pc scale, which is likely to originate in the decay of  supersonic gas motion in strong shocks \citep{Kauffmann2017a}. 
In such environment,   star formation in the CMZ clouds may be suppressed \citep{Kauffmann2017a}. 
Many CMZ molecular clouds have also been observed to have unusually shallow density gradients (and corresponding steep mass-size relations) compared  with most regions elsewhere in the Milky Way \citep{Kauffmann2017b}.
\citet{Lu2019a}  argued that the  star formation in the CMZ clouds  is inactive overall. 
The dense gas fractions of the other observed clouds except Sgr~C are smaller than 1 \% and the star formation rate (SFR) is  similarly low (Lu et al. 2019a).
They also suggested that the low SFR in the CMZ could be because there is less gas confined in gravitationally bound cores \citep{Lu2019a, Lu2019b}.
The cores may be prevented from gravitationally collapses by the strong turbulence in this region or if it started it  may have only recently started.
The extreme environment in the CMZ provides unique opportunities for studying star formation in the centers of external galaxies  in general.

The 50~MC has a string of three compact HII regions (CHII) and one ultra-compact HII region (UCHII) in G$-$0.02$-$0.07, or Sgr~A East A--D \citep[e.g.][]{Ekers1983, Goss1985, Yusef-Zadeh2010, Mills2011}. 
The HII regions appear to lie along a dense ridge of the 50~MC, the ``molecular ridge'' by \citet{Coil2000}.
The HII regions, Sgr~A East A--D (hereafter  we  refer to them as HII-A, HII-B, HII-C, and HII-D as in \citet{Tsuboi2019}, respectively, 
are thought to host a single late O-type or early B-type star  for each  and to  be at the age of $\sim$$10^4$  years, 
 with HII-D being the youngest given its small nebular size \citep{Yusef-Zadeh2010, Mills2011, Tsuboi2019}. 

The CMZ molecular clouds are known to contain strong shock waves \citep[e.g.][]{Tsuboi2012},
 which are  responsible to generate filamentary structures often observed in the Galactic disk 
clouds  \citep{Rathborne2014, Rathborne2015, Bally2014}. 
Similar filamentary structures have been found in the 50~MC \citep{Uehara2017} and G0.253+0.016 \citep{Rathborne2015}. 

\citet{Uehara2017, Uehara2021} identified 27 molecular-cloud filaments in the 50~MC and suggested that  filaments are  ubiquitous also in the molecular clouds  in the CMZ \citep{Andre2010}.
Furthermore,  \citet{Uehara2019} showed that the cloud-cloud collision (CCC) efficiently formed massive bound cores even if the slope of the core mass function (CMF)  was not  greatly altered  by CCC.
 Active star formation  is expected to occur 
 in these cold ($\sim$20~K) cores, including those that created the above-mentioned three CHIIs and one UCHII.  The cores will  then collapse and  evolve to warm ($\sim$100~K) hot molecular cores (HMCs).

In the standard evolutionary scenario of massive stars, 
high-mass starless cores (HMSCs) represent the earliest evolutionary stage of massive star formation \citep{Motte2018}.
In the next stage, high-mass protostellar objects (HMPOs) form in the HMSCs and the HMSCs evolve to HMCs. In the Galactic disk, HMCs have been observed in many molecular emission lines from millimeter to submillimeter wavelengths.
The HMC has diameters $\leq$ 0.1 pc, densities $\geq$ 10$^7$ cm$^{-3}$, and temperatures  $\geq$ 100 K. 
The lifetimes of HMCs are 10$^4$--10$^5$~yr \citep{Herbst2009, Battersby2017}.
The HMC is considered to represent the evolutionary stage in a massive star formation  in which protostars  grow through  active accretion of circumstellar material  \citep[e.g.][]{Kurtz2000,  Beuther2007}. 
Observations with a high angular resolution suggest that some HMCs are heated by embedded sources, which are  usually suspected to be HMPOs or massive young stellar objects (MYSOs) \citep[e.g.][]{Rolffs2011, Serra2012, Sanna2014, Silva2017}.
 As a result, HMPOs produce strong millimeter continuum and mid-infrared emission but no detectable centimeter emission \citep{Sridharan2002}.
 Since the centimeter emission is free-free emission from ionized gas, 
no significant centimeter emission implies that HMPOs have not yet reached the stage in which they produce Lyc photons  and ionize the surrounding material. 
Subsequently, a central massive protostar comes to produce a large quantity of ionizing radiation and to form a hyper-compact HII region (HCHII), which evolves to an HII region after UCHII and CHII \citep{DePree2004}.

Here we focus on the evolutionary stage between the massive molecular cores and UCHIIs. 
It is important for  understanding of  the early stage of massive star formation in the CMZ to look for the objects between the HMSC and UCHII stages. 

In this paper, we describe the radio continuum and spectral line observations of the 50~MC and data reduction in \S~2. 
We show the observational results and identify Hot Molecular Core Candidates (HMCCs) in the cloud in \S~3. 
We discuss results from the continuum and the SO, HC$^{15}$N, and CH$_3$OH data and discuss the role of the HMC in the early stage of massive star formation in the CMZ in \S~ 4.

\section{OBSERVATIONS AND DATA REDUCTION}

The observations of SO ({\it N}$_{\it J}$\,=\,2$_2$\,--\,1$_1$ at 86.09355 GHz), $^{34}$SO ({\it N}$_{\it J}$\,=\,2$_3$\,--\,1$_2$ at 97.71540\,GHz and {\it N}$_{\it J}$\,=\,5$_4$\,--\,4$_4$ at 96.78176\,GHz), HC$^{15}$N ({{\it J}=1$-$0} at 86.05497\, GHz), CH$_3$OH ({\it J$_{K_a, K_c}$}=2$_{1,1}$--1$_{1,0}$ E at 96.75551\,GHz), and H42${\rm \alpha }$ at 85.68818\, GHz were made as a part of the Atacama Large Millimeter/Submillmeter Array (ALMA) Cy~1 wide-field observation (2012.1.00080.S. PI M.Tsuboi). 
The observation consisted of a 136-pointing mosaic with the 12-m array and a 68-pointing mosaic  with the 7m array (ACA), 
covering a total of 330\arcsec\ $\times$ 330\arcsec\ area including both the 50~MC and Sgr A$^\ast$. 

The  frequency  channel  width   was  244  kHz,  corresponding to the  velocity  resolution   of  1.7  \,km\,s$^{-1}$ (488  kHz). 
The objects J0006$-$0623, J1517$-$2422, J717$-$3342, J1733$-$1304, J1743$-$3058, J1744$-$3116, and J2148+0657 were used as the phase calibrators.
The flux density scale was determined using Titan, Neptune, and Mars. 
We reduced the data using the standard packages of CASA  \citep{McMullin2007}.
The line emissions were separated from the continuum emission in the UV data using the CASA task UVCONTSUB.
For the line emissions, the UV data for each channel was CLEANed and Fourier-transformed to a map, and all the resultant maps were combined to three-dimensional data cubes in the right-ascension, declination, and frequency space.
The final images were made by applying natural weighting for the visibility (UV) data to obtain a better signal-to-noise ratio. The resultant synthesized beam size  was 2\farcs49$\,\times\,$1\farcs85  (PA=$-$89\fdg 70) for the continuum and SO, $^{34}$SO, and HC$^{15}$N spectral line images.
The CH$_3$OH lines  were detected in the other sideband. 
The synthesized beam size of the CH$_3$OH maps  was 2\farcs30$\,\times\,$1\farcs66  (PA=$-$86\fdg 39).
The typical 3$\sigma$ rms noise level  was 1.0 mJy\,beam\,$^{-1}$,\, or 35.4~mK, in all the maps.  The line profiles of a channel were integrated over 2 \,km\,s$^{-1}$ at both bands.

We  adopt 8.5 kpc as the distance to the Galactic center;  24\arcsec\  corresponds to about 1 pc at  the distance,  and thus our beam size corresponds to about 0.1 pc.
The field center  was ${\rm \alpha (J2000)}$=17$^{\rm h}$45$^{\rm m}$52${\fs}$0, ${\rm \delta (J2000)}$=$-$28$^\circ$59$'$30$\farcs$0.

\section{RESULTS AND HMCC IDENTIFICATION }
\subsection{86~GHz Continuum Emission}

\begin{figure*}[htbp]
\includegraphics*[bb= 0 0 800 600, scale=0.6]{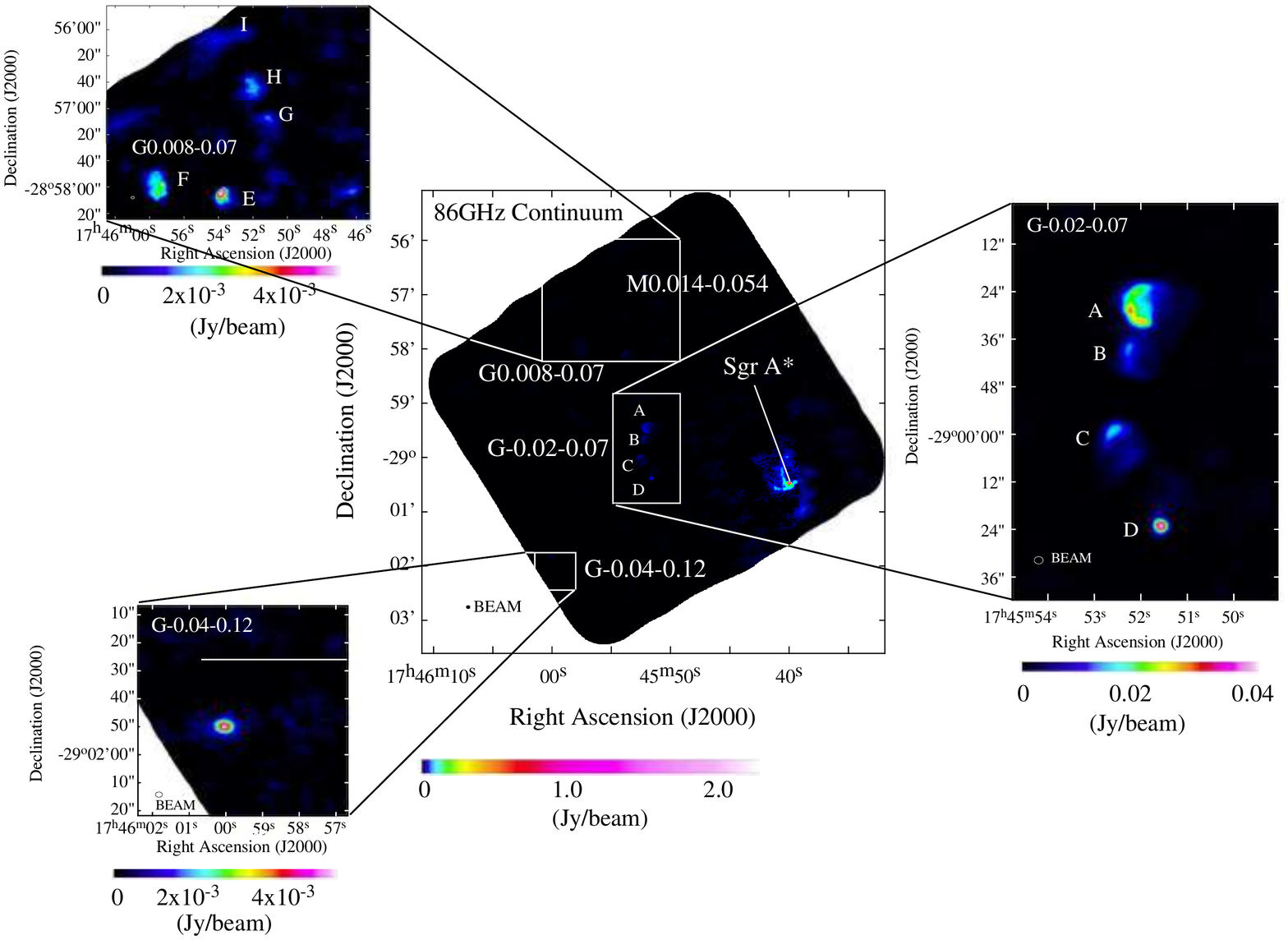}
\caption{(Center) 86-GHz continuum map of the observed area, together with the zoomed-up maps for the selected 3 locations: 
(Right)  8.4-GHz sources A, B, C, and D  \citep{Mills2011},
(Upper left) G0.008$-$0.07,
(Lower left)  G$-$0.04$-$0.12.
The intensity scale is indicated in the color bar below each panel. The angular resolution  is 2\farcs49$\,\times\,$1\farcs85  (PA=$-$89\fdg 70).
The beam size is indicated at the bottom left corner of each panel.
\label{Fig01}}
\end{figure*}

Figure\ref{Fig01} shows the 86-GHz continuum image of the 50~MC region (see also Figure 1 in \citet{Tsuboi2019}).
 Although a 96-GHz continuum image was obtained simultaneously, 
 we  do not use it in this continuum analysis because  it is essentially the same as the 86-GHz image, which alone  provides the sufficient sensitivity.
The sources HII-A, HII-B, HII-C, and HII-D (sources A, B, C, and D), which are prominent  in the centimeter continuum maps \citep[][]{Mills2011}, are also  prominent in the 86-GHz continuum image. 
The known centimeter-continuum faint sources in the region, G0.008$-$0.07 (sources E and F) and G$-$0.04$-$0.12 \citep[e.g.][]{Mills2011}, are also clearly detected.

\begin{figure*}[htbp]
\includegraphics*[bb= 0 450 550 800, scale=0.9]{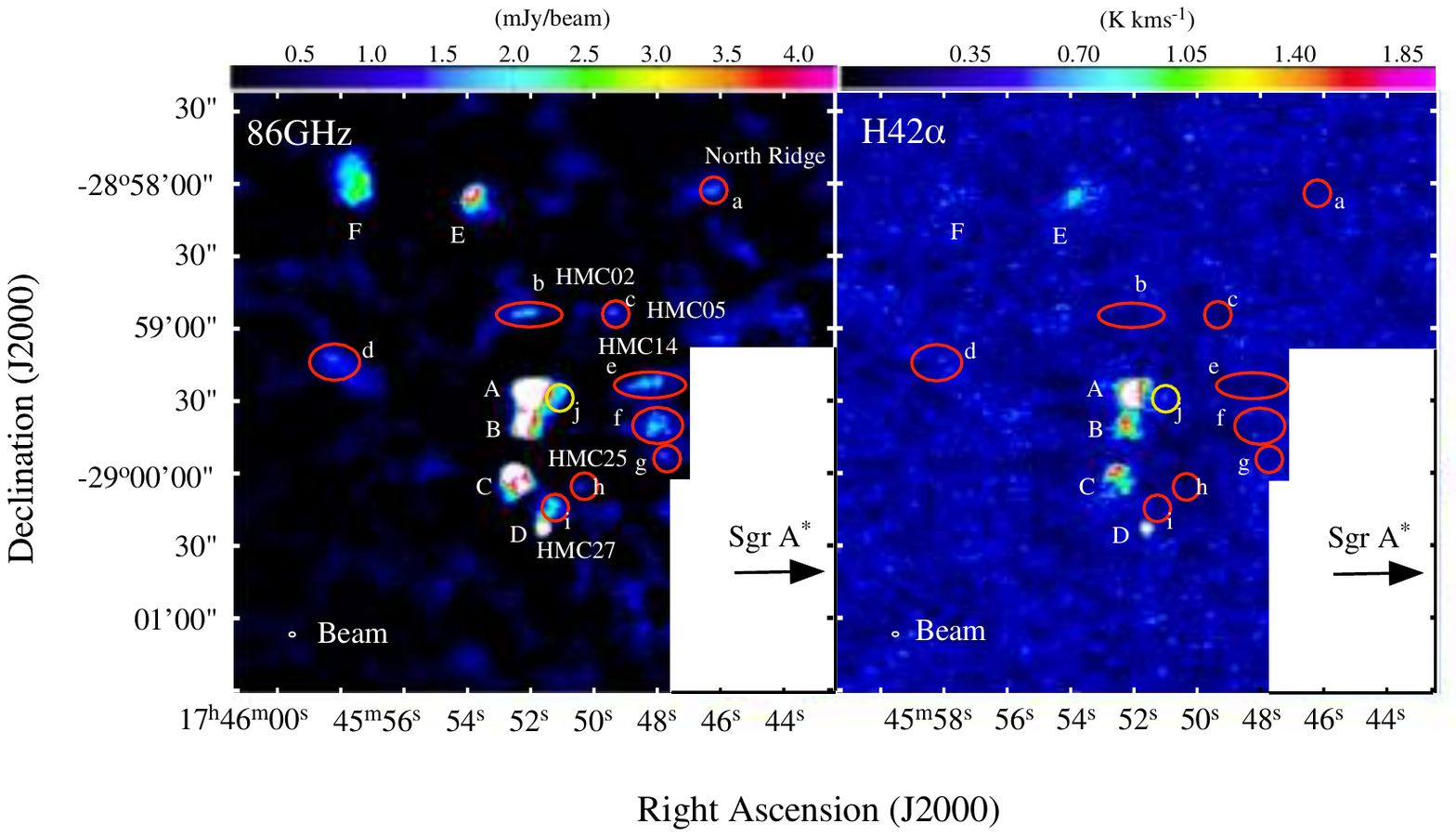}
\caption{ Observed (Left panel) 86-GHz continuum  and  (Right panel) H42$\alpha$ integrated maps, together with the intensity color-scale bars at the top of each panel. 
 The  prominent 8.4-GHz sources A, B, C, D, E, and F  \citep{Mills2011} are labeled.  The known sources a--i  in dust-continuum emission  
 and H42$\alpha$  source~j are indicated with red and yellow circles, respectively. 
The angular resolution  is 2\farcs49$\,\times\,$1\farcs85  (PA=$-$89\fdg 70).
The beam size is indicated at the bottom left corner of each panel.
\label{Fig02}}
\end{figure*}

Although the millimeter continuum emission is considered to  mainly originate from  ionized gas through the free-free emission mechanism, 
 a possibility of non-thermal origin characteristic in the Galactic center,  e.g., emission related with Sgr~A*, is not totally excluded. 
A reliable probe to distinguish the thermal and non-thermal origins is  hydrogen recombination lines; if they are detected, the millimeter continuum emission from the source is likely to be in thermal origin. 
In this case, one of the hydrogen recombination lines, the H42${\rm \alpha}$  line, has been detected toward  HIIs A--D, G0.008$-$0.07, and G$-$0.04$-$0.12 \citep[e.g.][]{Mills2011, Tsuboi2019}. Therefore, their emission is likely to be in thermal origin. 
Figure~\ref{Fig02} shows the distributions of the 86-GHz continuum  and H42${\rm \alpha}$ line  emissions. 
Although nine dust cores are detected  in the 86-GHz continuum emission (sources a--i in Figure~\ref{Fig02} Left panel; n.b., these sources are not detected in the H42${\rm \alpha}$ recombination line in Figure~\ref{Fig02} Right panel). 
A similar situation  has been also  reported in M0.014$-$0.054 \citep[][]{Tsuboi2021}. 
In addition,  \citet{Walker2018} have detected massive dust cores,  which  will eventually grow to HCHIIs, in the CMZ's dust ridge.

Source a corresponds to the HC$^{15}$N core located  45$\arcsec$ northeast of ``Northern Ridge'', which will be discussed in section 3.4.1. 
Sources b, c, e, h, and i are associated with HMCCs as mentioned later. 
Sources d, f, and g have no corresponding HMCCs.

 By contrast,  source j is  detected also in the H42${\rm \alpha }$ recombination line (see the right panel of Figure~\ref{Fig02}). Hence,  source j  should have ionized gas and be 
  regarded as a HCHII candidate.
 Table~\ref{Table01} lists these sources. 

\begin{figure}[htbp]
\includegraphics*[bb= 0 150 600 700, scale=0.3]{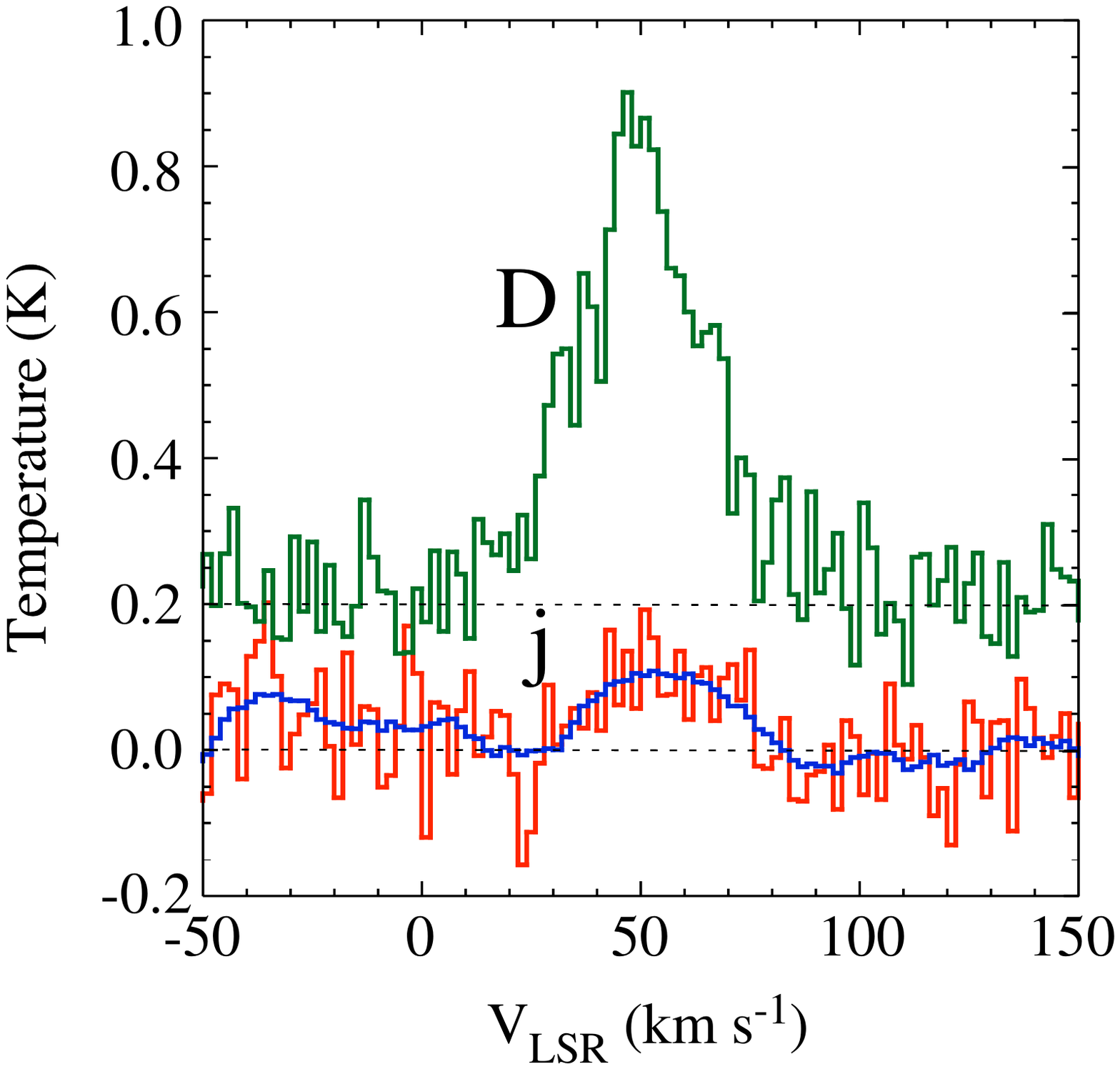}
\caption{ H42$\alpha$ line profile toward (green) HII~D and (red) source~j in Figure~\ref{Fig02}.
 The 11~ch smoothed intensities are overlaid on the latter in blue.
\label{Fig03}}
\end{figure}

Figure~\ref{Fig03}  shows the H42$\alpha$ recombination-line spectra toward  HII-D  and source j. 
The spectrum toward  HII-D has a single peak at around the LSR velocity of 50~km\,s$^{-1}$.
The spectrum toward source j also has a marginal peak at around the LSR velocity of 50~km\,s$^{-1}$.
The blue curve in the figure shows the 11-ch running mean of the spectrum. 
The peak is identified in the mean spectrum. 
The radio recombination lines from the HCHIIs are known to be extremely broad, typically $\Delta$V =40--50~km\,s$^{-1}$ and sometimes greater than 100~km\,s$^{-1}$\citep[e.g.][]{Sewilo2004, Sewilo2011}, and  the width tends to be broader at  lower-frequency transitions. 
However, the observed velocity width of the H42$\alpha$ recombination line of source j is only 30~km\,s$^{-1}$, which is greatly narrower than 
 those of typical HCHIIs. In addition, the location of source j is adjacent to  HII A. 
Therefore, we conclude that source j is a part of the shell-like structure of  HII-A.

If the UCHIIs and HCHIIs exist in the 50~MC, 
their flux densities are expected to be  200--700 mJy  on the basis of the typical flux density of the HCHIIs observed in the Galactic disk at 43 GHz \citep{Sewilo2011}. 
 The observed flux densities of  HII A--D  at 86~GHz were  426, 141, 172, and 90~mJy, respectively \citep{Tsuboi2019}.
Combining with the facts that their continuum emissions at 86 GHz are optically thin and have flat spectrum indexes ($\sim$$-$0.1) because they are free-free emission, 
 we  should be able to detect them at a similar intensity if the UCHIIs and HCHIIs exist in the 50~MC.  
However, we detected none.  Therefore, we conclude that no new UCHIIs  or HCHIIs exist in this region.

\begin{table*}[ht]

\begin{minipage}{\textwidth}
\caption{Hot Molecular Core Candidates  in the 50~MC selected  on the basis of the 86-GHz continuum and H42$\alpha$ line emissions}
\label{Table01} 
\begin{center}
\scalebox{0.65}[0.65] 
{

\begin{tabular}{lccccccccl}
\hline\hline
86GHz & \multicolumn{2}{c}{Position}    & Size & Flux             &Gas Mass                         & Gas Mass  & Mean number&  Mean number        & Note    \\
 Continuum    & Right Ascension & Declination &  Max $\times$ Min   & & T$\rm{_d}$=50K &T$\rm{_d}$=100K  &density T$\rm{_d}$=50K  &density T$\rm{_d}$=100K & \\
\#    & J(2000)              &  J(2000)     &  arcsec ($\arcsec$)      & (mJy) & M$_{dust}$  (M$_\odot$) & M$_{dust}$  (M$_\odot$) &  (cm$^{-3}$)&  (cm$^{-3}$)   &\\    
\hline

a & 17$^{\rm h}$45$^{\rm m}$46${\fs}$29 &-28$^\circ$58$'$03$\farcs$32 & 6.16 $\times$ 3.23 & 2.92& 3.52 $\times$ 10$^2$ & 1.80 $\times$ 10$^2$ &1.43  $\times$ 10$^6$ &7.31  $\times$ 10$^5$& north of North Ridge \\
b & 17$^{\rm h}$45$^{\rm m}$52${\fs}$12 &-28$^\circ$58$'$53$\farcs$63 & 15.39 $\times$  5.16 & 4.52 & 5.45 $\times$ 10$^3$ & 2.78 $\times$ 10$^3$  & 2.78 $\times$ 10$^6$ & 1.42 $\times$ 10$^6$ &  HMC02, C1P1,C1P2 \citep{Lu2019a} \\\
c & 17$^{\rm h}$45$^{\rm m}$49${\fs}$39 &-28$^\circ$58$'$53$\farcs$39 & 6.25 $\times$ 4.49 & 7,26&  8.76  $\times$ 10$^2$ &  4.47  $\times$ 10$^2$     &2.13 $\times$ 10$^6$ &1.09 $\times$ 10$^6$ & HMC05 \\
d & 17$^{\rm h}$45$^{\rm m}$58${\fs}$24 &-28$^\circ$59$'$12$\farcs$75 & 9.83 $\times$ 3.98 & 5.55 & 6.70 $\times$ 10$^2$ & 3.42 $\times$ 10$^2$    & 9.88 $\times$ 10$^5$ & 5.04 $\times$ 10$^5$ & \\
e & 17$^{\rm h}$45$^{\rm m}$48${\fs}$08 &-28$^\circ$59$'$22$\farcs$59 & 9.63 $\times$ 5.79 & 19.13 & 2.52 $\times$ 10$^2$ & 1.29 $\times$ 10$^2$   & 3.17 $\times$ 10$^6$ & 1.62 $\times$ 10$^6$ & HMC14\\
f & 17$^{\rm h}$45$^{\rm m}$48${\fs}$05 &-28$^\circ$59$'$40$\farcs$91 & 7.75 $\times$ 4.69 & 10.08  & 2.31 $\times$ 10$^3$ & 1.18 $\times$ 10$^3$     & 2.00 $\times$ 10$^6$ & 1.02 $\times$ 10$^6$& \\
g & 17$^{\rm h}$45$^{\rm m}$47${\fs}$81 &-28$^\circ$59$'$52$\farcs$50 & 7.28 $\times$ 4.37 & 8.18   & 1.22 $\times$ 10$^3$ & 6.21 $\times$ 10$^2$ & 2.00 $\times$ 10$^6$ & 1.02 $\times$ 10$^6$& \\
h & 17$^{\rm h}$45$^{\rm m}$50${\fs}$45 &-29$^\circ$00$'$05$\farcs$83 & 2.96 $\times$2.32 & 1.42  & 9.87 $\times$ 10$^2$ & 5.04 $\times$ 10$^2$    &1.98 $\times$ 10$^6$ &1.01 $\times$ 10$^6$ & HMC24 or HMC25 \\
i & 17$^{\rm h}$45$^{\rm m}$51${\fs}$34 &-29$^\circ$00$'$15$\farcs$20 & 9.44 $\times$ 4.80 & 13.06  & 1.71 $\times$ 10$^2$ & 8.73 $\times$ 10$^1$  & 3.43 $\times$ 10$^6$ & 1.75 $\times$ 10$^6$ & HMC27\\
j & 17$^{\rm h}$45$^{\rm m}$51${\fs}$08 &-28$^\circ$59$'$27$\farcs$82 & 2.73 $\times$ 1.56 & 2.64  & 1.58 $\times$ 10$^3$ & 8.04 $\times$ 10$^2$ & 1.30 $\times$ 10$^6$ & 9.54 $\times$ 10$^5$ &  north of HMC15 \\

\hline
H42$\alpha$ &&&  & (mJy\,km\ s$^{-1}$)   &&&&\\
HCHII & 17$^{\rm h}$45$^{\rm m}$51${\fs}$02 &-28$^\circ$59$'$29$\farcs$48 & 2.16 $\times$ 1.58 & 7.48 &--- &--- &---
&--- &source~j H42$\alpha$ emission\\

\hline
\vspace{3mm}

\end{tabular}
}
\\
\end{center}
\end{minipage}
\end{table*}


\subsection{Molecular Line Emission as HMC Tracers}

The HMC is generally characterized by a high gas temperature exceeding 100 K and rich chemistry observable in molecular emission lines in mm and sub-mm wavelengths.
Molecular emission lines such as SO, SO$_2$, CH$_3$OH, and CH$_3$CN  are often detected  in the spectra of HMCs. 
In the 50~MC, the SO  ({\it N}$_{\it J}$\,=\,2$_2$\,-\,1$_1$) emission line is clearly detected 
in  spectra of the HMCCs. 
We find that although the {\it N}$_{\it J}$\,=\,2$_3$\,-\,1$_2$ emission line of $^{34}$SO  is also detected, 
 the highly excited  {\it N}$_{\it J}$\,=\,5$_4$\,-\,4$_4$ emission line is not detected with a significance of 3$\sigma$.
 Then, we  further analyze the SO molecular line  in order to obtain the spatial and velocity distributions of the molecular gas in the HMCCs. 

The CH$_3$OH lines  can also be used as tracers of HMCs. 
We detected six lines of  CH$_3$OH in the field of the 50~MC; among them, we here focus on {\it J$_{K_a, K_c}$}=2$_{1,1}$-1$_{1,0}$ A$^{--}$ at 96.75828\,GHz, which is not blended with other lines above the  3$\sigma$ noise level. 
 Four other lines of  CH$_3$OH, i.e., ({\it J$_{K_a, K_c}$}=2$_{-1,2}$-1$_{-1,1}$ E: 96.73936\,GHz), ({\it J$_{K_a, K_c}$}=2$_{0,2}$-1$_{0,1} $A$^+$: 96.74138\,GHz),  ({\it J$_{K_a, K_c}$}=2$_{0,2}$-1$_{0,1} $E: 96.74455\,GHz), and ({\it J$_{K_a, K_c}$}=2$_{1,1}$-1$_{1,0} $E: 96.75551\,GHz),  are blended with each other and hence are not used in our analysis. 
The other line, CH$_3$OH ({\it J$_{K_a, K_c}$}=6$_{ -2, 5}$- 7$_{ -1, 7}$E: 85.56808 GHz),  is detected  as an absorption line. It  is known to become  a Class-II maser line in some conditions \citep{Cragg2005} and hence is ignored here.
 The observed frequency bands include several  {\it v$_t$}=1 high-excitation transitions of  the above-mentioned CH$_3$OH lines, which would be emitted from  the cores with a high gas temperature \citep[e.g.][]{Barnes2019}. However, they are not detected in  our observed dust cores and HMCCs.
 
The other line of our interest is the HC$^{15}$N emission line. It is optically thin,  has a high critical density of $\sim$10$^7$\, cm$^{-3}$  \citep[e.g.][]{Rolffs2011}, 
 and  has been used as a probe for  high-density and warm cores. 
 \citet{Boonman2001} reported that the HCN emission line is enhanced in the dense regions that  are at the stage evolving from gravitationally bound cores to  HMCs. 
Moreover, \citet{Stephan2018} modeled the spatio-temporal evolution of the chemistry of HMCs and their simulation showed   significant HC$^{15}$N and CH$_3$OH line emissions from warm core.
Line emissions from HMCs are not decreased  during  the evolution of the cores. 
 Since the abundance ratio of $^{14}$N/$^{15}$N  varies greatly between 70 and more than 1000 as in prestellar cores \citep[e.g.][]{Ikeda2002},  
the HC$^{15}$N emission often becomes weak. 
The line profiles of HMCCs at the SO emission line are similar to the HC$^{15}$N emission line\citep{Schilke2001}. 
 Consequently, HC$^{15}$N is a good tracer of  a dense and warm core. 

Many theoretical and observational studies have been conducted on HMCs;
 they are generally categorized into two models: a core model with a dust sublimation zone  in the vicinity of the central source (Central Source Model: CSM) and  another core model in which  some external source warms the core (External Source model: ESM) \citep[e.g.][]{Kauffman1998}.
In the HMC in the CSM,  HMPO or earlier HCHII is placed in the center.
The ESM is often  preferred with observational results of HMCs with UCHIIs.
In the ESM model, 
HMCs may be formed in situations where some shock compresses the molecular gas, resulting in  gravitational collapse of the dense core. 
 \citet[][]{Nomura2004} studied the evolution of the molecular abundance in HMCs and the abundance ratio of the radius.
SO molecules generally increase in abundance as approach the central source.
It is  known that the abundance  of SO is a decreasing function of  the radius.
The abundances of SO and CH$_3$OH are enhanced by ice evaporation and shock, both of which are triggered  by molecular outflows \citep[e.g.][]{Tak2003}.
Then,  the intensity of the emission line may vary  even  when the  temperature is the same.
Specifically, in the CMZ, the abundance  of SO may have increased due to other shocks.
Hence,  a HMC can be present near the peak, and a weak filamentary structure may  be observed in the vicinity.
We should distinguish the SO emission of HMCCs from others such as turbulence and shock caused by CCC. 

 \citet{Tsuboi2015a} found  a half shell-like feature with a high temperature ratio of T(SiO)/T(H$^{13}$CO$^+$) in the 50~MC. 
 Given that the abundance of  SiO  is increased by a C-shock in molecular clouds  whereas that of  H$^{13}$CO$^+$  is not affected by the shock \citep[e.g.][]{Amo2011}, 
the feature would be evidence of shock wave propagation in the cloud. 
\citet{Barnes2019}  reported signs of embedded star formation in  Clouds~D and E/F of a part of the CMZ
and detected the SO  ({\it N}$_{\it J}$\,=\,3$_4$\,-\,4$_3$) emission, notably from  Cloud~E.
 Their images were compact  to the extent that the target was not sufficiently resolved  with the spatial resolution  of their observation of  Cloud~E  (1\farcs27$\,\times\,$0\farcs90  (PA=0\fdg 0)),  whereas the peak in their images of SO ({\it N}$_{\it J}$\,=\,5$_6$\,--\,4$_5$)  was only moderately compact with weak environment emission.
 \citet{Barnes2019} suggested that the cores in  Clouds~D and E/F  had evolved both physically and chemically and that 
 molecules such as SO  had probably originated from regions that harbored embedded star formation (e.g., due to strong shock).

Therefore, comparisons among the SO, CH$_3$OH and HC$^{15}$N lines, which will be discussed later, 
 can provide clear indication of whether the HMCCs have denser and warmer conditions than the cores observed with the CS and H$^{13}$CO$^+$ emissions.

\subsection{Identification of HMCCs}
\subsubsection{HMCCs}
\begin{subfigures}

Figures~\ref{Fig04a}a and \ref{Fig04b}b show the velocity channel maps of the SO emission superimposed on the 86-GHz continuum map. 
Each channel map is integrated over 2.0\,km\,s$^{-1}$ width centered on the  velocity indicated in each panel of the figures. 
We identify many filament-like structures  over  a velocity range of {\it V}$_{\rm LSR}$=$-$14 to 84 \,km\,s$^{-1}$ in the SO emission line. 
There are many various peaks with small sizes and broad velocity widths in the filaments.
Many of the peaks  appear to be connected by weak bridge components in both spatial and velocity domains. 
HMCs are thought to have a tendency to be located at such peaks and to be buried into the surrounding cool static ambient gas.
In order to identify HMCCs,  it is necessary to distinguish between warm HMCCs and cold envelopes,  using their line profiles.


\begin{figure*}[t]
\includegraphics*[bb= 0 100 700 800, scale=0.75]{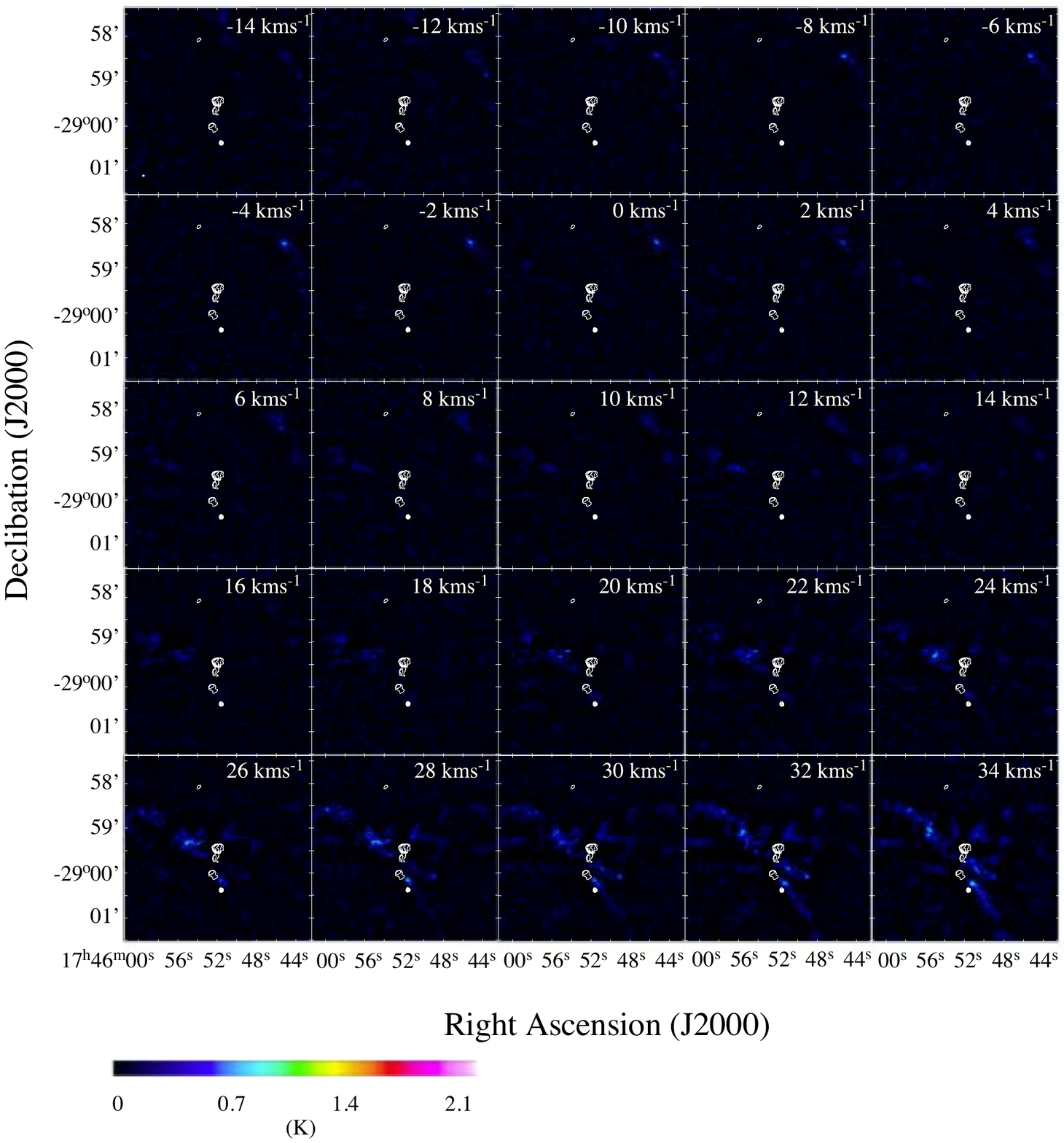}
\label{Fig04a}
\caption{SO velocity-divided color-image maps with  a uniformly-weighted angular resolution of 2\farcs49$\,\times\,$1\farcs85  (PA=$-$89\fdg 70).
The center velocity in V$_{LSR}$ (km\,s$^{-1}$)  is shown  at the upper-right corner of each map.
The peak specific-intensity is 2.25~K.
The typical  rms noise level is 35.4~mK.
The 86-GHz continuum image is  overlaid with  linearly-spaced white line contours with  the levels from  10\% to 90\%  with intervals of 10~points.
}
\end{figure*}
\begin{figure*}[t]
\includegraphics*[bb= 0 100 700 800, scale=0.75]{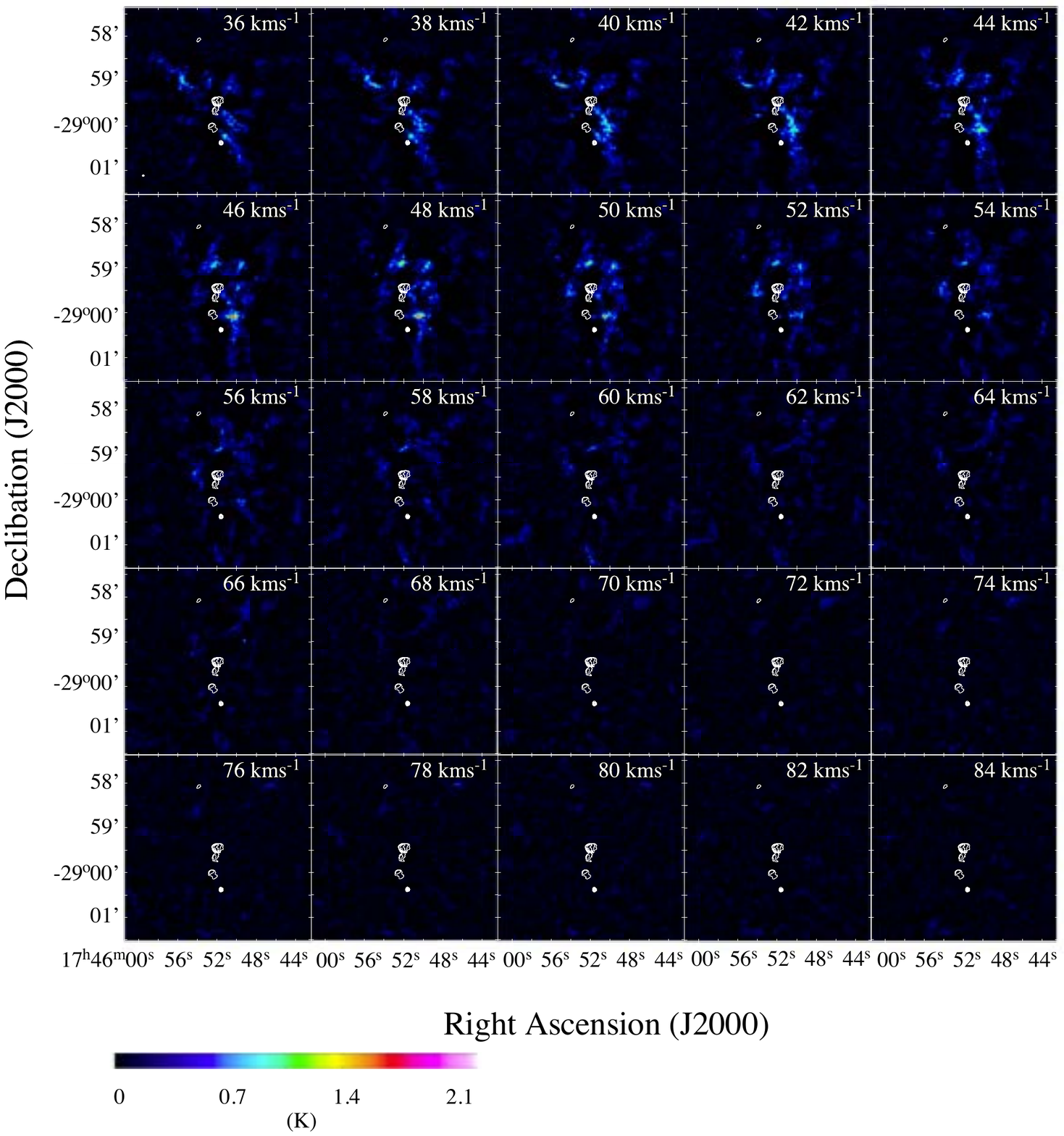}
\label{Fig04b}
\caption{continued}
\end{figure*}
\end{subfigures}

In the typical HMC object Orion KL, 
the line profile of the SO emission line consists of  distinct two components, ``hot core'' and  ``plateau'' \citep[e.g.][]{Plambeck1982, Wright1996}, in addition to
 considerably weaker components, ``compact ridge'' and ``extended ridge'', whose  contributions  are secondary or less and are usually ignored. 
 The ``hot core'' and  ``plateau'' components are generally considered to correspond to the HMC and cold envelope, respectively \citep[e.g.][]{Wright1996}.
  \citet{Tak2003} showed that  the SO ({\it N}$_{\it J}$\,=\,6$_6$\,--\,5$_5$ and 8$_7$\,--\,7$_6$) emission profiles  from the regions of HMCs had high and low-velocity components and suggested that 
 they correspond to, respectively, the ``hot core'' and  ``plateau'' mentioned above.
Assuming that the SO emission line profiles   of HMCs generally have the ``hot core'' and  ``plateau'',  
we  divide the observed  line profiles  of the regions of HMCCs into the HMC and cold envelope components including filaments in this section.
We note that
the SO emission from filaments, which appear at a velocity band from V$_{LSR}$=30~km\,s$^{-1}$ to 60~km\,s$^{-1}$  (Figure~\ref{Fig04a}),
 is mostly weak and hence that
 the HMC components are detectable in  high-intensity regions only.

\subsubsection{Identification by ``clumpfind'' and Visual  Inspection}

 One of the popular algorithms to identify HMCCs  is ``clumpfind''  \citep{Williams1994}, which we adopt in this section.
We  first validate how suitable ``clumpfind'' is  for our purpose.
The ``clumpfind'' decomposes structures  with a set of elliptical Gaussians. 
Although the SO emission may not  be strictly a combination of Gaussian structures,
 the structure of the line from the HMCCs  will not  be anyhow resolved with a sufficiently high resolution in our data, given that  our beam is only slightly smaller than the spatial extent of the HMCCs.

\citet{Li2020}  recently made a quantitative comparison of the performance of popular algorithms for molecular-gas-clump identification, including  GaussClumps \citep{Stutzki1990}, clumpfind, Fellwalker \citep{Berry2015}, Reinhold developed by Kim Reinhold, and Dendrograms \citep{Rosolowsky2008}.
Designing simulated clumps of various sizes, peak brightness, and crowdedness,  
\citet{Li2020}  concluded that  Fellwalker, Dendrograms, and GaussClumps performed better with regard to detection completeness  and also found that the average deviations in clump parameters  gradually increase with any of the algorithms as the size and Signal-to-Noise Ratio of clumps increase. 
\citet{Li2020} also showed that ``clumpfind'' identified the cores with sufficient precision, although the score on some tests  with ``clumpfind'' is not better than the others.
They suggested that it was difficult to identify cores in an automated way.  
Bearing in mind the points raised by \citet{Li2020}, we manually identify  HMCCs from the searched HMCC1s (only idenfied by ``clumpfind''), using ``clumpfind'',   in the following procedures. 

\begin{enumerate}

\item To find  HMCCs, 
we use the ``clumpfind'' software  to identify  clumps on the SO, $^{34}$SO,  HC$^{15}$N, and CH$_3$OH emission maps.
We use 20$\sigma$ and 10$\sigma$ as the lowest contour (threshold) and contour spacing, respectively, for the first three emission lines, 
 and 40$\sigma$ and  20$\sigma$, respectively, for the CH$_3$OH emission line.
The 1$\sigma$ level  is  35.4 mK in any of the maps.
These  parameters are selected in such a way that it would be easier  to distinguish the static gas  from the low and high-velocity
components  of the HMCs  with the criteria of a small radius and weak intensity \citep{Tak2003}.
Moreover,  HMCC1s are identified  with criteria of small radius  (r $<$ 10$\arcsec$) and weak intensity (T $>$ 700~mK).
The derived parameters of the HMCC1s are summarized in Table~\ref{Table02}.

\item If one HMCC1 is located within 5$\arcsec$of another HMCC1 and the velocity difference between the two HMCC1s is smaller than 10.0 km\,s$^{-1}$, we  regard the two HMCC1s as the same HMCC1.
\item The identified HMCC1s are classified into the three groups of isolated HMCC1s, HMCC2s, and HMCC3s.
The HMCC2 has two velocity components with similar isolated positions, and the HMCC3 has three or more velocity peaks \citep[e.g.][]{Serra2012}.
 A HMCC3 would be more reliable as a  HMCC than  an isolated HMCC1 and  HMCC2 because  the spectra of HMCCs have been reported to  have usually several peaks in the molecular emission lines, e.g., SO and CH$_3$OH ones.
 Thus, we first identify HMCCs  in the regions of the HMCC3s and  then examine the HMCC2s and  HMCC1s, using ``clumpfind''.  Figure~\ref{Fig05} schematically illustrates how HMCC1s are identified  using ``clumpfind'' \citep{Williams1994} on the SO emission maps with a certain threshold. 
Scattering HMCC1 is assumed to be due to physical and chemical reaction in the HMC  because \citet{Serra2012} showed that there are two chemical groups (Type II and Type III) in the SO and CH$_3$OH distributions.

\begin{figure}[htbp]
\includegraphics*[bb= 65 300 500 650, scale=0.45]{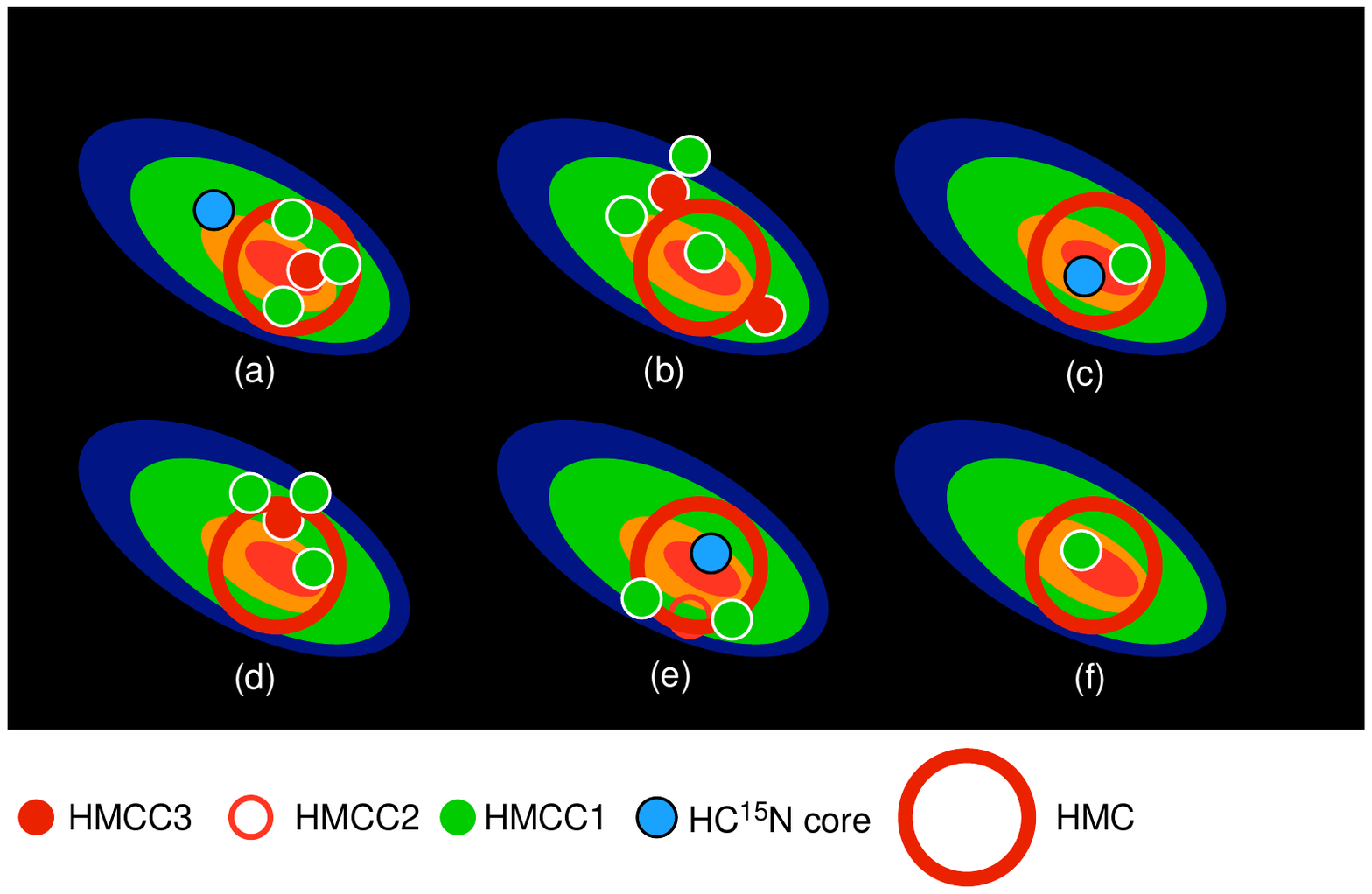}
\caption{Identification of HMCCs.
 HMCC1s are identified  using ``clumpfind'' \citep{Williams1994} on the SO, $^{34}$SO, and HC$^{15}$N, and CH$_3$OH emission maps with a certain threshold for each line.
If one HMCC1 is located within 5$\farcs$00 of another HMCC1 and the velocity difference between the two HMCC1s is smaller than 10.0 km\,s$^{-1}$, we  regard the two HMCC1s as the same HMCC1.
 A HMCC2 has two velocity components with similar positions, and  a HMCC3 has three or more velocity peaks.
HMCCs are mainly identified  into one of the following six types: 
(a) Peak in the integrated intensity map  having a HMCC3, 
(b) Peak located between two HMCC3s or more than three HMCC3s,
(c) Peak having both HMCC1 and  HC$^{15}$N cores,
(d) Peak located near HMCC3,
(e) Peak having a HC$^{15}$N core and located near HMCC2,
(f) Peak  identified as a HMCC1  with  visual  inspection.

\label{Fig05}}
\end{figure}


Tables~\ref{Table02} and \ref{Table03} list the identified HMCCs. 
The peak positions  of the SO, $^{34}$SO, HC$^{15}$N, and CH$_3$OH emission lines agree with  one another  for some but not all HMCC1s.
The  peak positions of the $^{34}$SO emission line of the HMCC3s are well correlated to those  of the SO emission line.
 By contrast, those  of the HC$^{15}$N and CH$_3$OH emission lines of the HMCC1s do not  show a good correlation with those  of the SO emission line. 
The positions of the HMCCs cannot be  precisely determined within the uncertainty of 5$\arcsec$ because of the differences in the peak locations of the observed emission lines.
To identify HMCCs  definitively,  visual confirmation is necessary.

\item Although mass and size measurements  with ``clumpfind'' are reported to be broadly consistent with those  with other methods in general \citep[e.g.][]{Kauffmann2010a, Kauffmann2010b}, those  with ``clumpfind'' in our work  turns out not to be sufficiently  so \citep{Pineda2009}. 
 Since our results show several peaks in the HMCCs,  the masses of the entire HMCCs are not able to be estimated  with the results  with ``clumpfind''.
We  compile  the final list of the HMCCs by  selecting visually the ones  for which a separation between the peaks of emission lines is  smaller than 5$\arcsec$ in HMCC3s or HMCC2s, using the channel maps and/or  integrated intensity maps averaged over 10 \,km\,s$^{-1}$.
Figure~\ref{Fig06} shows a peak velocity map of the SO emission.
At 15$\sigma$ or higher, 
the distribution of the SO molecular emission is limited, 
and a considerable amount of filament structure can be eliminated. 
These are considered to indicate the HMC and its envelope component. 
The identified HMCCs are mainly classified according to their locations and velocities out of  the eight regions named ``Northern Ridge'', ``North'', ``Northeast'', ``Northwest'', ``East'', ``West'', ``Southeast'', and ``Southwest''.
The HMCC in ``Northern Ridge'' are regarded as typical HMCCs as  described later. 
\item Using the HMC in ``Northern Ridge'' as a template, 
we  identify 28 HMCCs including 19 HMCC3s in the area excluding ``Northern Ridge''. 
They are  referred to as HMCs~01--28 hereafter.
Table~\ref{Table03} tabulates their positions, sizes, peak intensities, peak velocities, line widths, and integrated intensities.
 Note that the sizes are estimated  with 2-dimensional Gaussian fitting with CASA.
The estimation of these physical quantities will be  further investigated in the next section. 
 The fact that the typical size of the HMCCs, r$\sim$0.1~pc or smaller, is as small as the  beam-size, 3$\arcsec$, at the source distance of  our observations implies that the calculated quantities have large  uncertainties.

\end{enumerate}

\begin{figure*}[htbp]
\includegraphics*[bb= 0 300 600 730, scale=0.8]{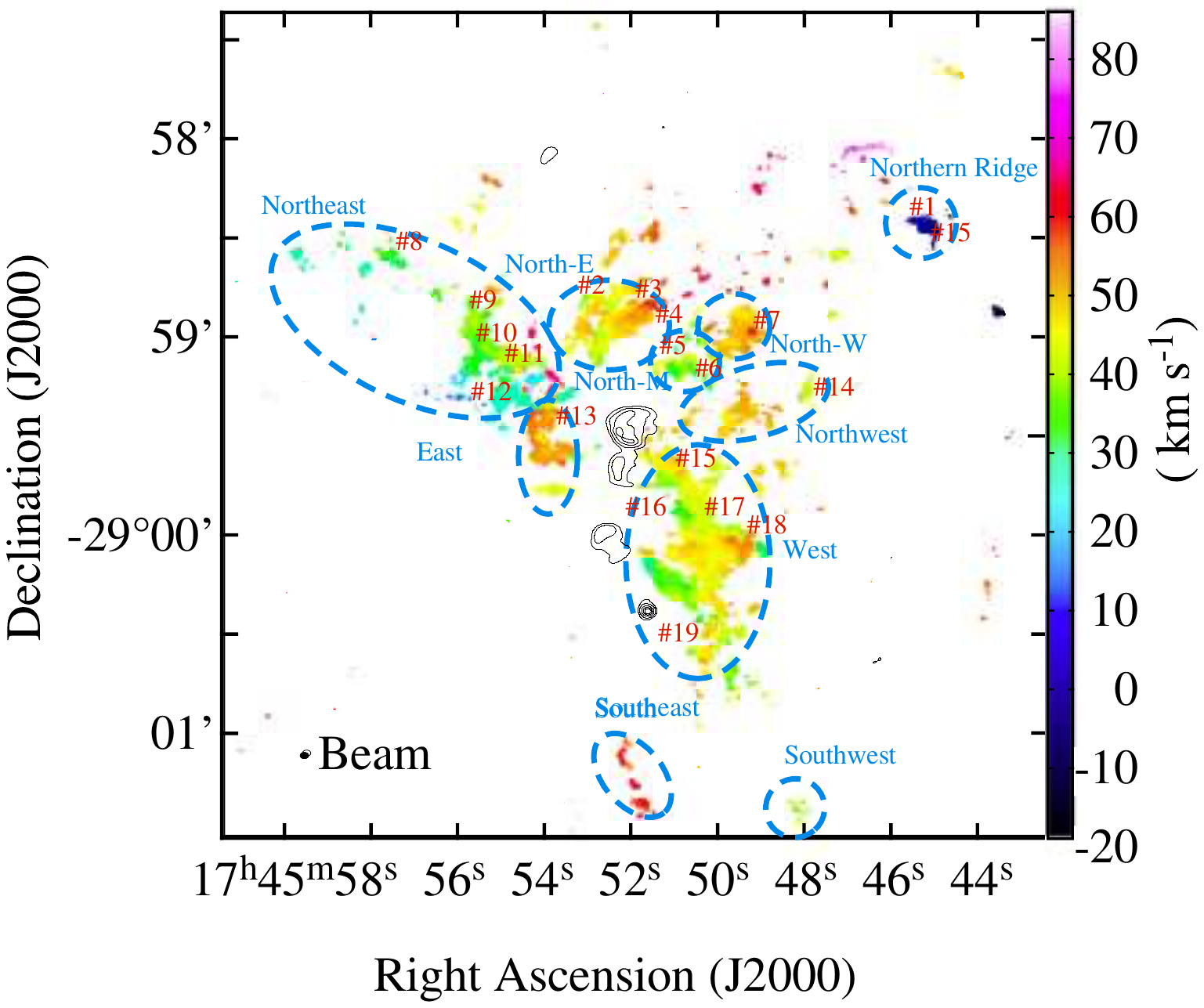}
\caption{Peak velocity shift of the SO emission  from the average of the region. 
Only the regions with a velocity larger than 15$\sigma$  (1$\sigma$  is 35.4~mK) are plotted. 
The 86-GHz continuum map, which has  a uniformly weighted angular resolution of 2\farcs49$\,\times\,$1\farcs85  (PA=$-$89\fdg 70),
 is  overlaid  with linearly-spaced  contours with levels from 10 to 90\%  of the peak 41.1~mJy\,beam$^{-1}$, or 1.23~K, with intervals of 10 points.
Beam size is indicated at the bottom left corner of each panel. 
The identified regions for clusters of the HMCCs and some selected HMCCs are indicated. 
\label{Fig06}}
\end{figure*}

\begin{figure*}
\includegraphics*[bb= -50 230 700 750, scale=0.7]{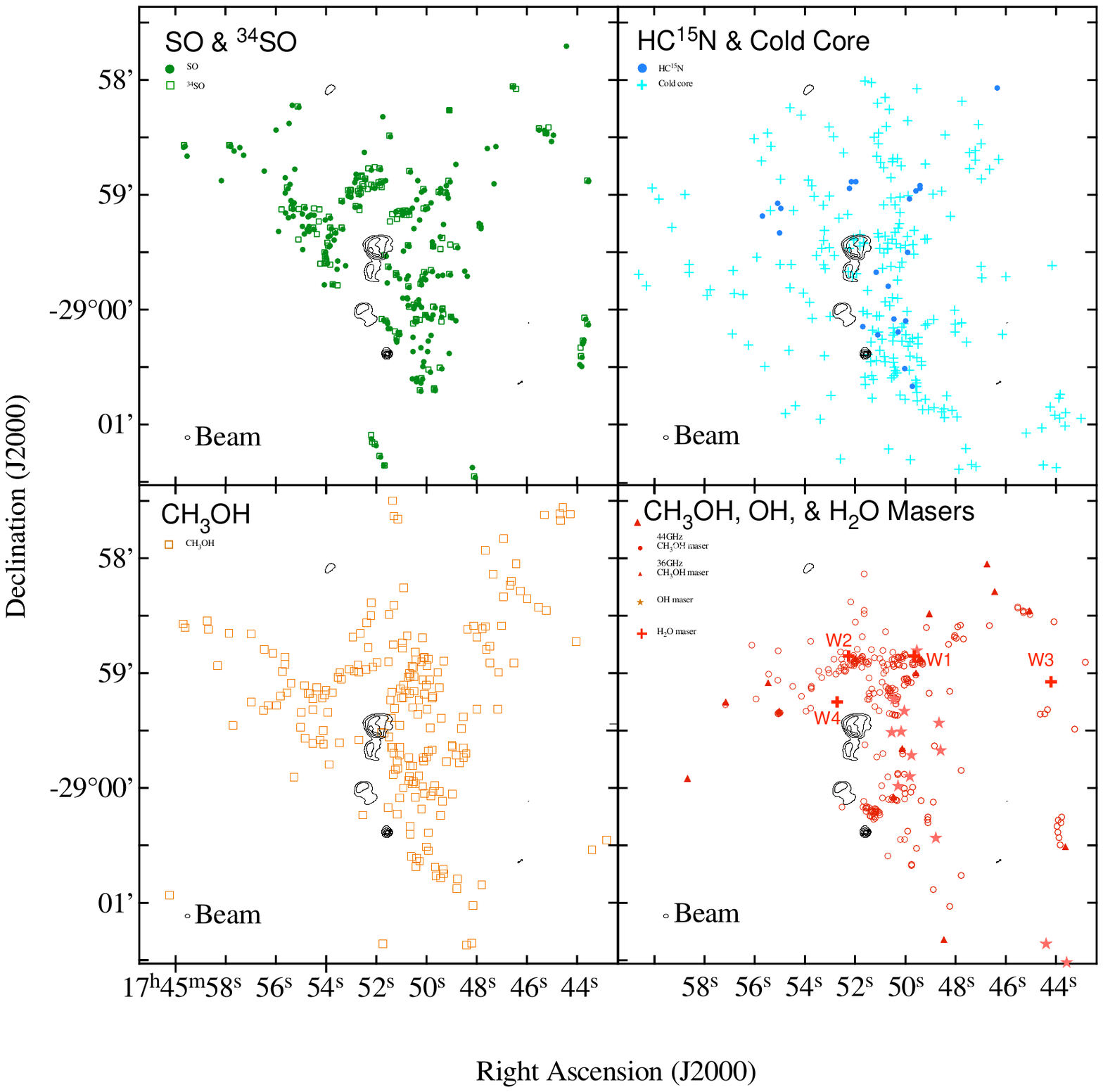}
\caption{
Positions of the the identified HMCC1s, molecular-cloud cores, and some masers superposed with the 86GHz continuum map at a uniformly weighted angular resolution of 2\farcs49$\,\times\,$1\farcs85  (PA=$-$89\fdg 70). 
The first two groups are identified  with the clumpfind algorithm in this work and the last one is taken from literature (see below). 
The  beam size is indicated as an oval on the bottom left corner of each panel. 
The 86-GHz continuum image is  given by the contours with the levels from 10\% to 90\%  of the peak intensity (41.1 mJy beam$^{-1}$, or 1.23 K, in T$_{\rm B}$) with intervals of 10 points.
Upper left: Green filled and open circles indicate the positions of the HMCC1s identified with the SO and $^{34}$SO  emissions, respectively.  
Upper right: Blue filled circles indicate  the positions of dense cores with the HC$^{15}$N emission (this work), whereas 
blue crosses indicate cold molecular cores identified by \citep{Uehara2019}.
Bottom left: Orange open squares indicate  the positions of the HMCC1s identified with the CH$_3$OH emission.
Bottom right: Red open circles, open triangles and red  asterisks indicate  the positions of 44~GHz Class-I CH$_3$OH masers, 36~GHz Class-I CH$_3$OH masers \citep{Yusef-Zadeh2013, McEwen2016}, OH masers \citep{Sjouwerman2008, Pihlstrom2011,Cotton2016}, and H$_2$O masers (W1--W4) \citep{Lu2019a}, respectively.
\label{Fig07}}
\end{figure*}

Figure~\ref{Fig07} shows the positions of the HMCCs identified  with the SO, $^{34}$SO, and CH$_3$OH emission lines superposed with the 86-GHz continuum contour map, for which 
the angular resolution is 2\farcs49$\,\times\,$1\farcs85  (PA=$-$89\fdg 70), where   uniformly weighted visibility data are selected.
The figure also includes   the positions of the dense molecular-cloud cores identified with the HC$^{15}$N emission lines  (our work and that by \citet{Uehara2019}) and those of the Class-I CH$_3$OH masers at 44GHz,  
Class-I CH$_3$OH masers at 36GHz \citep{Yusef-Zadeh2013, McEwen2016}, 
OH masers \citep{Sjouwerman2008, Pihlstrom2011,Cotton2016}, and H$_2$O masers \citep{Lu2019a}.
We find that the spacial distribution of the positions of the HMCC1s identified with the SO and $^{34}$SO emission lines are similar to the results identified by ``clumpfind'' with the HC$^{15}$N and CH$_3$OH line emissions. 

We examine  the identified HMCCs  with the SO emission  with regard to the following four  points.
\begin{itemize}
\item Whether the position of the identified HMC  with the SO emission  agrees with that of a HMCC3 within 5$\arcsec$ or not?
\item Whether the HMCC is located between two HMCC3s or not? 
\item Whether the HMCC is apparently associated with an HC$^{15}$N core or not?  
\item Whether the HMCC is associated with CH$_3$OH HMCC1 and maser spots or not?
\end{itemize}
The positions of the HMCC1s identified  with the HC$^{15}$N, CH$_3$OH, and SO emission lines are found to be well correlated.  
By contrast, the positions of the CS cold cores \citep{Uehara2019}  agree with three HMCC1s, that is, only 0.3\% of the cold cores in the region (3/1061) \citep{Uehara2021}. 
The results of the positional agreement are  included in Table~\ref{Table03}.

\begin{figure*}[htbp]
\includegraphics*[bb= -50 310 700 730, scale=0.7]{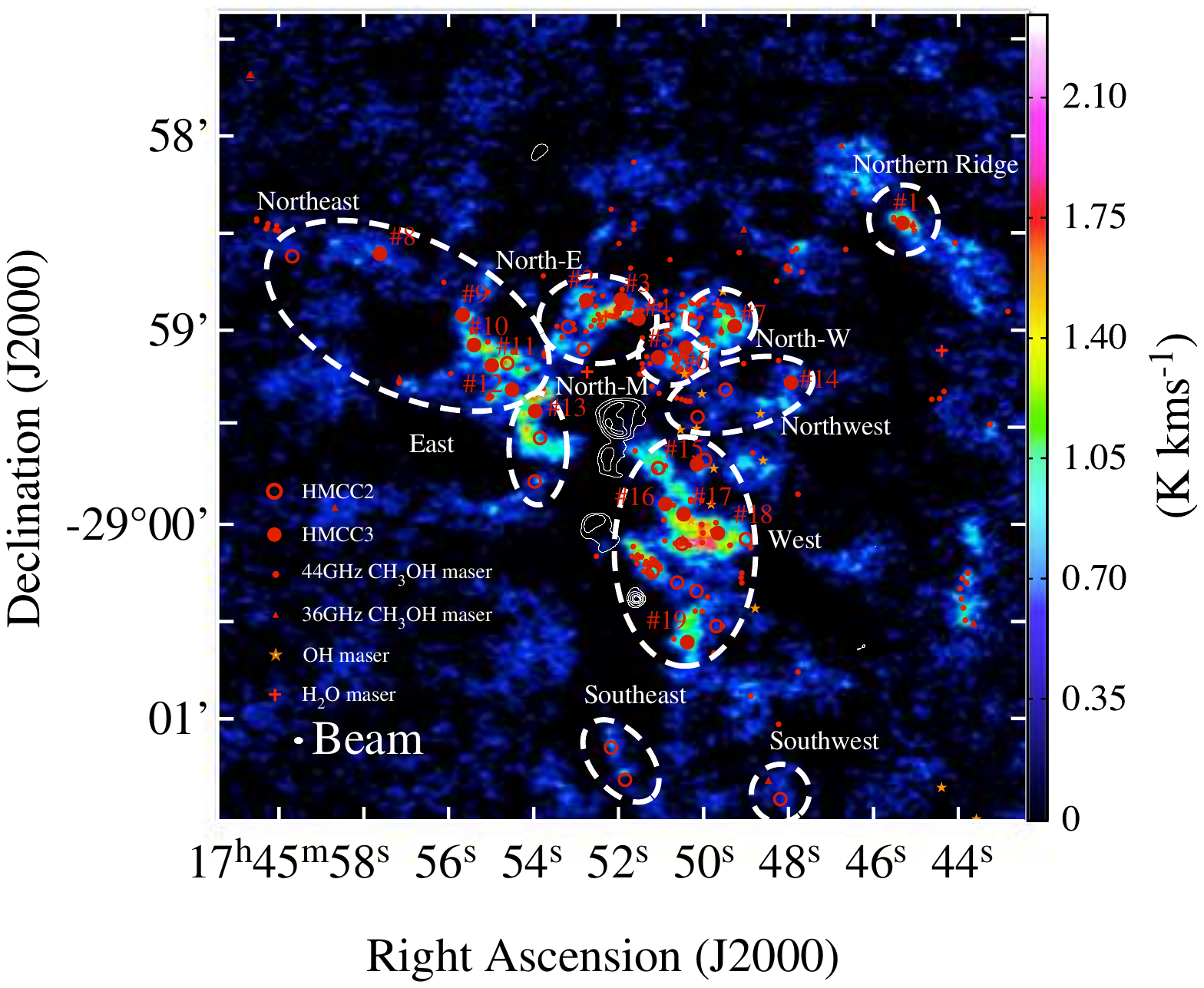}
\caption{Velocity-integrated (moment 0, from $-$14 to 86\,km\,s$^{-1}$) intensity color-scale map of the SO  emission line  with a uniformly weighted angular resolution of 2\farcs49$\,\times\,$1\farcs85  (PA=$-$89\fdg 70).
The peak specific intensity is 12.32~K\,km\,s$^{-1}$.
The typical  rms noise level is 0.18~K\,km\,s$^{-1}$, or
1~mJy\,beam$^{-1}$\,km\,s$^{-1}$.
White contours show the 86~GHz continuum image  with  the levels from 10\% to 90\%  with intervals of 10~points.
The beam size is indicated at the bottom-left corner.
Red small open circles, red filled triangles, and orange  asterisks indicate 44-GHz Class-I CH$_3$OH masers, 36-GHz Class-I CH$_3$OH masers  \citep{Yusef-Zadeh2013, McEwen2016}, OH masers \citep{Sjouwerman2008, Pihlstrom2011, Cotton2016}, and H$_2$O masers \citep{Lu2019a}, respectively.
Red filled and large open circles indicate HMCC3s and HMCC2s, respectively.
\label{Fig08}}
\end{figure*}

Figure~\ref{Fig08} shows the integrated intensity map of the SO emission line  superposed on the 86-GHz continuum map  and also plots 
the positions of the HMMC2s and HMMC3s  identified  with the SO, $^{34}$SO, HC$^{15}$N, and CH$_3$OH emission lines using  ``clumpfind''  
and those 
 of the Class-I CH$_3$OH masers \citep{Yusef-Zadeh2013, McEwen2016} and OH masers \citep{Sjouwerman2008, Pihlstrom2011,Cotton2016}.  

Class-I CH$_3$OH masers in star formation region are known to have a tendency to be associated with shock in protostellar outflow.
On the other hand, the ubiquity of the Class-I CH$_3$OH masers in the CMZ  suggests that Class-I CH$_3$OH masers in this region may be in a different origin, perhaps  large-scale shocks from turbulence.
Hence, it is necessary to carefully scrutinize the HMCCs that are apparently associated with  Class-I CH$_3$OH masers before making conclusive identification. 

 OH masers are useful  for distinguishing the two possibilities about its origin of outflows and large-scale shocks.  
Most of the OH masers in the 50~MC are a transition of 1720~MHz whereas the majority of the rest is a transition of 1612~MHz.
The OH masers associated with massive star formation regions, known as interstellar OH masers,
 are strong predominantly  in the mainline transitions, i.e., 1665 and 1667 MHz. 
 By contrast,   supernova remnants are only associated with  1720~MHz OH masers, 
which trace the interaction between supernova remnants and surrounding dense molecular cloud \citep[e.g.][]{Frail1996}.
Finally, OH masers associated with evolved stars often show double-horned spectral profiles at 1612~MHz.
 We conclude that there  are no OH masers in the 50~MC associated with massive star formation.

 H$_2$O masers also need attention.
 Four H$_2$O masers are detected in the area \citep{Lu2019a}.
W2 and W4 are associated with star formation,  whereas W1 and W3 have Asymptotic Giant Branch (AGB) star counterparts.

%
\begin{table}
\caption{ Number of 50~MC Hot Molecular Core Candidates \label{Table02}}
\begin{center}

\begin{tabular}{ccccc}
\hline\hline
HMCC &  SO & $^{34}$SO & HC$^{15}$N & CH$_3$OH  \\
\hline

clumpfind raw       & 168 & 22 & 64& 216   \\
Identified HMCC1       &110 & 21 & 39& 84   \\
HMCC2       & 39 & 16 & - & -   \\
HMCC3       & 19 & 8 & - & -   \\
HMCC       & 28(17) & 4 &10 & 24   \\

\hline

\end{tabular}
\end{center}
\scriptsize{
Note: In the clumpfind identification, we  adopt a threshold of 20$\sigma$ and contour spacing of 10$\sigma$ with  a rms value of 35.4~mK for SO, $^{34}$SO, and HC$^{15}$N. 
For CH$_3$OH, we adopt a threshold of 40$\sigma$ and contour spacing of 20$\sigma$.
The number inside the  parentheses  indicates the  that of the HMCCs identified only from HMCC3s.
The numbers  in the $^{34}$SO, HC$^{15}$N, and CH$_3$OH  columns  indicate those of the HMCs associated with a HMC of SO.
\\
}
\end{table}
%
%
%
%
\small
\begin{table*}[p]
\centering
\rotatebox{90}{
\begin{minipage}{\textwidth}
\caption{Physical Properties of the Hot Molecular Cores in the 50~MC  investigated with the SO emission}
\label{Table03}
\begin{center}
\scalebox{0.49}[0.49] 
{
\begin{tabular}{lcccccccccccccccccccl}
\hline
\hline
 &
 & SO& HC$^{15}$N &CH$_3$OH&CH$_3$OH &H$_2$O& 86~GHz & \multicolumn{2}{c}{Position}   & Size
 & \multicolumn{3}{c}{SO}  &
 & \multicolumn{3}{c}{$^{34}$SO}  &
 & Note  \\
 Region
 & HMC\#&HMCC3\# & Core &  HMCC1 & masers &masers& Dust 
 &  Right Ascension &  Declination &   Maj$\times$Min &
 peak S$_\nu$ &  {\it V}$_{\rm LSR}$ & 
 $\Delta${\it V}$_{\rm FWHM}$ &  $\int S_{\nu} dv$ & 
 peak S$_\nu$ &  {\it V}$_{\rm LSR}$ & 
 $\Delta${\it V}$_{\rm FWHM}$ &  $\int S_{\nu} dv$ & \\
 &  & & & & & &
 &J(2000) &  J(2000) &   arcsec($\arcsec$) &
(K) &  (km\,s$^{-1}$) & 
 (km\,s$^{-1}$) &  (K \,km\,s$^{-1}$) & 
 (K) &  (km\,s$^{-1}$) & 
 (km\,s$^{-1}$) &  (K \,km\,s$^{-1}$) & \\

\hline
Northern Ridge
&  &\#1 & $\surd$ &  $\surd$&  $\surd$ &   &  $\surd$
&17$^{\rm h}$45$^{\rm m}$45${\fs}$20 &-28$^\circ$58$'$26$\farcs$36  
&7$\farcs$64$\times$4$\farcs$17
&  0.40 $\pm$ 0.02 & -2.6 $\pm$ 0.4 & 18.0 $\pm$ 0.9 & 7.62 $\pm$ 0.48
&  0.19$\pm$ 0.03 & -3.8 $\pm$ 1.1 & 15.9 $\pm$ 3.1 &  3.27 $\pm$ 0.77  
&$^{(1)}$ \\
North-E
&01 &\#2 &    & &  $\surd$ & &  
&17$^{\rm h}$45$^{\rm m}$52${\fs}$82 &-28$^\circ$58$'$50$\farcs$61  
&12$\farcs$17$\times$7$\farcs$25
&  0.27 $\pm$ 0.02 & 47.0 $\pm$ 0.5 & 17.4 $\pm$ 1.3 & 4.98 $\pm$ 0.48
&  0.13$\pm$ 0.02 & 47.4 $\pm$ 1.0 & 16.4 $\pm$ 2.5 &  2.24 $\pm$ 0.44  
&$^{(2)}$ \\
&02 &(\#3, \#4 ) &  $\surd$ & $\surd$ &  $\surd$ &  $\surd$ &  $\surd$
&17$^{\rm h}$45$^{\rm m}$52${\fs}$67 &-28$^\circ$58$'$59$\farcs$26  
&12$\farcs$22$\times$6$\farcs$07
& 0.37 $\pm$ 0.02 & 40.9 $\pm$ 0.3 & 14.4 $\pm$ 0.8 & 5.76 $\pm$ 0.44
& 0.15 $\pm$ 0.02 & 41.4$\pm$ 0.9 & 12.4 $\pm$ 2.2 &  2.02 $\pm$ 0.48  
& $^{(3)}$\\
&03 &   &   &  $\surd$  &  $\surd$ &  &
&17$^{\rm h}$45$^{\rm m}$52${\fs}$06 &-28$^\circ$58$'$54$\farcs$07  
&24$\farcs$64$\times$7$\farcs$32
&  0.56 $\pm$ 0.03 & 50.9 $\pm$ 0.3 & 16.2 $\pm$ 0.9 & 9.73 $\pm$ 0.68
&  0.30$\pm$ 0.02 & 51.4 $\pm$ 0.6 & 14.2 $\pm$ 1.3 &  4.52 $\pm$ 0.55  
& $^{(3)}$\\\
North-M
&04 &(\#5, \#6) & &  $\surd$  &  $\surd$ &  &
&17$^{\rm h}$45$^{\rm m}$50${\fs}$71 &-28$^\circ$59$'$08$\farcs$27  
&8$\farcs$53$\times$3$\farcs$88
&  0.35 $\pm$ 0.03 & 36.4 $\pm$ 0.6 & 15.9 $\pm$ 1.5 & 5.98 $\pm$ 0.75
&  0.18 $\pm$ 0.03 & 33.9 $\pm$ 1.1 & 15.5 $\pm$ 2.6 &  2.97 $\pm$ 0.66  
& $^{(3)}$, $^{(4)}$      \\
North-W
&05 &(\#7) & $\surd$  &  $\surd$ &   &  &  $\surd$ 
&17$^{\rm h}$45$^{\rm m}$49${\fs}$52 &-28$^\circ$58$'$59$\farcs$26  
&8$\farcs$68$\times$4$\farcs$95
& 0.61 $\pm$ 0.03 & 49.7 $\pm$ 0.2 & 10.0 $\pm$ 0.6 & 6.44 $\pm$ 0.48
&  0.13 $\pm$ 0.03 & 52.5 $\pm$ 1.3 & 15.3 $\pm$ 3.5 &  2.07 $\pm$ 0.59  
&$^{(3)}$ \\
NorthEast
&06 &\#8 &   & & &  &
&17$^{\rm h}$45$^{\rm m}$57${\fs}$65&-28$^\circ$58$'$36$\farcs$05 
&15$\farcs$62$\times$6$\farcs$74
&  0.31 $\pm$ 0.03 & 34.1 $\pm$ 0.3 & 7.9 $\pm$ 0.8 & 2.59 $\pm$ 0.33
&  0.16 $\pm$ 0.04 & 33.8 $\pm$ 0.7 & 6.7 $\pm$ 1.7 &  1.14 $\pm$ 0.37
&$^{(2)}$ \\
&07 &\#9 &  &  $\surd$ &  &  &
&17$^{\rm h}$45$^{\rm m}$55${\fs}$59 &-28$^\circ$58$'$56$\farcs$14  
&10$\farcs$70$\times$6$\farcs$24
&  0.55 $\pm$ 0.03 & 38.5 $\pm$ 0.3 & 9.2 $\pm$ 0.6 & 5.36 $\pm$ 0.48
&  0.24 $\pm$ 0.02& 38.0 $\pm$ 0.4 & 8.6 $\pm$ 1.0 &  2.17 $\pm$ 0.59  
&$^{(3)}$ \\
&08 &\#10 & &  $\surd$ &  &  &
&17$^{\rm h}$45$^{\rm m}$55${\fs}$38 &-28$^\circ$59$'$03$\farcs$41 
&12$\farcs$99$\times$4$\farcs$48
&  0.49 $\pm$ 0.03 & 36.8 $\pm$ 0.4 & 15.1 $\pm$ 1.0 & 7.84 $\pm$ 0.66
&  0.22 $\pm$ 0.02 & 36.3 $\pm$ 0.5 & 11.1 $\pm$ 1.1 &  2.61 $\pm$ 0.35  
&$^{(3)}$ \\
&09 & (\#11) & $\surd$ &  $\surd$ &  &  &
&17$^{\rm h}$45$^{\rm m}$54${\fs}$99 &-28$^\circ$59$'$05$\farcs$49 
&14$\farcs$36$\times$4$\farcs$56
&  0.78 $\pm$ 0.04 & 39.2 $\pm$ 0.2 & 6.8 $\pm$ 0.4 & 5.69 $\pm$ 0.50
&  0.34 $\pm$ 0.03 & 39.5 $\pm$ 0.3 & 5.7 $\pm$ 0.7 &  2.06 $\pm$ 0.31  
&$^{(3,4)}$ \\
&10 &  & $\surd$ &  $\surd$ &  &  &
&17$^{\rm h}$45$^{\rm m}$54${\fs}$99 &-28$^\circ$59$'$20$\farcs$38 
&14$\farcs$76$\times$13$\farcs$81
&  0.48 $\pm$ 0.02 & 24.9 $\pm$ 0.3 & 14.2 $\pm$ 0.7 & 7.23 $\pm$ 0.50
&  0.26 $\pm$ 0.02 & 25.5 $\pm$ 0.5 & 13.6 $\pm$ 1.2 &  3.84 $\pm$ 0.44  
&  $^{(3,5)}$    \\
&11 & &  &  $\surd$ &  &  &
&17$^{\rm h}$45$^{\rm m}$54${\fs}$43 &-28$^\circ$59$'$07$\farcs$57  
&7$\farcs$71$\times$4$\farcs$29
&  0.47 $\pm$ 0.04 & 40.0 $\pm$ 0.3 & 8.1 $\pm$ 0.8 & 4.04 $\pm$ 0.51
&  0.15 $\pm$ 0.02 & 41.6 $\pm$ 0.6 & 8.7 $\pm$ 1.3 &  1.36 $\pm$ 0.28  
&$^{(2,5)}$ \\
East
&12 &(\#12, \#13?) &  &  $\surd$  & &  &
&17$^{\rm h}$45$^{\rm m}$54${\fs}$22 &-28$^\circ$59$'$21$\farcs$08  
&6$\farcs$43$\times$2$\farcs$00
&  0.29 $\pm$ 0.02 & 28.6 $\pm$ 0.5 & 13.0 $\pm$ 1.1 & 3.97 $\pm$ 0.46
&  0.14 $\pm$ 0.01 &28.6 $\pm$ 0.3 & 6.9 $\pm$ 0.7 &  0.99 $\pm$ 0.14
&$^{(2,5)}$ \\
&13 &(\#13?) & &  & &  &
&17$^{\rm h}$45$^{\rm m}$54${\fs}$12 &-28$^\circ$59$'$30$\farcs$77 
&7$\farcs$60$\times$4$\farcs$91
&  0.23 $\pm$ 0.03 & 32.5 $\pm$ 1.1 & 16.0 $\pm$ 2.6 & 3.98 $\pm$ 0.84
&  0.08 $\pm$ 0.02 & 35.4 $\pm$ 1.5 & 14.0 $\pm$ 3.9 &  1.25 $\pm$ 0.42  
&$^{(2,5)}$ \\
NorthWest
&14 &\#14 & & & &  & $\surd$
&17$^{\rm h}$51$^{\rm m}$51${\fs}$56 &-29$^\circ$59$'$37$\farcs$92  
&5$\farcs$06$\times$3$\farcs$00
&  0.40 $\pm$ 0.04 & 45.4 $\pm$ 0.5 & 9.1$\pm$ 1.1 & 3.86 $\pm$ 0.61
& 0.17 $\pm$ 0.03 & 47.4 $\pm$ 0.8 & 10.3 $\pm$ 1.9 &  1.89 $\pm$ 0.46  
&$^{(2)}$ \\
West
&15 &  &  &  $\surd$ &    $\surd$ &  &  
&17$^{\rm h}$45$^{\rm m}$51${\fs}$56 &-29$^\circ$59$'$37$\farcs$70  
&3$\farcs$49$\times$3$\farcs$25
&  0.38 $\pm$ 0.03 & 42.6 $\pm$ 0.5 & 12.7 $\pm$ 1.2 & 5.18 $\pm$ 0.66
&  0.17 $\pm$ 0.03 & 42.4$\pm$ 0.7 & 12.9$\pm$ 1.6 &  2.39 $\pm$ 0.40  
&  \\
&16 & & $\surd$ &  $\surd$ &  $\surd$? &  &
&17$^{\rm h}$45$^{\rm m}$51${\fs}$24 &-28$^\circ$59$'$40$\farcs$81 
&10$\farcs$83$\times$6$\farcs$35
&  0.37 $\pm$ 0.04 & 42.2 $\pm$ 1.0 & 15.8 $\pm$ 2.2 & 6.17 $\pm$ 1.16
&  0.17 $\pm$ 0.03 & 39.1 $\pm$ 0.7 & 6.7 $\pm$ 1.5 &  1.23 $\pm$ 0.37  
&  \\
&17 &\#15 &  &  $\surd$ &    $\surd$ &  &
&17$^{\rm h}$45$^{\rm m}$50${\fs}$05 &-28$^\circ$59$'$43$\farcs$68  
&7$\farcs$99$\times$4$\farcs$01
&  0.26 $\pm$ 0.02 & 39.3 $\pm$ 0.5 & 12.0 $\pm$ 1.2 & 3.36 $\pm$ 0.46
&  0.10 $\pm$ 0.02 & 38.5 $\pm$ 1.1 & 11.6 $\pm$ 2.8 &  1.25 $\pm$ 0.42 
&$^{(3)}$ \\
&18 & & $\surd$ &  $\surd$ &  &  &
&17$^{\rm h}$45$^{\rm m}$50${\fs}$74 &-28$^\circ$59$'$48$\farcs$09  
&8$\farcs$98$\times$4$\farcs$92
&  0.56 $\pm$ 0.02 & 41.9 $\pm$ 0.2 & 10.7 $\pm$ 0.5 & 6.33 $\pm$ 0.40
&  0.28 $\pm$ 0.02 & 42.6 $\pm$ 0.3 & 8.5 $\pm$ 0.6 &  2.50$\pm$ 0.24  
&$^{(5)}$ \\
&19 &\#16 & &  $\surd$  &  $\surd$ &  &
&17$^{\rm h}$45$^{\rm m}$50${\fs}$76 &-28$^\circ$59$'$53$\farcs$97  
&10$\farcs$71$\times$6$\farcs$00
&  0.37 $\pm$ 0.03 & 35.1 $\pm$ 0.5 & 14.0 $\pm$ 1.1 & 5.45 $\pm$ 0.57
&  0.10 $\pm$ 0.01 &34.0 $\pm$ 0.8 & 14.3 $\pm$ 1.8 & 1.56 $\pm$ 0.26  
&$^{(2)}$ \\
&20 &  &  &  $\surd$  &  $\surd$ &  &
&17$^{\rm h}$45$^{\rm m}$50${\fs}$32 &-28$^\circ$59$'$53$\farcs$97  
&10$\farcs$28$\times$6$\farcs$22
&  0.60 $\pm$ 0.04 & 40.2 $\pm$ 0.3 &8.0 $\pm$ 0.7 & 5.10 $\pm$ 0.55
&  0.25 $\pm$ 0.02 & 40.2 $\pm$ 0.4 & 7.5 $\pm$ 0.9 &  1.95 $\pm$ 0.29  
& $^{(4)}$\\
&21 &\#17 & &  $\surd$  &  &  &
&17$^{\rm h}$45$^{\rm m}$50${\fs}$50 &-28$^\circ$59$'$58$\farcs$13  
&3$\farcs$12$\times$2$\farcs$39
&  0.48 $\pm$ 0.02 & 41.3 $\pm$ 0.3 & 14.0 $\pm$ 0.7 & 7.09 $\pm$ 0.48
& 0.20 $\pm$ 0.03 & 41.8 $\pm$ 0.7 & 10.3 $\pm$ 1.7 &  2.15 $\pm$ 0.48 
&$^{(3)}$ \\
&22 &  & &  $\surd$ &  $\surd$ &  &
&17$^{\rm h}$45$^{\rm m}$49${\fs}$81 &-28$^\circ$59$'$59$\farcs$17  
&11$\farcs$16$\times$3$\farcs$80
&  0.32 $\pm$ 0.02 & 42.0 $\pm$ 0.7 & 20.1 $\pm$ 1.7 & 6.96 $\pm$ 0.77
&  0.12 $\pm$ 0.02 & 38.1 $\pm$ 0.8 & 8.4 $\pm$ 1.9 &  1.08 $\pm$ 0.31  
&$^{(3,5)}$ \\
&23 & &  &  $\surd$ &  $\surd$ &  &
&17$^{\rm h}$45$^{\rm m}$50${\fs}$00 &-29$^\circ$00$'$00$\farcs$90  
&8$\farcs$17$\times$6$\farcs$35
&  0.42 $\pm$ 0.02 & 46.2 $\pm$ 0.5 & 19.9 $\pm$ 1.1 & 8.89 $\pm$ 0.66
& 0.18 $\pm$ 0.04 & 41.1 $\pm$ 1.0 & 10.2 $\pm$ 2.4 &  1.96 $\pm$ 0.62  
&$^{(3,5)}$ \\
&24 & \#18& $\surd$ &  $\surd$ &  $\surd$ &  &$\surd$ 
&17$^{\rm h}$45$^{\rm m}$49${\fs}$81 &-29$^\circ$00$'$04$\farcs$36  
&6$\farcs$71$\times$5$\farcs$45
&  0.33 $\pm$ 0.02 & 42.0 $\pm$ 0.7 & 20.0 $\pm$ 1.6 & 6.94 $\pm$ 0.75
&  0.18 $\pm$ 0.04 & 48.0 $\pm$ 0.6 & 13.4 $\pm$ 1.5 &  2.53 $\pm$ 0.37  
&$^{(3,4)}$ \\
&25 & & $\surd$ & & $\surd$ &  &  $\surd$? 
&17$^{\rm h}$45$^{\rm m}$50${\fs}$40 &-29$^\circ$00$'$05$\farcs$74  
&7$\farcs$53$\times$4$\farcs$18
&  0.89 $\pm$ 0.05 & 46.5 $\pm$ 0.4 & 12.0 $\pm$ 0.8 & 11.35 $\pm$ 1.05
& 0.40 $\pm$ 0.03 & 47.8 $\pm$ 0.3 & 9.0 $\pm$ 0.7 &  3.86 $\pm$ 0.42  
&$^{(3,5)}$ \\
&26 &  &   &$\surd$ & $\surd$ &  &
&17$^{\rm h}$45$^{\rm m}$49${\fs}$45 &-29$^\circ$00$'$07$\farcs$13  
&15$\farcs$10$\times$7$\farcs$87
&  0.33 $\pm$ 0.03 & 49.8 $\pm$ 0.9 & 19.9 $\pm$ 2.4 & 7.03 $\pm$ 1.05
&  0.17 $\pm$ 0.02 & 51.3 $\pm$ 1.0 & 15.7 $\pm$ 2.4 &  2.75 $\pm$ 0.55  
&$^{(3,4,5)}$ \\
&27 & & $\surd$ & $\surd$ &  $\surd$ &  &  $\surd$
&17$^{\rm h}$45$^{\rm m}$51${\fs}$16 &-29$^\circ$00$'$12$\farcs$32  
&10$\farcs$47$\times$5$\farcs$69
&  0.45 $\pm$ 0.03 & 37.4 $\pm$ 2.5 & 22.5 $\pm$ 5.8 & 10.68 $\pm$ 2.96
&  0.23 $\pm$ 0.02 & 34.5 $\pm$ 0.8 & 16.7 $\pm$ 1.8 & 4.19 $\pm$ 0.59  
&$^{(3,4)}$ \\
&28 &\#19 & & $\surd$ &  &  &
&17$^{\rm h}$45$^{\rm m}$50${\fs}$29 &-29$^\circ$00$'$37$\farcs$25  
&13$\farcs$14$\times$6$\farcs$58
&  0.34 $\pm$ 0.01 & 45.5 $\pm$ 0.3 & 16.4 $\pm$ 0.8 & 5.95 $\pm$ 0.37
&  0.18 $\pm$ 0.02 & 43.9 $\pm$ 0.9 & 14.2 $\pm$ 1.6 &  2.68 $\pm$ 0.40  
&$^{(2)}$ \\

\hline
%
\vspace{3mm}

\end{tabular}
}
\end{center}
\footnotetext{
Note: (Column 3) Indicating  the SO HMCC3 number, where its position should satisfy the criterion of falling  within 5$\farcs$00 of the corresponding HMC position; if not,  the number is displayed in a pair of parentheses.   (Column 4) Check mark means that the position of the HC$^{15}$N core  agrees with that of a HMC within 5$\farcs$00. (Column 5) Check mark means CH$_3$OH masers found within 5$\farcs$00.  Peak intensities and Integrated intensities of the SO and $^{34}$SO lines are obtained  for the 3$\farcs$00 beam.
The  rms noise level is 35.4~mK. The numbers in the ``Note'' column indicates one of the following:
(1) being at Northern Ridge e.g. \citet{Takekawa2017a},
(2) being weak,
(3) constituting a cluster?,
(4) having absorption at the red or blue wing,
(5) having two velocity components.
\\
}
\end{minipage}
} 
\end{table*}

\normalsize

\subsection{Individual HMCCs}

Figures~\ref{Fig09a}--\ref{Fig09f}  show the enlarged integrated intensity maps  in the SO emission line  around the 28 identified HMCCs, 
 which are named HMC01--HMC28.
The HMCCs are  categorized according to their positions into the eight regions: ``Northern Ridge'', ``North'', ``Northeast'', ``Northwest'', ``East'', ``West'', ``Southeast'', and ``Southwest''.
Each has a characteristic distribution; for example,  the distributions of  HC$^{15}$N cores and CH$_3$OH masers are varied.
The  velocity  range is  10 \,km\,s$^{-1}$, and that set so as to include the velocities of the HMC. 
There are 19 identified HMCC3s, which are numbered HMCC3\#1--HMCC3\#19.
Those are indicated as \#1--\#19 in Figures~\ref{Fig09a}--\ref{Fig09f}.  

\subsubsection{Northern Ridge}
Figure~\ref{Fig09a} shows the Northern Ridge region (see also  Figure~\ref{Fig08}), 
  where 
  only one peak  is identified in the SO and HC$^{15}$N distributions. 
The peak of SO emission in the center of  the figure is a HMCC3 \#1 designated as  HMC \#1 (Table~\ref{Table03}) and several CH$_3$OH masers  exist in its close vicinity.
HMC \#1 is a typical HMC because the SO emission concentrates within 5$\arcsec$ and is apparently associated with CH$_3$OH masers.
HMC \#1 has also the same characteristics as HMC in M0.014-0.054 of \citet{Tsuboi2021}.
A HC$^{15}$N core is located 30$\arcsec$ northeast of HMCC3 \#1 and is associated with a HMCC1.

This HMCC seems to be a part of ``Northern Ridge'' or HCN$-$0.009$-$0.044, which is located  in the northeast side of Sgr~A~East with a filamentary structure \citep{Takekawa2017a, Takekawa2017b}.
However, ``Northern Ridge'' is considered not to be physically related to the 50~MC  on the basis of different positions and LSR velocities as  apparent in Figure~\ref{Fig06}.
A detailed report on the CCC of Northern Ridge in the Galactic Center Arc is  presented by \citet{Tsuboi2021}.

\begin{subfigures}

\begin{figure}
\includegraphics*[bb=80 250 580 650, scale=0.60]{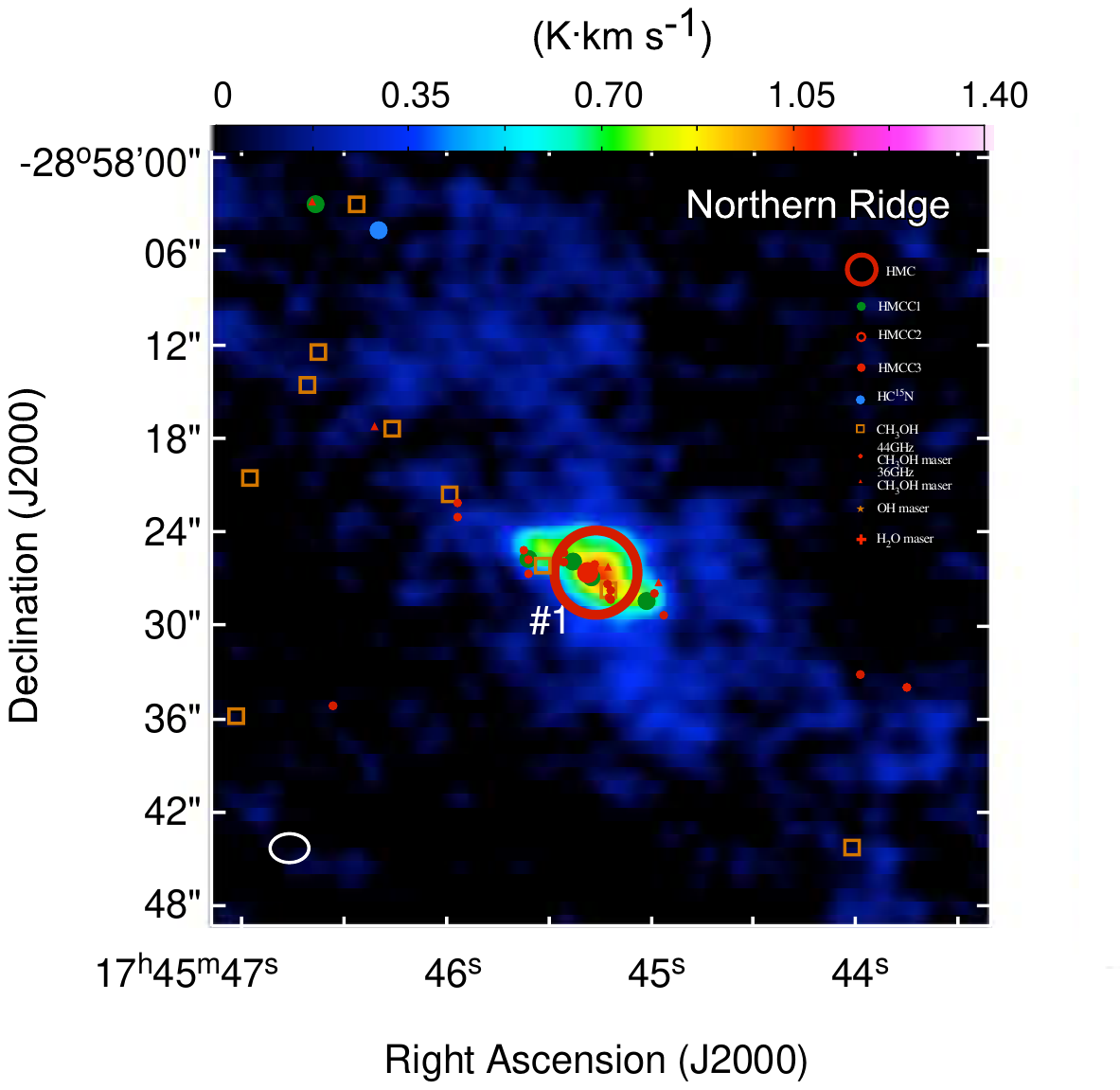}
\caption{Northern Ridge:
 Color map of the SO velocity-integrated intensity at  a velocity of 0 \,km\,s$^{-1}$. 
The peak intensity is given in Table~\ref{Table02}.
The typical  rms noise level is 18.0~mK km\,s$^{-1}$. 
Max integrated intensities are normalized by 1.19~K km\,s$^{-1}$ for SO.
Beam size is indicated at the left bottom corner.
Blue circles, orange squares, green filled circles, red open circles, and red filled circles  indicate HC$^{15}$N cores, CH$_3$OH, HMCC1, HMCC2, and HMCC3, respectively.
Small red filled circles, red filled triangles, orange filled  asterisks, and red crosses indicate 44-GHz Class-I CH$_3$OH masers, 36-GHz Class-I CH$_3$OH masers, OH masers, and H$_2$O masers, respectively.
\label{Fig09a}}
\end{figure}

\subsubsection{North: HMCs01--05}

Figure~\ref{Fig09b} shows five HMCCs in  ``North.''
This region encompasses  six HMCC3s designated as \#2 to \#7  and  is divided into three sub-regions according to their spatial locations and velocities: ``North-E'', ``North-M'', and ``North-W''.

Region  ``North-E'' is located at the east of North and has three HMCC3s of \#2, \#3, and \#4.
HMC01  encompassing HMCC3 \#2 is a weak HMC with a velocity of 40 \,km\,s$^{-1}$  and with a few CH$_3$OH masers nearby.
HMC02 seems to be a single HMCC identified as HMCC1s \#3 and \#4 with  velocities of 50 \,km\,s$^{-1}$ and 60 \,km\,s$^{-1}$, respectively;
we conjecture that it is because the HC$^{15}$N emission has several cores with a few CH$_3$OH masers and an H$_2$O maser (W2)  \citep{Lu2019a} in the center of the bright peak between HMCC1s \#3 and \#4.
HMC03 has  neither HMCC3  nor HC$^{15}$N core associated and is located between HMC01 and HMC02  with a few CH$_3$OH masers nearby.
This region ``North-E''  at 40 \,km\,s$^{-1}$ and 50 \,km\,s$^{-1}$ has a nested relation to each other.

Region  ``North-M'' is a group with a velocity of 40 \,km\,s$^{-1}$, which is different from the velocities of 50 \,km\,s$^{-1}$ for North-E and -W.
HMC04 is located between two HMCC3s \#5 and \#6 with a separation of 5$\arcsec$ and is apparently associated with several CH$_3$OH masers.

HMC05 belongs to  the group,   ``North-W'', with a velocity of 50 \,km\,s$^{-1}$.
 HMC05  encompasses three or four HC$^{15}$N cores with CH$_3$OH masers.
Also adjacent to HMC05 are an OH maser and H$_2$O maser (W1) associated with an AGB star.


\begin{figure}
\includegraphics*[bb= 130 250 500 740, scale=0.7]{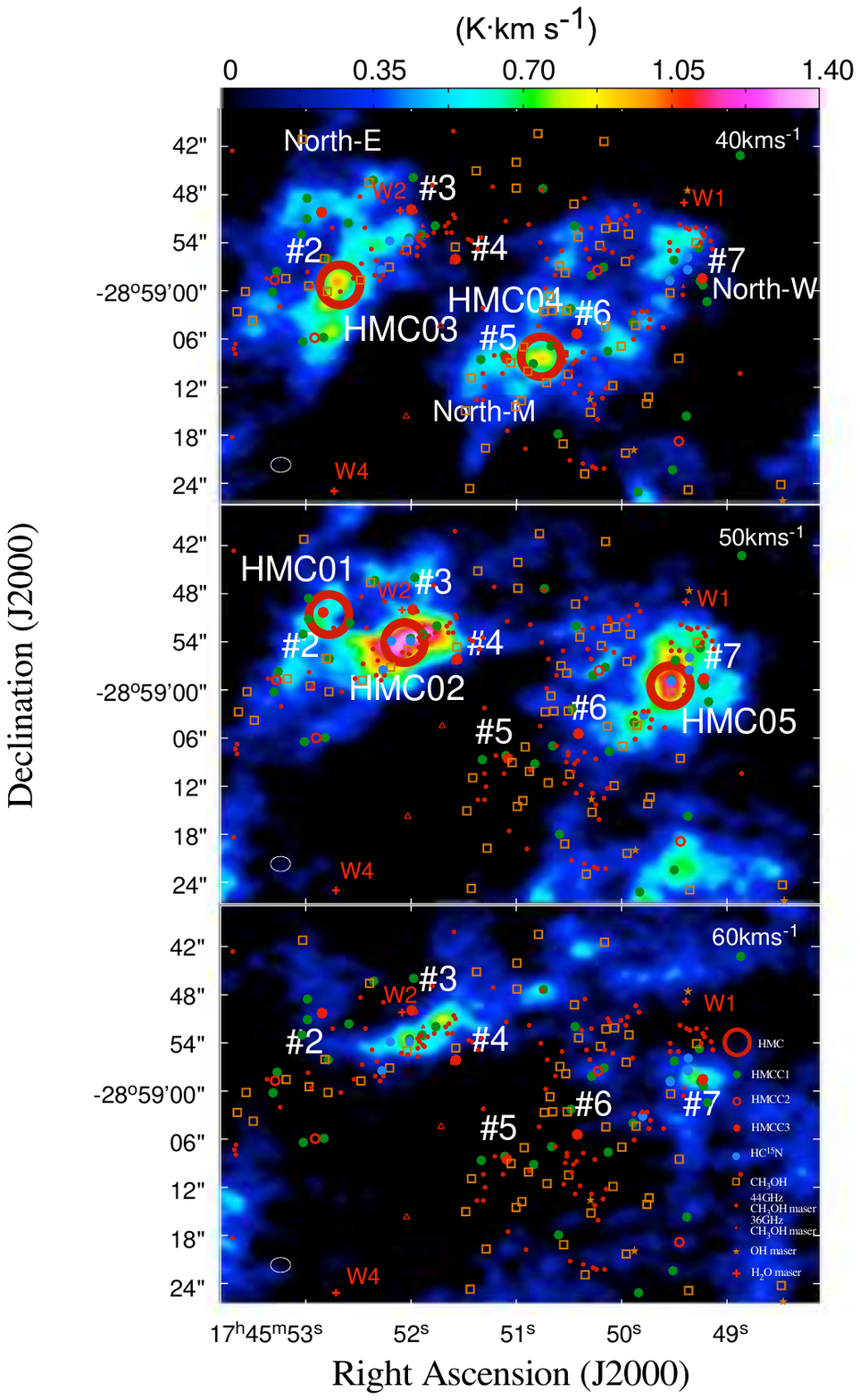}
\caption{Region North:
The SO velocity-integrated map at  velocities of 40, 50, and 60 \,km\,s$^{-1}$. Color scales and notation are identical with those in Figure~\ref{Fig09a}.
\label{Fig09b}}
\end{figure}

\subsubsection{Northeast: HMCs06--11}
Figure~\ref{Fig09c} shows six HMCCs in region ``Northeast''.
These HMCCs are located along a filament  running in the northeast to southwest  directions and there are five HMCC3s designated as \#8 to \#12, four HC$^{15}$N cores, and a few CH$_3$OH masers in this region.
The images at the LSR velocities of 30 and 40 \,km\,s$^{-1}$ are similar.
HMC06 has a weak peak of HMCC3 \#8 with a velocity of 30--40 \,km\,s$^{-1}$.
HMC07 has a moderate intensity corresponding to HMCC3 \#9 with a velocity of 40 \,km\,s$^{-1}$ and no CH$_3$OH masers nearby.
HMC08  corresponds to HMCC3 \#10 adjacent to an HC$^{15}$N core and with no CH$_3$OH masers nearby.
HMC09 has a peak velocity of 40 \,km\,s$^{-1}$ with no HMCC1 but HMCC3 \#11 located  7$\arcsec$ south, and two HC$^{15}$N cores  exist within 5$\arcsec$.
 HMC10 has a velocity of 30 \,km\,s$^{-1}$ with an HC$^{15}$N core.
HMC10 do not locate on the line of HMCs07, 08, 09, and 11.
HMC11 has a weak peak  near a HMCC2, which is located  5$\arcsec$southeast.


\begin{figure}
\includegraphics*[bb= 80 150 500 780, scale=0.55]{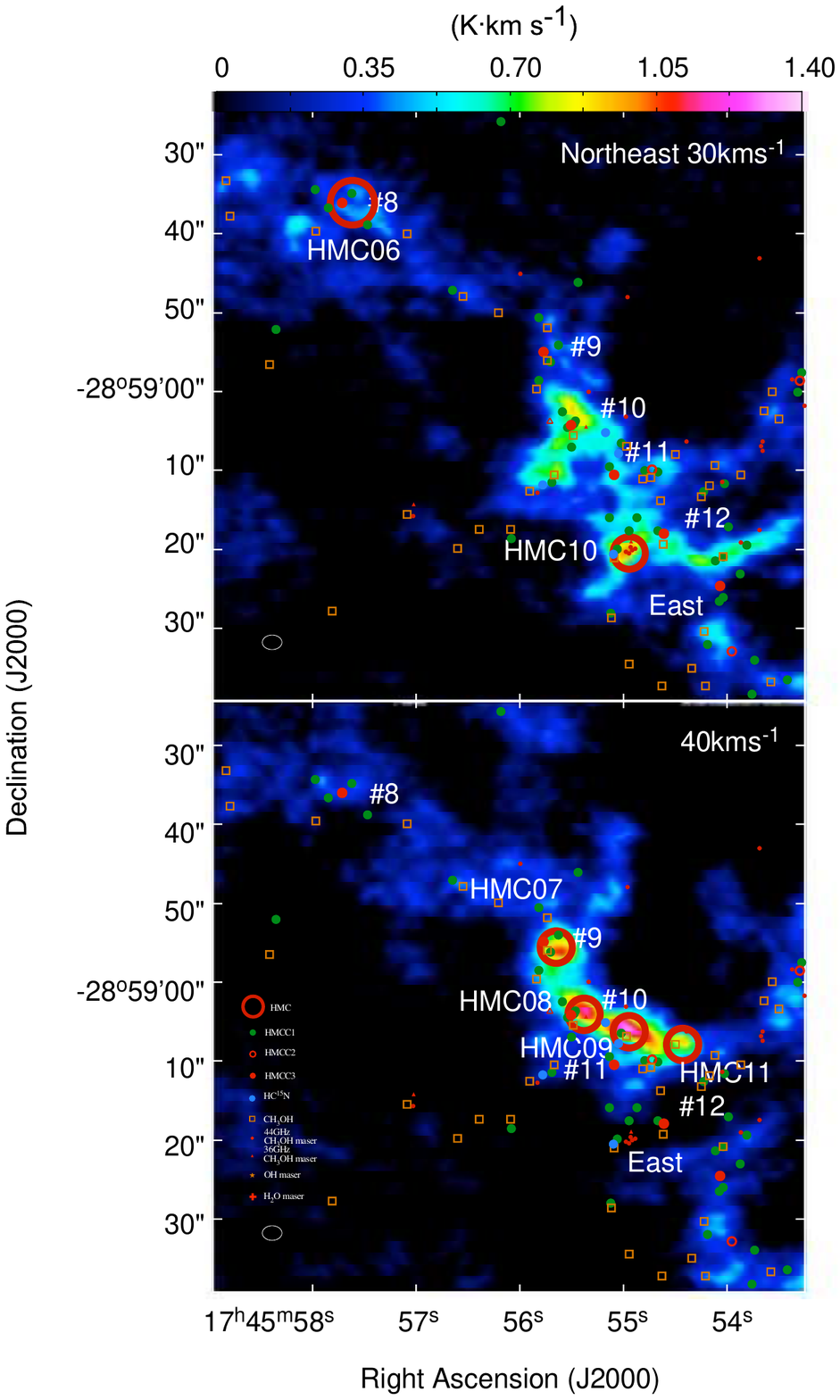}
\caption{NorthEast:
The SO velocity-integrated map at  velocities of 30 and 40 \,km\,s$^{-1}$. Color scales and notation are identical with those in Figure~\ref{Fig09a}. 
\label{Fig09c}}
\end{figure}

\subsubsection{East: HMCs12, 13}
Figure~\ref{Fig09d} shows two HMCCs in region ``East''.
This region is adjacent to region Northeast  but has a different velocity.
A weak source, HMC12  corresponding to HMCC3 \#12, and HMC13  corresponding to another HMCC3 designated as \#13 are located in the north of the crescent structure with a diameter of 15$\arcsec$ at the velocities of 30 and 50 \,km\,s$^{-1}$, respectively.
To the south of HMC13, a few HMCC1s are distributed in southern part of the crescent structure, where there is a HMCC2.

H$_2$O maser W4 is located  15$\arcsec$ east of HMC13.
This H$_2$O maser is not associated with the 50~MC  given that  the H$_2$O maser has a peak at 156.0\,km\,s$^{-1}$.


\begin{figure}[htb]
\includegraphics*[bb= 170 150 500 825, scale=0.7]{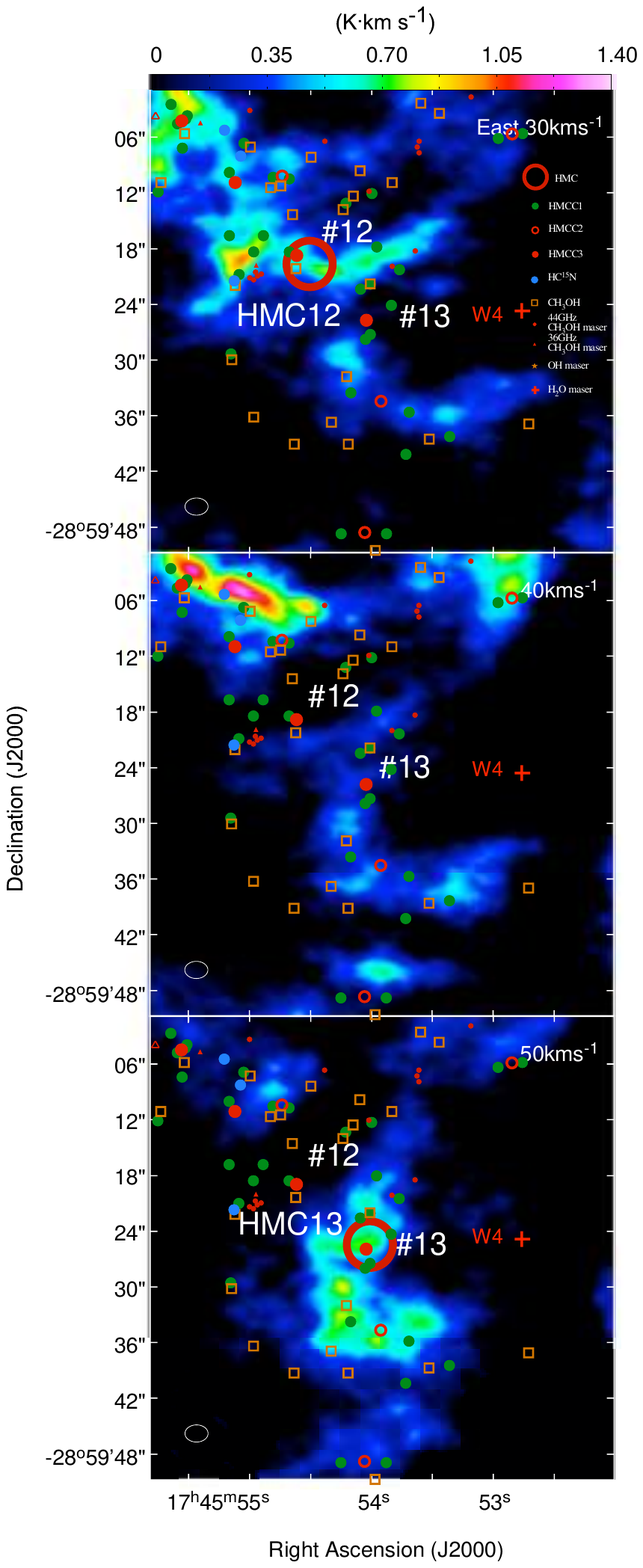}
\caption{East:
The SO velocity-integrated map at  velocities of 30, 40, and 50 \,km\,s$^{-1}$. Color scales and notation are identical with those in Figure~\ref{Fig09a}.
\label{Fig09d}}
\end{figure}

\subsubsection{Northwest: HMC14}
Figure~\ref{Fig09e} shows a HMC in region ``Northwest''.
This region harbors a weak condensation, HMC14  corresponding to a HMCC3 designated as \#14 with a velocity of 40 \,km\,s$^{-1}$.
There  are no HC$^{15}$N cores and are a few CH$_3$OH masers in the region.


\begin{figure}
\includegraphics*[bb= 100 150 500 770, scale=0.55]{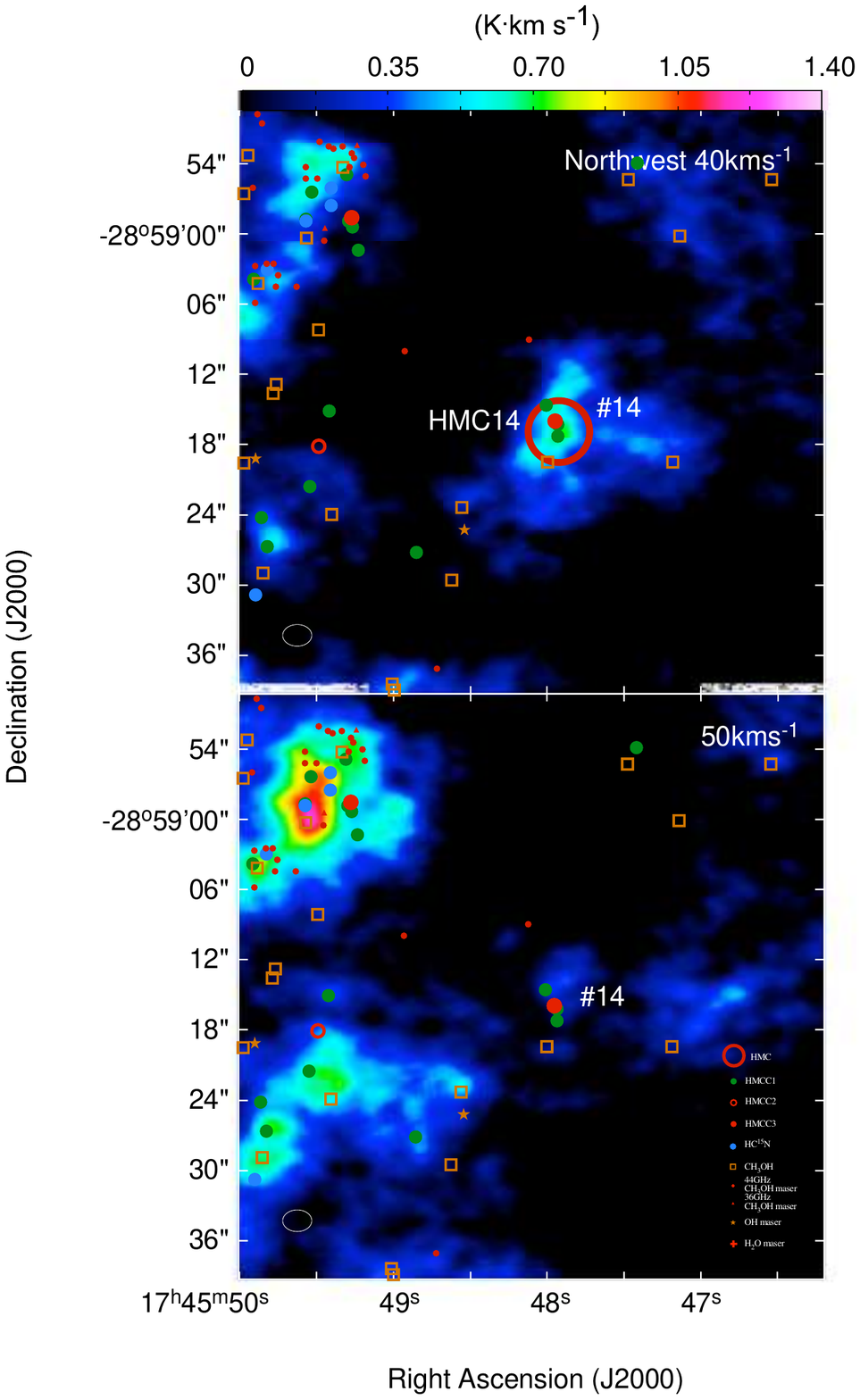}
\caption{NorthWest:
The SO velocity-integrated map at  velocities of 40 and 50 \,km\,s$^{-1}$. Color scales and notation are identical with those in Figure~\ref{Fig09a}.
\label{Fig09e} }
\end{figure}


\subsubsection{West: HMCs15--28}
Figure~\ref{Fig09f} shows 14 HMCCs (\#15--\#28) in region ``West''.
A half of the HMCCs in this region are located in the upper half (i.e., northern) area with a velocity of 30--40 \,km\,s$^{-1}$,  whereas the other half (i.e., southern) is in the lower half area with a velocity of 50 \,km\,s$^{-1}$.

In the northern area, there is a prominent structure of a filament running northeast-southwest  that is dominant at  a velocity of 40 \,km\,s$^{-1}$; it encompasses  diffuse HMC15, HMC16 with an HC$^{15}$N core, strong HMC18 with an HC$^{15}$N core, HMC20 with two CH$_3$OH masers, and HMC22 with a CH$_3$OH maser.
HMC17 is located away from the filament and corresponds to a  HMCC3 \#15.  
Diffuse HMC19, corresponding to HMCC3 \#16, is  also located off  the filament.
HMC21, corresponding to HMCC3 \#17,  appears to be linked to HMC19 and belongs to a different filament  from the one that encompasses HMCs15--20 and HMC22.

In the southern area, or the central part  in the figures, there is  weak HMC23  corresponding to HMCC3 \#18  with  a CH$_3$OH maser nearby, HMC24 with an HC$^{15}$N core nearby, and weak HMC26 with a CH$_3$OH maser nearby, where may  bridge HMC22 and HMC23.
HMC25 is the strongest HMC in this region with no counterpart HMCC3 but a HMCC2 is located within 5$\arcsec$.
HMCs23, 24, 25, and 26  apparently constitute a cluster.

HMC27 is located  to the east of the center of the figures and has  unique features; it has no HMCC3 counterparts but a HMCC2 and are apparently associated with two HC$^{15}$N cores  and many CH$_3$OH masers within 5$\arcsec$.
We note that there are  HMCC1s with different velocities  apparently associated with a HC$^{15}$N core  and a number of CH$_3$OH masers  at 10$\arcsec$ northeast of HMC27 on panels of 30--50 \,km\,s$^{-1}$ in Figure~\ref{Fig09f}.
This region has weak SO emission with two velocity components and broad velocity width. 
It is not clear  whether each of these HMCC1s  is a HMC or a part of it. 
HMC27 is located at the northern tip of another filament  running northeast to southwest directions, which is connected to HMC28.
The southern part of this filament is diffuse, encompassing  several isolated HC$^{15}$N cores and a HMCC1.
 HMC28 at the bottom of the figures is in a diffuse component and corresponds to HMCC3 \#19.


\begin{figure*}
\includegraphics*[bb= 80 120 800 490, scale=0.7]{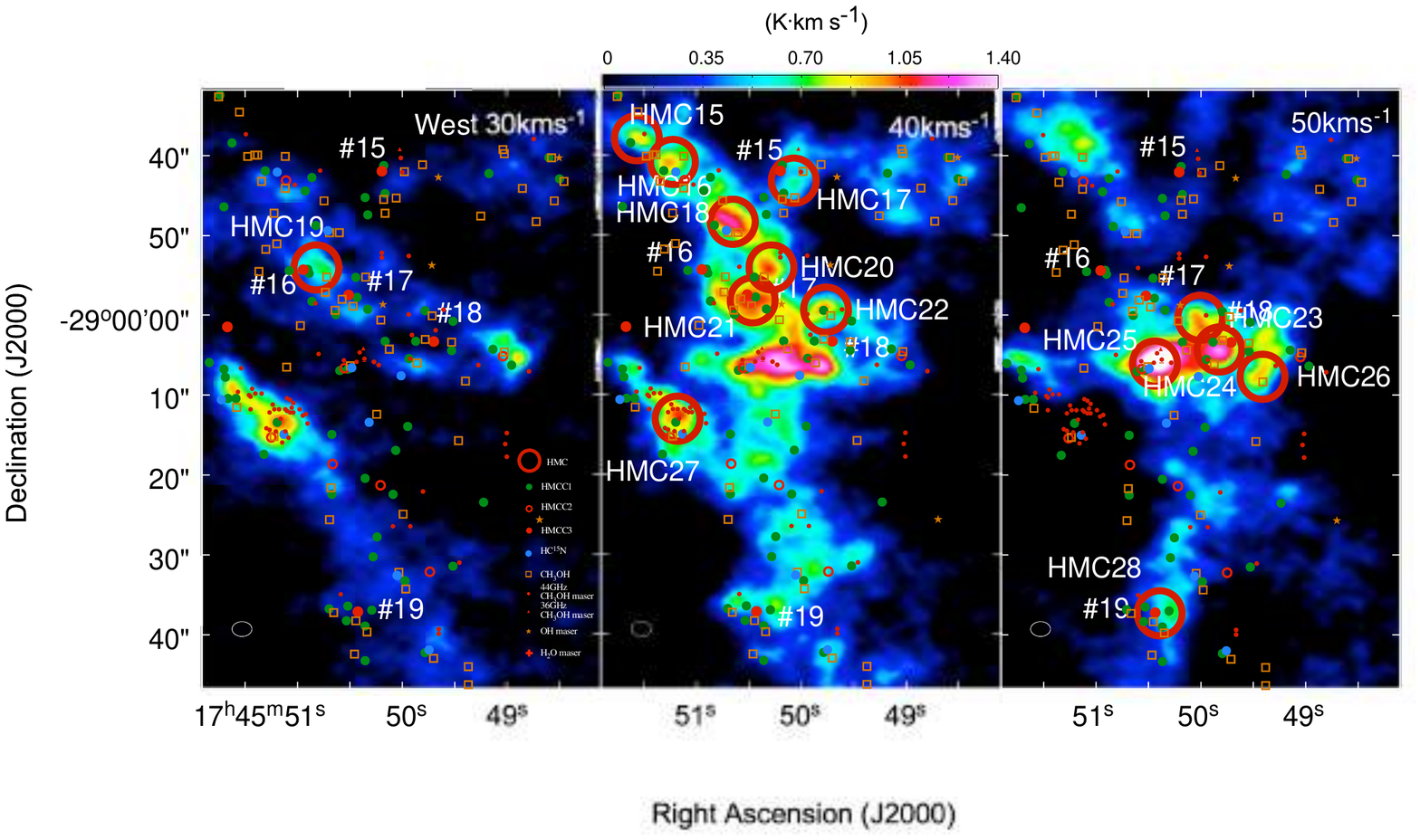}
\caption{West:
The SO velocity-integrated map at  velocities of 30, 40, and 50 \,km\,s$^{-1}$. Color scales and notation are identical with those in Figure~\ref{Fig09a}.
\label{Fig09f} }
\end{figure*}

\subsubsection{Southeast and Southwest}
There  are no  HMCCs identified in regions ``Southeast'' and ``Southwest''.
Region Southeast has two weak HMCC2s along a north-to-south filament with a velocity of 60 \,km\,s$^{-1}$.
There are four weak SO peaks  along a 25$\arcsec$ north-to-south elongated structure, 
 but there  are no peaks of HC$^{15}$N  or CH$_3$OH.

Region Southwest has two weak SO peaks, which are located about 1$'$ away from region Southeast and has a different velocity  of 30 \,km\,s$^{-1}$.

To summarize the results of the eight regions,
although some HMCCs are clearly identified in  some channel maps, 
they usually  are unclear in the velocity-integrated maps. 
Some of the HMCCs have good positional correlations between their  SO emission  and HC$^{15}$N and CH$_3$OH emissions, whereas
others do not.
These results are discussed in detail in \S4.1.

\end{subfigures}

\subsection{Masses of HMCCs}
\subsubsection{Masses derived from the 86-GHz continuum emission}
The 86-GHz continuum emission of the HMCCs is considered to originate  in the dust thermal emission (\S 3.1.).  
 Then, the mass of the gas  can be calculated  according to the following equation \citep{Hildebrand1983}:
\begin{equation}
\mathrm{M_d}={{{4a\rho D^2 F_{\nu}} \over {3 B_{\nu}(T)Q_{125}}}( \lambda/125) ^{\beta}},
\end{equation}
where  $F_{\nu}$ is the flux density, $B_{\nu}$ is the the black-body intensity, $\beta$ is the is the wavelength dependence index,
$a$ is the mean grain radius, $\rho$ is the density of dust grain material, \rm{$D$} is the distance to the cloud, $Q_{125}$ is the grain emissivity at 125 \micron, and $\lambda$ is the observing wavelength in micron.
We adopt the  nominal value of 2.0 as the power-law index  $\beta$, as used in ${S_\nu \propto \nu^{2+\beta}}$ in the Rayleigh-Jeans domain \citep{Hildebrand1983}.
Using the typical values for  $\rho$=3 g~cm$^{-3}$, {\it a}=0.1 \micron, ${Q_{125}}$ = 3/4000 \citep{Hildebrand1983}, and ${\it T{\rm_d}=}$50~K and 100~K,
and assuming  the typical gas-to-dust mass ratio  of 100, the mass of the gas,  M$_{dust}$,  is 
 derived  to be 1.26$\pm$1.55 $\times$$10^3$ M$_\odot$ for  {\it T}$_{\mathrm{d}}$=50~K and 6.45 $\pm$ 8.81 $\times$$10^2$ M$_\odot$ for  {\it T}$_{\mathrm{d}}$=100~K.
Accordingly, the maximum masses of dust cores are 5.45 $\times$$10^3$ M$_\odot$  and 2.78 $\times$$10^2$ M$_\odot$ at  {\it T}$_{\mathrm{d}}$=50~K and   {\it T}$_{\mathrm{d}}$=100~K, respectively, whereas
the minimum masses of dust cores are 1.71 $\times$$10^2$ M$_\odot$  and 8.73 $\times$$10^1$ M$_\odot$, respectively.
 Table~\ref{Table01} summarizes the result.

In the following discussion, we  propagate uncertainties of a factor of 2 in the dust masses, which originate from the uncertainties in   the dust opacity,  gas-to-dust ratio, dust emission flux, and  distance  to the source, as discussed below.
For dust cores with significant internal heating with potentially higher dust temperatures (e.g.,  higher than {\it T}$_{\mathrm{d}}$=50~K and   {\it T}$_{\mathrm{d}}$=100~K, respectively), the derived masses  would be systematically  lower than the true values.
The derived densities of the dust cores depend on the angular sizes and all the quantities that determine the dust masses. 
The measured angular sizes usually have uncertainties of 20\%. 
We propagate these random errors but exclude the systematic error in the dust temperature and obtain an uncertainty of a factor of 2 for the masses. 
Similar to the masses, for dust cores with  assumed  {\it T}$_{\mathrm{d}}$=50~K and   {\it T}$_{\mathrm{d}}$=100~K, respectively, the densities  would be systematically lower by a factor of 2--3.

\subsubsection{Masses derived from the SO emission}

Under local thermodynamic equilibrium (LTE) conditions, 
	the column density of a linear molecule can be  given by \citep[e.g.][]{Goldsmith1999}, 
\begin{eqnarray}
	\mathrm{N(H_2)} &=&\frac{3k}{8\pi^3\nu\mu^2S}\frac{Q_\mathrm{rot}}{g_{J+1}}\times
	\frac{\mathrm{exp}\left(\frac{E_\mathrm{up}}{k T_\mathrm{ex}}\right)}{J(T_\mathrm{ex})} \nonumber \frac{1}{X(^{32}\mathrm{SO})} \\
	& &\times\frac{1}{J(T_\mathrm{ex})-J(T_\mathrm{bg})}\frac{\tau_{^{32}\mathrm{SO}}}{1-\mathrm{exp}(-\tau_{^{32}\mathrm{SO}})}\int S_\mathrm{\nu}~dv \label{eq-lte-density}.
\end{eqnarray}
where $\nu$ is the frequency of the SO line,
$T_\mathrm{ex}$ is the excitation temperature of SO molecule,
$T_\mathrm{bg}$ is the cosmic background,
$\tau_{^{32}SO}$ is the optical depth of $^{32}$SO,
{\it S} is the line strength,
SO abundance is X(SO),
$k$ is the Boltzmann constant,
$Q_\mathrm{rot}$ is the rotational partition function,
the radiation field  $J(T)$ is described by the Planck function $B_\nu(T)$,
the total degeneracy for an an energy level of a transition is given by the product of rotational $g_J$ degeneracies for rotational molecular transitions,
$S_\nu$ is the flux densisty,
and $\mu$ is the permanent dipole moment (1.550D for $^{32}$SO). 
For the N$_J$=2$_2$--$1_1$ transition of $^{32}$SO, 
we have the line strength $S = \frac{J+1}{2J+3} = \frac{2}{5}$ and  degeneracy $g_\mathrm{J+1} = 2J+3 = 5$, where $J$ is the rotational quantum number of the lower state. 
$E_\mathrm{{up}}$ is the energy of the upper state,  $E_\mathrm{{up}}$=19.31~K. 
In the calculation, a single excitation temperature was assumed to be 50~K or 100~K \citep{Tak2003}.
We take the partition function values {\textit{Q$_{\mathrm{rot}}$}}(\textit{T}) = 125 and 250 from the nonlinear plot of The Cologne Database for Molecular Spectroscopy (CDMS)  for temperatures of 50~K  and 100~K, respectively.  
We assume  a SO abundance of X(SO) = $10^{-9}$,  given that the SO relative abundance varies in  a range of X(SO) $\sim$ (0.5--4)$\times$10$^{-9}$ for various types of HMCCs and has a median value of  1.1--1.3 $\times$ 10$^{-9}$ \citep{Li2015, Zinchenko2018}.


Assuming  $^{32}$SO/$^{34}$SO=22.5  \citep{Ho1983},  
we have the following intensity ratio, 
\begin{equation}
\frac{\mathrm{I}(^{34}\mathrm{SO(2_3-1_2)})}{\mathrm{I}(^{32}\mathrm{SO(2_3-1_2)})}=\frac{1-\mathrm{exp}(-\tau_{^{32}\mathrm{SO}}/22.5)}{1-\mathrm{exp}(-\tau_{^{32}\mathrm{SO}})},
\end{equation}
where $\tau$${_{^{32}SO}}$ is the optical depth of the ${^{32}}$SO(2$_3$-1$_2$) emission. 
Our observations show I($^{32}$SO(2$_2$--1$_1$))/I($^{34}$SO(2$_3$--1$_2$))  $\sim$2.0.
The (2$_2$--1$_1$)/(2$_3$--1$_2$) intensity ratio of $^{32}$SO molecules at both the temperatures is 0.40 from the results of RADEX \citep[][]{Tak2007}.
As a result, the intensity ratio of I($^{32}$SO({$2_3-1_2$}))/I($^{34}$SO({$2_3-1_2$}))  is $\sim$5.0.
Consequently,   $\tau_\mathrm{{SO}} $ is calculated to be 4.8, 
 which is moderate optically thick.
From the results of RADEX, 
the optical depth ratio of $\tau_{^{32}\mathrm{SO}(2_2--1_1)}$/$\tau_{^{32}\mathrm{SO}(2_3--1_2)}$ is estimated to be $\sim$1/3, and thus the optical depth of  $^{32}$SO(2$_2$-1$_1$)  is $\sim$1.6.
Therefore,
the  correction factor for the LTE mass to correct for the optical depth is  derived to be
%
\begin{equation}
\frac{\tau_{^{32}\mathrm{SO}}}{1-\mathrm{exp}(-\tau_{^{32}\mathrm{SO}})} \sim\, 1.7.   
\end{equation}
%
The HMCC radius is derived from 
\begin{equation}
\mathrm{r(pc)} = \frac{\sqrt{\mathrm{Maj (pc)} \times \mathrm{Min (pc)}}}{2} = \sqrt{\mathrm{r_{Maj (pc)}} \times \mathrm{r_{Min (pc)}}}.
\end{equation}

In Table~\ref{Table04}, we show the H$_2$ column density,  mean number density,  LTE mass derived from the number density assuming a spherical morphology, and   virial mass  derived from the linewidth $\Delta$$\textit{V}$ tabulated in Table~\ref{Table03}. 
In consequence, the column densities of the HMCCs are calculated to be N(SO)=6.21 $\pm$ 2.11$\times$$10^{15}$cm$^{-2}$ for  {\it T}$_{\mathrm{ex}}$=50~K and 4.95 $\pm$ 1.68$\times$$10^{15}$cm$^{-2}$ for  {\it T}$_{\mathrm{ex}}$=100~K.
The resultant hydrogen molecule number densities are n(H$\rm{_2}$)=1.41 $\pm$ 0.61 $\times$$10^7$cm$^{-3}$  and 1.12 $\pm$ 0.49 $\times$$10^7$cm$^{-3}$, respectively.
The LTE masses are estimated to be 1.17 $\pm$ 1.10 $\times$$10^4$ M$_\odot$  and 9.36 $\pm$ 8.80 $\times$$10^4$ M$_\odot$, respectively. 
These values are larger than the mean masses of the H$^{13}$CO$^{+}$ and C$^{34}$S bound cores of 960 $\pm$ 850 M$_\odot$ and 1300 $\pm$ 890 M$_\odot$, respectively  \citep[][]{Uehara2019}.
The virial masses are estimated to be 9.57 $\pm$ 7.30 $\times$$10^3$M$_\odot$, and accordingly
the ratios of the LTE mass to the virial mass  are 1.412 $\pm$ 0.923  and 1.125$\pm$ 0.736, respectively. 
We find that 15 HMCCs including the Northern Ridge HMCC have  ratios of $\leq$1. 
The other 14 HMCCs have supercritical masses with a ratio of  $\geq$1, where  the self-gravity of the HMCCs  is likely to be dominant over  the internal gas pressure. 
The bound HMCCs will efficiently  form massive protostars if the cores are heated. 
We note that this result and following discussion would hold regardless of inclusion or exclusion of  the Northern Ridge HMC.


The  uncertainties of  the excitation temperatures of the SO molecules,  line fluxes,  line widths, and  angular sizes,  all propagate into that of the derived  LTE masses. 
In our estimate, we have considered  a large uncertainty of a factor of 2 for the LTE masses  though not considering the systematic error in the excitation temperature. 
In reality,  the abundance X(SO) may be systematically overestimated by a factor of 3 or larger. 

 The errors of the line widths (20\%) and  angular sizes (20\%)  propagate into that of the derived virial masses of the HMCCs. 
Whereas our estimate has considered  a large uncertainty of a factor of 2 or larger,
the masses of the HMCCs may be still systematically overestimated by a factor of 3 or larger. 
It is unclear whether at 0.1 pc the magnetic field is as well as 1.0~pc in the CMZ, 
where the magnetic field at 1 pc scales in CMZ is suggested to be $\sim$5~mG with large uncertainties \citep[e.g.][]{Pillai2015}.
The support against gravitational collapse from the magnetic field would be significant but it is difficult to estimate the strength of the magnetic field in the HMCCs.
We note that we ignore the effect of the magnetic field in this paper.

\small
\begin{table*}[p]
\centering
\rotatebox{90}{
\begin{minipage}{\textwidth}
\caption{ Masses of the  SO Hot Molecular Cores in the 50~MC}
\label{Table04}
\begin{center}
\scalebox{0.65}[0.65] 
{
\begin{tabular}{lccccccccccccl}
\hline\hline
 &  &  & 50K 
 &  &  & 100K 
 &  &  &  &  &  & & \\
Region & HMCC\# & N(H$_2$) & N(SO) & n(H$_2$) 
& N(H$_2$)  & N(SO) & n(H$_2$) 
& LTE Mass& LTE Mass  & virial Mass 
& \underline{{\rm{LTE Mass}(50K)}}  &\underline{{\rm{LTE Mass(100K)}}} &Note \\
& &  (cm$^{-2}$) &(m$^{-2}$) &(cm$^{-3}$) 
& (cm$^{-2}$) &(cm$^{-2}$)  &(cm$^{-3}$) 
& (50K)  (M$_{\odot}$) &(100K) (M$_{\odot}$)  
&  (M$_{\odot}$) & Virial Mass  &Virial Mass&  \\  
\hline
Northern Ridge &  & 7.61E+24 & 7.61E+15 & 2.10E+07 & 6.07E+24 & 6.07E+15 & 1.67E+07 & 6.97E+03 & 5.55.E+03 & 1.08.E+04 & 0.646 & 0.515 & C\\
North-E & 1 & 4.97E+24 & 4.97E+15 & 8.23E+06 & 3.96E+24 & 3.96E+15 & 6.56E+06 & 1.26E+04 & 1.00.E+04 & 1.61.E+04 & 0.783 & 0.624 & C\\
 & 2 & 5.76E+24 & 5.76E+15 & 1.04E+07 & 4.59E+24 & 4.59E+15 & 8.29E+06 & 1.23E+04 & 9.78.E+03 & 1.11.E+04 & 1.107 & 0.882 & S\\
 & 3 & 9.72E+24 & 9.72E+15 & 1.13E+07 & 7.75E+24 & 7.75E+15 & 8.98E+06 & 5.03E+04 & 4.01.E+04 & 2.82.E+04 & 1.784 & 1.423 & S\\
North-M & 4 & 5.98E+24 & 5.98E+15 & 1.62E+07 & 4.77E+24 & 4.77E+15 & 1.29E+07 & 5.68E+03 & 4.53.E+03 & 9.41.E+03 & 0.603 & 0.481 & C\\
North-W & 5 & 6.44E+24 & 6.44E+15 & 1.53E+07 & 5.13E+24 & 5.13E+15 & 1.22E+07 & 7.95E+03 & 6.33.E+03 & 3.76.E+03 & 2.113 & 1.684& S \\
Northeast & 6 & 2.59E+24 & 2.59E+15 & 3.92E+06 & 2.06E+24 & 2.06E+15 & 3.12E+06 & 7.82E+03 & 6.23.E+03 & 2.93.E+03 & 2.664 & 2.124 & S\\
 & 7 & 5.36E+24 & 5.36E+15 & 1.02E+07 & 4.27E+24 & 4.27E+15 & 8.12E+06 & 1.03E+04 & 8.19.E+03 & 4.83.E+03 & 2.127 & 1.695 & S\\
 & 8 & 7.83E+24 & 7.83E+15 & 1.60E+07 & 6.24E+24 & 6.24E+15 & 1.27E+07 & 1.31E+04 & 1.04.E+04 & 1.29.E+04 & 1.016 & 0.810 & S\\
 & 9 & 5.69E+24 & 5.69E+15 & 1.09E+07 & 4.53E+24 & 4.53E+15 & 8.71E+06 & 1.07E+04 & 8.52.E+03 & 2.91.E+03 & 3.666 & 2.922 & S\\
 & 10 & 7.23E+24 & 7.23E+15 & 7.87E+06 & 5.76E+24 & 5.76E+15 & 6.28E+06 & 4.23E+04 & 3.37.E+04 & 1.29.E+04 & 3.278 & 2.613 & S\\
 & 11 & 4.03E+24 & 4.03E+15 & 1.09E+07 & 3.22E+24 & 3.22E+15 & 8.70E+06 & 3.83E+03 & 3.06.E+03 & 2.19.E+03 & 1.752 & 1.397 & S\\
East & 12 & 3.96E+24 & 3.96E+15 & 1.72E+07 & 3.16E+24 & 3.16E+15 & 1.37E+07 & 1.46E+03 & 1.17.E+03 & 4.74.E+03 & 0.308 & 0.246 & C\\
 & 13 & 3.98E+24 & 3.98E+15 & 1.01E+07 & 3.17E+24 & 3.17E+15 & 8.08E+06 & 4.27E+03 & 3.40.E+03 & 5.66.E+03 & 0.754 & 0.601 & C\\
Northwest & 14 & 3.85E+24 & 3.85E+15 & 1.54E+07 & 3.07E+24 & 3.07E+15 & 1.23E+07 & 1.68E+03 & 1.34.E+03 & 3.04.E+03 & 0.553 & 0.441 & C\\
West & 15 & 5.17E+24 & 5.17E+15 & 1.46E+07 & 4.12E+24 & 4.12E+15 & 1.16E+07 & 4.53E+03 & 3.61.E+03 & 5.92.E+03 & 0.766 & 0.611 & C\\
 & 16 & 6.16E+24 & 6.16E+15 & 2.85E+07 & 4.91E+24 & 4.91E+15 & 2.27E+07 & 2.00E+03 & 1.60.E+03 & 3.80.E+03 & 0.527 & 0.420 & C\\
 & 17 & 3.36E+24 & 3.36E+15 & 6.30E+06 & 2.68E+24 & 2.68E+15 & 5.02E+06 & 6.63E+03 & 5.28.E+03 & 6.83.E+03 & 0.970 & 0.773 & C\\
 & 18 & 6.33E+24 & 6.33E+15 & 1.74E+07 & 5.04E+24 & 5.04E+15 & 1.39E+07 & 5.82E+03 & 4.64.E+03 & 4.01.E+03 & 1.453 & 1.159 & S\\
 & 19 & 5.45E+24 & 5.45E+15 & 1.27E+07 & 4.34E+24 & 4.34E+15 & 1.02E+07 & 6.91E+03 & 5.51.E+03 & 7.68.E+03 & 0.899 & 0.717 & C\\
 & 20 & 5.10E+24 & 5.10E+15 & 9.89E+06 & 4.06E+24 & 4.06E+15 & 7.89E+06 & 9.40E+03 & 7.50.E+03 & 3.00.E+03 & 3.137 & 2.501 & S\\
 & 21 & 7.08E+24 & 7.08E+15 & 1.38E+07 & 5.64E+24 & 5.64E+15 & 1.10E+07 & 1.30E+04 & 1.04.E+04 & 8.82.E+03 & 1.474 & 1.175 & S\\
 & 22 & 6.95E+24 & 6.95E+15 & 1.66E+07 & 5.54E+24 & 5.54E+15 & 1.32E+07 & 8.47E+03 & 6.75.E+03 & 1.97.E+04 & 0.430 & 0.343 & C\\
 & 23 & 8.88E+24 & 8.88E+15 & 1.92E+07 & 7.08E+24 & 7.08E+15 & 1.53E+07 & 1.32E+04 & 1.05.E+04 & 1.41.E+04 & 0.936 & 0.746 & C\\
 & 24 & 6.93E+24 & 6.93E+15 & 1.78E+07 & 5.53E+24 & 5.53E+15 & 1.42E+07 & 7.28E+03 & 5.80.E+03 & 1.17.E+04 & 0.621 & 0.495 & C\\
 & 25 & 1.13E+25 & 1.13E+16 & 3.14E+07 & 9.04E+24 & 9.04E+15 & 2.50E+07 & 1.02E+04 & 8.17.E+03 & 4.69.E+03 & 2.183 & 1.741 & S\\
 & 26 & 7.02E+24 & 7.02E+15 & 1.00E+07 & 5.60E+24 & 5.60E+15 & 7.99E+06 & 2.40E+04 & 1.91.E+04 & 2.61.E+04 & 0.918 & 0.732 & C\\
 & 27 & 1.07E+25 & 1.07E+16 & 2.15E+07 & 8.51E+24 & 8.51E+15 & 1.71E+07 & 1.83E+04 & 1.46.E+04 & 2.31.E+04 & 0.789 & 0.629 & C\\
 & 28 & 5.94E+24 & 5.94E+15 & 9.94E+06 & 4.74E+24 & 4.74E+15 & 7.92E+06 & 1.48E+04 & 1.18.E+04 & 7.74.E+03 & 1.908 & 1.521 & S\\
 
 \hline
Average &  & 6.21E+24 & 6.21E+15 & 1.41.E+07 & 4.95E+24 & 4.95E+15 & 1.12E+07 & 1.17.E+04 & 9.36.E+03 & 9.57.E+03 & 1.412 & 1.125 &S=14\\
SD &  & 2.10E+24 & 2.10E+15 & 6.09.E+06 & 1.68E+24 & 1.68E+15 & 4.86E+06 & 1.10.E+04 & 8.80.E+03 & 7.30.E+03 & 0.923 & 0.736 & C=15 \\
\hline

\footnotetext{Column ``Note'':
S: supercritical mass, i.e., the LTE mass is larger than the virial mass. 
 In this case, the self-gravity of the HMC can overwhelm the gas pressure.
Cloud contraction under these circumstances efficiently forms high-mass protostars  nearly isothermally even if the core is still being heated.  
C: LTE mass is comparable with or  smaller than the virial mass. 
 In this case, the HMC is either critical or subcritical. \\
}
\end{tabular}
}
\end{center}

\end{minipage}
} 
\end{table*}
\normalsize


\section{DISCUSSION}

\subsection{Hot Molecular Cores, UCHIIs, and Masers}

The lifetimes of HMCs are 10$^4$--10$^5$~yr \citep{Herbst2009, Battersby2017}.
 Those of UCHIIs are $\sim$10$^4$ yr \citep{Yusef-Zadeh2010, Tsuboi2019} and $\sim$10$^5$~yr to reach a radius of a parsec and to break its parent molecular cloud \citep{Akeson1996, Churchwell2002, MacLow2007}, respectively.
 The fact that the HMC exists up to the Hollow HMC (HHMC) stage, which is explained in the next paragraph,  in the 50~MC implies that  $\sim$10$^5$ yr has passed since star formation was triggered by CCC.


From a viewpoint of chemical  evolution,  there is a stage, the Hollow HMC (HHMC), between the HMC and HII region \citep{Stephan2018}.
The ``Hollow Hot Core'' has the same density structure as the HMC, but it contains a cavity ionized by  a HCHII or  UCHII  at the center.  
The HHMC has the HCHII or  UCHII in the last stage of the HMC \citep[e.g.][]{Furuya2011, Rolffs2011, Serra2012, Fuente2018}. 
 As such, the evolutionary stage of the HMC is divided into  three stages, ``Dense Core'', ``Hot Core'', and ``Hollow Hot Core.'' 
In the  ``Dense Core'' stage,  a warm dense dust core or HC$^{15}$N core exist at the center. 
 In the ``Hot Core''  stage, the core becomes  warm and dense  with  a HMPO, which is associated with outflow and disk.
In the ``Hollow Hot Core'' stage, the core  consists of either a HCHII or  UCHII, probably associated with outflow and disk \citep[e.g.][]{DePree2000, Tanaka2016}.

The correlation  among the HMCCs,  dust cores,  HC$^{15}$N cores, and  CH$_3$OH masers with our data presented in Table~\ref{Table01} and Table~\ref{Table03} provides a statistical basis for estimating the relative timescales of the three stages.
The ratios of the HMCCs associated with  dust cores, HC$^{15}$N cores, and CH$_3$OH masers in our data of 28 HMCCs  are 18\% (6/28),
 32\% (9/28), and  54\% (15/28), respectively.
The ratio of the HMCCs associated with both  HC$^{15}$N  and  CH$_3$OH masers is only 18\% (5/28), and
 that with both the H$_2$O and  CH$_3$OH masers is only 4\% (1/28).
In  the early evolutionary stage,  the HMC is assume to be still associated with a dust core or HC$^{15}$N core, but the dust core soon dissipates.
In  the intermediate evolutionary stage, the outflow from a protostar in the evolving HMC excites  CH$_3$OH Class-I masers before  a HCHII is formed, but  CH$_3$OH Class-II masers are not yet pumped.

 Among the samples included in our data, 
H$_2$O and Class-I CH$_3$OH masers  were detected  in the region of HMC02  whereas no H$_2$CO masers  were detected \citep{Lu2019b}.
And SiO (v=2; J=2$-$1)  has not  been detected in our observational bands, either. 
Some indications that   H$_2$O masers are produced in disks have been also reported \citep[e.g.][]{Garay1999, Beuther2002}.
Detection of H$_2$O masers  suggests that the region is in an early stage of massive star formation,
 although we cannot  constrain exactly which stage  of the HMC the region is in.

\begin{figure*}[htb]
\includegraphics*[bb= 50 500 800 850, scale=0.9]{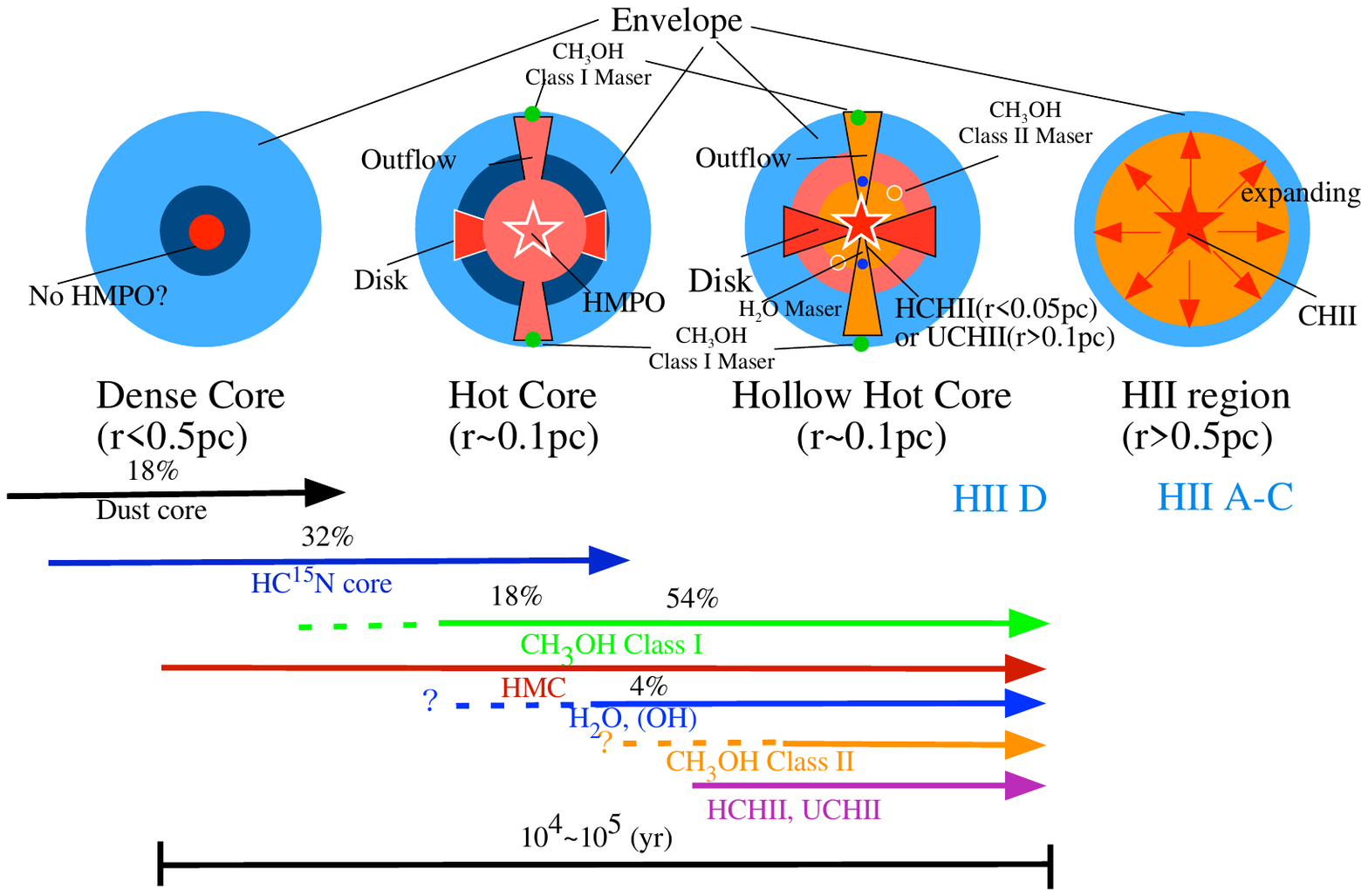}
\caption{
 Proposed evolutionary sequence  in the 50~MC for the HMCs,  HC$^{15}$N cores,  CH$_3$OH Class-I and Class-II masers, H$_2$O masers, OH masers, and  HII regions.
The ``Dense Core'' corresponds to the HC$^{15}$N core.
The evolutionary stages of the identified HMCs in our work include the ``Dense Core'',  ``hot core'', and   ``Hollow Hot Core''.
}
\label{Fig10}
\end{figure*}

 \citet{Beuther2002} presented a comparison of Class-II CH$_3$OH (6.7 GHz) and H$_2$O (22.2 GHz) masers in a sample of 29 massive star-forming regions. 
Whereas all masers are associated with massive mm cores,  only 3 out of 18 CH$_3$OH masers and 6 out of 22 H$_2$O masers are associated with cm emission, the fact of which likely indicates the presence of a recently ignited massive star \citep {Beuther2002}.
The Class-II CH$_3$OH masers are associated with deeply embedded HMPOs or HCHIIs and are excited by warm dust \citep{Pestalozzi2002, Walsh2003, Minier2005}. 
The Class-II CH$_3$OH and H$_2$O masers require a similar environment to be excited, but with the different excitation processes (radiative pumping for CH$_3$OH and collisional pumping for H$_2$O) \citep[e.g.][]{Garay1999, Beuther2002}.
 \citet{Beuther2002} suggested that the kinematic structures in the different maser species show no recognizable patterns, and  failed to draw a firm conclusion as to whether the features are produced in disks, outflows or expanding shock waves.
Therefore in the absence of  Class-II CH$_3$OH masers and H$_2$O masers except in one example, 
the HMCCs in the 50~MC are likely to be in the early stage of the HMC evolution before the formation of HCHIIs,.
As mentioned the above, our proposed evolutionary sequence of the HMCs in the 50 MC is summarized in Figure~\ref{Fig10}.

If the HMC evolves to  a HHMC with HCHIIs and UCHIIs,  the observed HHMCs in the 50~MC provide a statistical basis for estimating the relative timescales of these two evolutionary stages.
 \citet{Wilner2001}  argued that the number ratio of the HMCs to UCHIIs is about 50\% and suggested that this number ratio tracks the relative lifetimes of the two kinds of objects in W49A,  $\tau_{\rm Hot Core}$ $\sim$ 0.5 $\times$ $\tau_{\rm UCHII}$.
Moreover, \citet{Furuya2005} found that the HMC stage should last less than  one-third of the UCHII stage in G19.61$-$0.23, $\tau_{\rm Hot Core}$ $\sim$ 1/3 $\times$  $\tau_{\rm UCHII}$.
On the  contrary,  \citet{Miyawaki2002} presented the result with the SO emission lines from more HMCs adjacent to the UCHIIs in W49A  than those studied by \citet{Wilner2001} and concluded $\tau_{\rm Hot Core}$ $>$ $\tau_{\rm UCHII}$.
The differences between these objects indicates the differences of the evolutionary stage of HMCs that are formed at approximately the same time.

This work, as well as that by  \citet {Lu2019a}, shows that there are no UCHIIs  or HCHIIs in the HMCCs although  some HMCCs are located adjacent to  UCHIIs.
Many CH$_3$OH Class-II masers are known to be associated with HCHIIs but are no longer associated with  UCHII regions \citep {Phillips1998, Walsh1998}.
We conclude that most of the HMCCs in the 50~MC are in the stages between the ``Dense Core'' and ``Hot Core''.

\subsection{Relation between the Mass  and Size}

 We have concluded that 
the HMCCs in the 50~MC are in the early evolutionary stage (previous section). 
We have derived he masses  from dust emission,  LTE masses, and  virial masses  with certain uncertainties  in \S3.5.1 and \S3.5.2.
 In this section, we discuss the relation between the masses and other parameters, in particular the size, of the HMCCs in this stage.

\begin{figure}
\includegraphics*[bb= 0 150 500 620, scale=0.4]{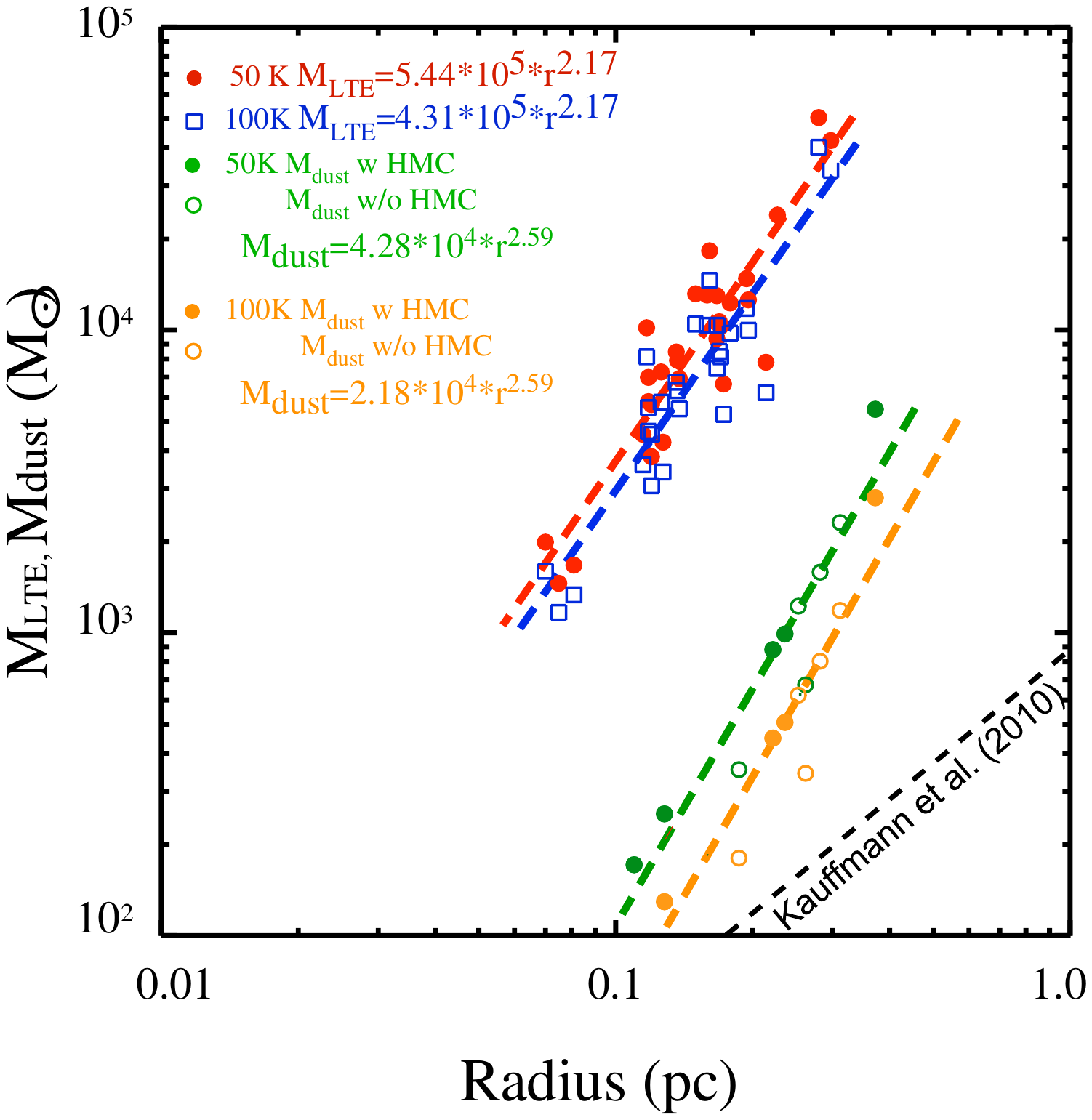}
\caption{
Mass-size relation of the HMCCs. 
Red filled circles and blue open squares indicate the LTE masses calculated  with  {\it T}$_{\mathrm{ex}}$ = 50~K and 100~K, respectively.
Green filled and open circles indicate the dust cores with {\it T}$_{\mathrm{d}}$=50~K associated with  a HMCC and no HMCC, respectively. 
Orange filled and open circles are the same as respective green circles except that the assumed temperature is  {\it T}$_{\mathrm{d}}$=100~K.
Black dashed line  shows the mass and size relation of cloud fragments in several molecular clouds over a wide range  of radius 0.05 $\leq$ r $\leq$ 3~pc \citep{Kauffmann2010a, Kauffmann2010b}. 
\label{Fig11}}
\end{figure}

Figure~\ref{Fig11} shows the mass-radius relations for the identified HMCCs and  dust cores in this work.
The masses of the HMCCs and  dust cores are  found to be proportional to their radii with very small scatters; the former is
 M$_{\rm LTE}$/(M$_\odot$)=5.44 $\times $10$^5$ (r/(pc))$^{2.17}$ 
and  4.31 $\times $10$^5$ (r/(pc))$^{2.17}$ for {\it T}$_{\mathrm{ex}}$ = 50~K and 100~K, respectively,
and the latter is M$_{\rm dust}$/(M$_\odot$)=4.28 $\times $10$^4$ (r/(pc))$^{2.60}$  and 
2.18 $\times $10$^4$ (r/(pc))$^{2.60}$ for {\it T}$_{\mathrm{d}}$=50~K and 100~K, respectively.  
 For the same radius, the mass of the dust core is an order of magnitude smaller than that of the HMCC.
The power-law index of the mass-radius relation for the dust cores is slightly larger than that for the HMCCs.
 Comparison of the mass of the dust core associated with  a HMCC to the mass of the HMCC yields a suggestion that 
X(SO) might be as small as $\sim$10$^{-10}$, which is an order of magnitude smaller than our adopted value of 10$^{-9}$ (in \S3.5.2).
These relations are considerably different from that of molecular clumps, M(M$_\odot$) $\leq$ 870 (r/(pc))$^{1.33}$ over a wide range of radius of 0.05 $\leq$ r $\leq$ 3 pc by \citet{Kauffmann2010b}, which is similar to the famous Larson's relation \citep{Larson1981}.

\begin{figure}
\includegraphics*[bb= 0 150 500 620, scale=0.4]{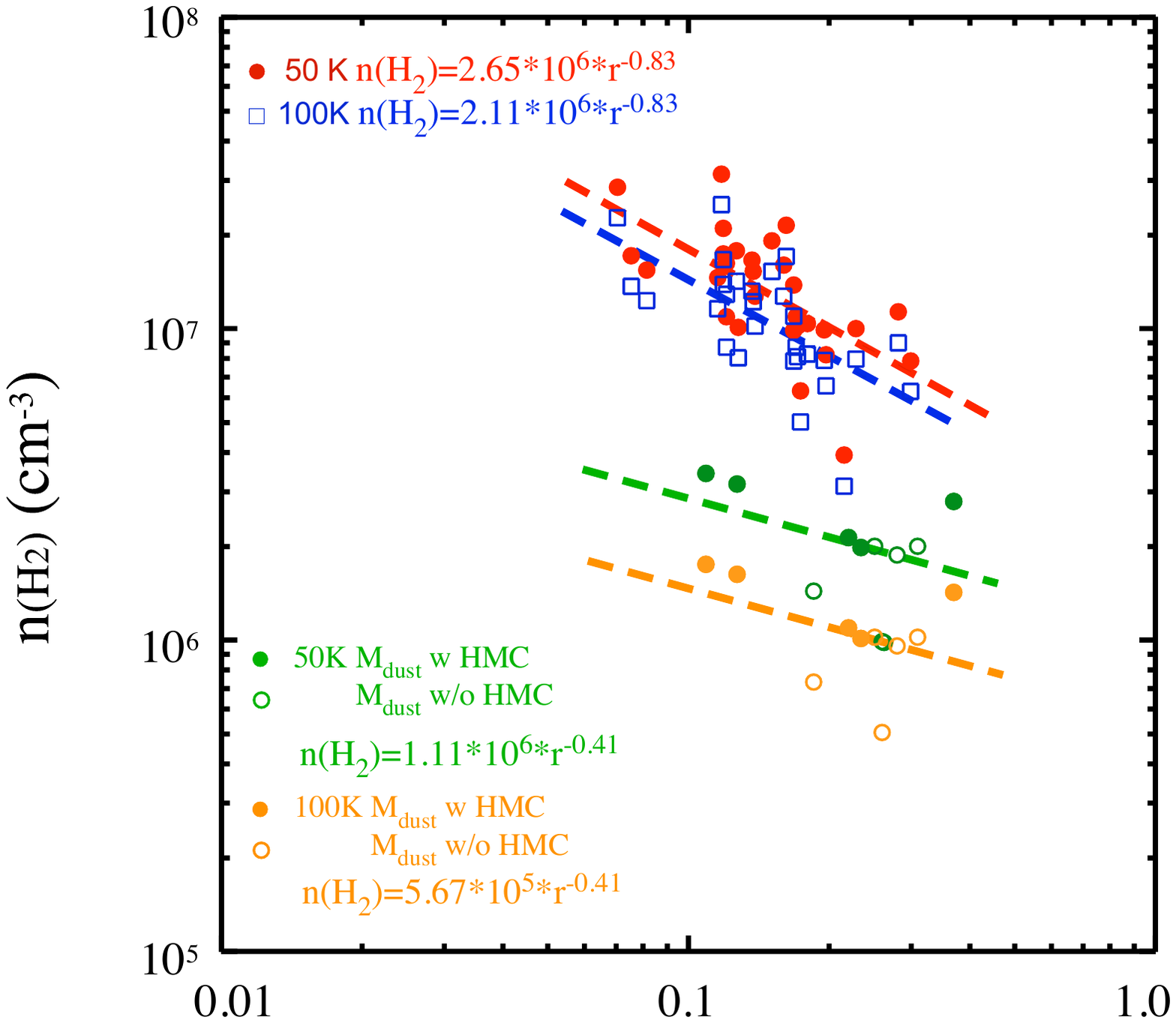}
\caption{
Relation between the mean hydrogen number density  and size  of the HMCCs of our sample. 
The symbols and colors, as well as the assumed temperatures, are the same as those in Figure~\ref{Fig11} except for the difference in the units of the vertical axis.   

\label{Fig12}}
\end{figure}

Figure~\ref{Fig12} shows the relation between the mean number density and radius  for the HMCCs and  dust cores; the former is
n({\rm H$_2$}) = 2.65 $\times $10$^6$ (r/(pc))$^{-0.83}$ and n({\rm H$_2$}) = 2.11 $\times $10$^6$ (r/(pc))$^{-0.83}$ for {\it T}$_{\mathrm{ex}}$ = 50~K and 100~K, respectively, and the latter is
 n({\rm H$_2$}) = 1.11 $\times $10$^6$ (r/(pc))$^{-0.41}$  and 
n({\rm H$_2$}) = 5.67 $\times $10$^5$ (r/(pc))$^{-0.41}$ for  {\it T}$_{\mathrm{d}}$=50~K and 100~K, respectively.  
The number density is found to be a decreasing function of the  radius. 
The slopes of the mean density derived from the dust emission are  less steep than those of the HMCCs.
If the molecular gas is compressed in one dimension, the number density is expected to obey the proportional relation n({\rm H$_2$}) $\propto$ r$^{-1}$.  Hence, the derived slopes for the HMCCs suggest  that some mechanism is likely to be in operation to compress molecular gas in one dimension to make the dense HMCCs.

Since the dynamic range of the column density, N(SO),   which can be traced  with the molecular emission line SO, is narrow,
 M$_{\rm LTE}$  $\propto$ r$^2$$\times$N (SO) $\propto$ r$^ 2$ (Figure~\ref{Fig11}) and n({\rm H$_2$})$\sim$M$_{\rm LTE}$/r$^3$ $\propto$ r$^{ -1}$ (Figure~\ref{Fig12}). Then, one might suspect that the slopes for the HMCCs in Figures~\ref{Fig11} and \ref{Fig12} might be an artifact. 
However, the four dust cores with the HMCCs show the slope $\sim$r$^{2.5}$, which is similar  to those of the HMCCs. Admittedly, the result is not yet conclusive, given that the number of our sample of HMCCs is limited and
 only four HMCCs are associated with the dust cores.

The free-fall time, {\rm t$_{\mathrm {ff}}$},  of the HMCC is
\begin{equation}
\mathrm{t_{ff}}  = \sqrt{\frac{\mathrm{3 \pi}}{\mathrm{32G\rho}}}  ,
\end{equation}
where G is the gravitational constant and $\rho$ is the mean mass density.
The mean free-fall time of the HMCC is calculated to be 1.16 $\pm$ 0.33 $\times$ 10$^4$~yr.
 This value is comparable  with the lifetime of the HMCC.

\begin{figure}
\includegraphics*[bb= 0 150 500 620, scale=0.4]{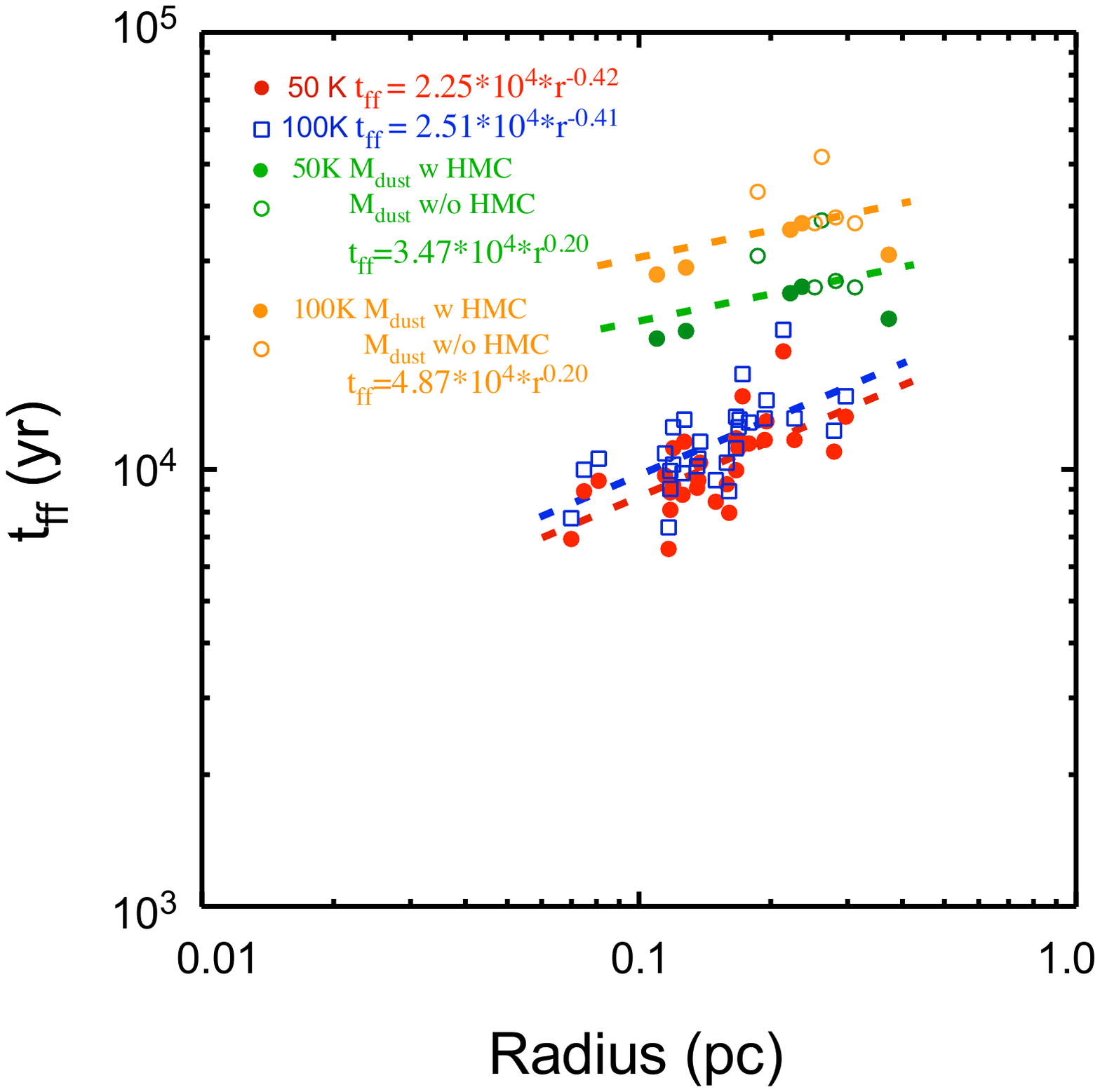}
\caption{
Relation between the free-fall time  and size  of the HMCCs of our sample. 
The symbols and colors, as well as the assumed temperatures, are the same as those in Figure~\ref{Fig11} except for the difference in the units of the vertical axis.   

\label{Fig13}}
\end{figure}

Figure~\ref{Fig13}  plots the free-fall time derived from equation~6 vs. radius of the HMCCs and the dust cores; the former is
${\mathrm{t_{ff}}}$/(yr) = 2.25 $\times$ 10$^4$  (r/(pc))$^{0.42}$ and
 2.51 $\times$ 10$^4$  (r/(pc))$^{0.41}$ for {\it T}$_{\mathrm{ex}}$ = 50~K and 100~K, respectively, and the latter is
${\mathrm{t_{ff}}}$/(yr)  = 3.47 $\times$ 10$^4$  (r/(pc))$^{0.20}$  and 
 4.87 $\times$ 10$^4$  (r/(pc))$^{0.20}$ for    {\it T}$_{\mathrm{d}}$=50~K and 100~K, respectively.  
The slope of the dust core is  less steep than that of the HMCCs. The difference in the free-fall times  must be due to the difference in the masses.

\begin{figure}
\includegraphics*[bb= 0 150 550 650, scale=0.4]{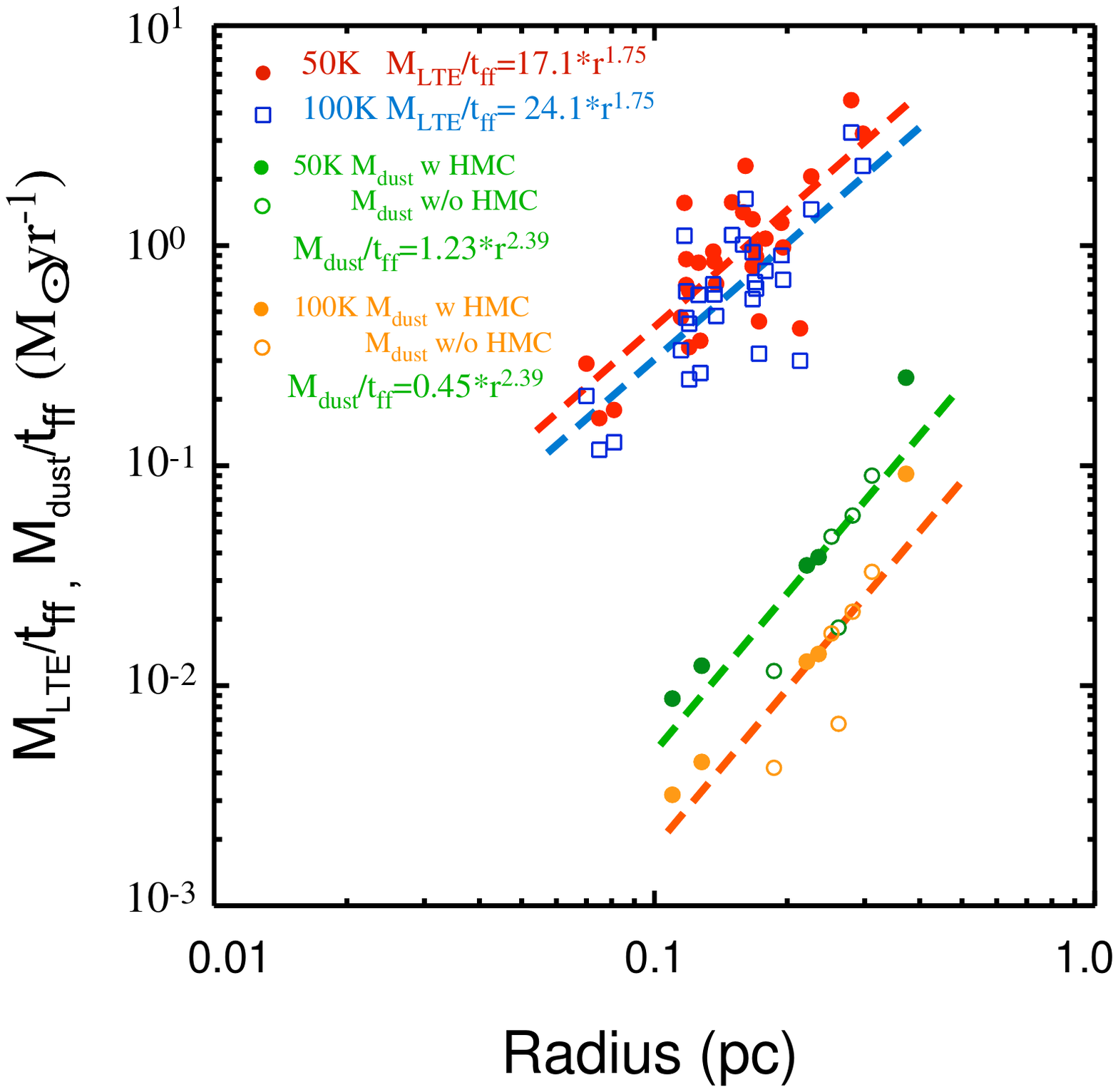}
\caption{
Relation of the ratio of the LTE mass to the free fall time of the HMCCs and the dust cores against the radius. 
The symbols and colors, as well as the assumed temperatures, are the same as those in Figure~\ref{Fig11} except for the difference in the units of the vertical axis.   

\label{Fig14}}
\end{figure}

Figure~\ref{Fig14} shows the ratio of the LTE mass to the free fall time of the HMCCs and the dust cores as a function of the radius; the former is
M$_{\rm LTE}$/${\mathrm{t_{ff}}}$  = 17.1 $\times$ (r/(pc))$^{1.75}$ and
 24.1 $\times$ (r/(pc))$^{1.75}$ for {\it T}$_{\mathrm{ex}}$ = 50~K and 100~K, respectively, and the latter is
M$_{\rm dust}$/${\mathrm{t_{ff}}}$  = 1.23 $\times$(r/(pc))$^{2.39}$  and 
 0.45 $\times$(r/(pc))$^{2.39}$ for T$_{\rm{d}}$=50~K and 100~K, respectively.
The mass accretion rate, ${\mathrm{\dot{M}}}$, is calculated as
\begin{equation}
\mathrm{\dot{M} (M_\odot/yr)} ={\mathrm{f_{infall}}}\ \frac{\mathrm{M}_{\mathrm{LTE}}(\mathrm{M}_\odot)}{\mathrm{t_{ff}}(yr)},
\end{equation}
where ${\mathrm{f_{infall}}}$ is the factor  in $v_{\mathrm{in}}$=${\mathrm{f_{infall}}}$$\times$$v_{\mathrm{ff}}$ 
 for the velocities of infall ($v_{\mathrm{in}}$) and free-fall ($v_{\mathrm{ff}}$).
The value of  ${\mathrm{f_{infall}}}$  was estimated to be 0.2--0.3 \citep{Wyrowski2012}.
 For r=0.1~pc, ${\mathrm{\dot{M}}}$ reaches (6.1--9.1)$\times$ 10$^{-2}$~M$_\odot$yr$^{-1}$  and  
(8.6--12.8) $\times$ 10$^{-2}$~M$_\odot$yr$^{-1}$ for the HMCCs with  {\it T}$_{\mathrm{ex}}$ = 50~K and 100~K, respectively.
 The mass accretion rate ${\mathrm{\dot{M}}}$ is required to be at least 3 $\times$10$^{-3}$~M$_\odot$yr$^{-1}$ to form massive stars \citep{Fazal2008}.
A massive star has, in general, a Keplerian disk with Class-II CH$_3$OH masers \citep{Beltran2016}.
\citet{Sanna2019} showed that the accretion disk is truncated between 2000 and 3000 au,  has a mass of about one-tenth of the central star mass, and is infalling toward the central star at a high rate of 6 $\times$10$^{-1}$~M$_\odot$yr$^{-1}$.
It is, however, unclear what kind of structure is inside  the HMC  at several thousands au and in the massive-protostar disc \citep{Kratter2006, Goddi2011}. 
At least, our estimated accretion rate is sufficient for massive star formation.

\begin{figure}
\includegraphics*[bb= 0 150 550 600, scale=0.40]{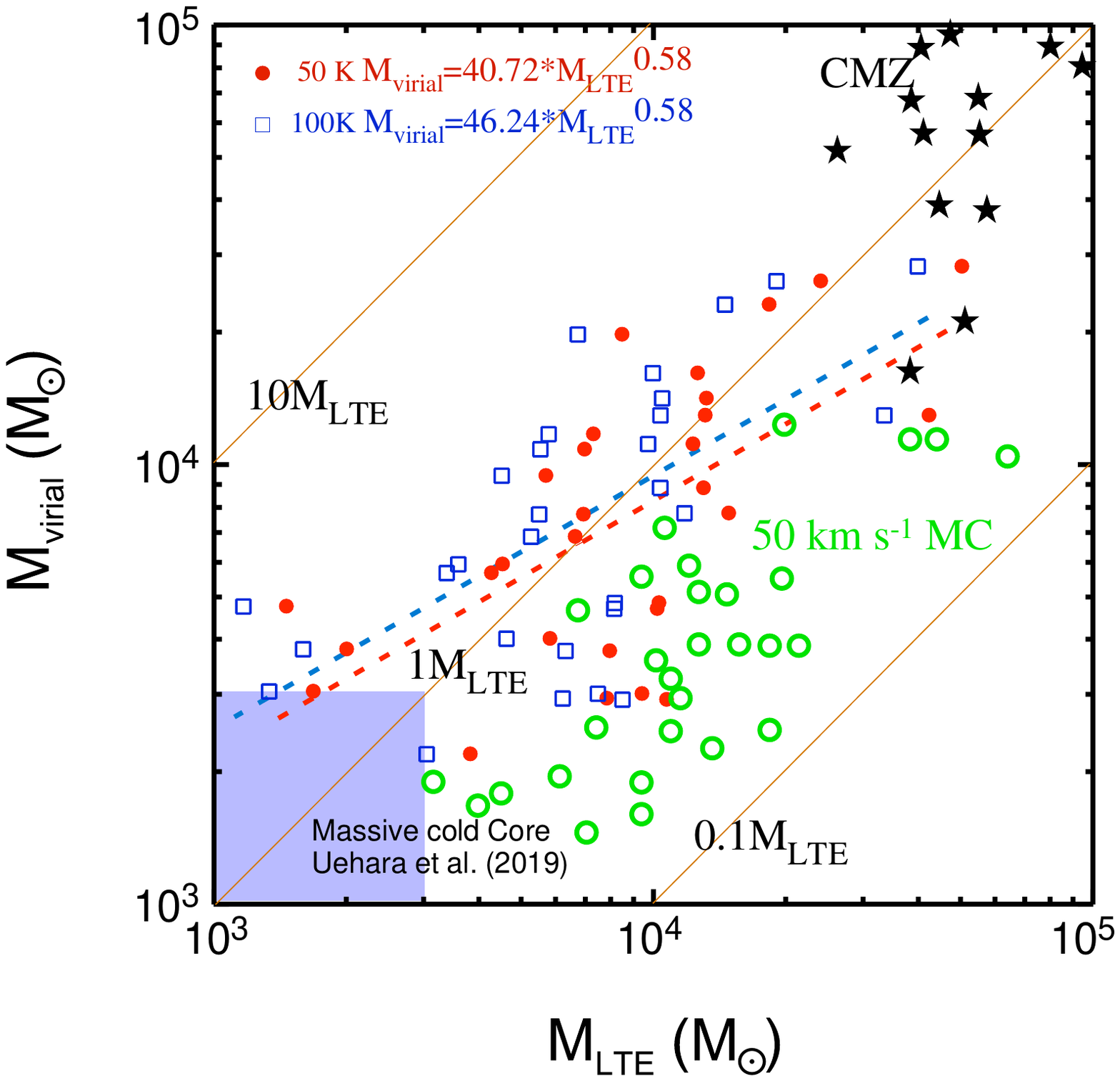}
\caption{
Relation between the LTE mass  and virial mass  of the HMCCs of our sample for the assumed temperatures of 
  {\it T}$_{\mathrm{ex}}$ = 50~K (in red filled circles) and 100~K (blue open squares). 
Orange lines show the relations M$_{\mathrm{virial}}$=10~M$_{\mathrm{LTE}}$, 1~M$_{\mathrm{LTE}}$, and 0.1M$_{\mathrm{LTE}}$.
Green circles and asterisks indicate the results of the molecular clumps in the 50~MC \citep{Tsuboi2012} and those
 in the CMZ \citep{Miyazaki2000}, respectively.
Purple square  at the bottom of the left corner indicates the range of the mass of massive cold cores \citep{Uehara2019}.
\label{Fig15}}
\end{figure}

Figure~\ref{Fig15} shows the scattered plot between the LTE mass  and virial mass  of the HMCCs:
M$_{\mathrm{virial}}$ = 40.72 $\times$ M$_{\mathrm{LTE}}$$^{0.58}$  
and  46.24 $\times$ M$_{\mathrm{LTE}}$$^{0.58}$ for the {\it T}$_{\mathrm{ex}}$ = 50~K and 100~K, respectively.
The LTE masses of the HMCCs are almost proportional to the virial masses, as expected.
\citet{Tsuboi2012}  presented the LTE mass vs. virial mass relation of the molecular clumps in the 50~MC, 
 showing the virial masses are about one order of magnitude smaller than the LTE masses  for a similar mass range to that for the HMCCs of our samples.
\citet{Miyazaki2000} showed the LTE masses of the molecular clumps in the CMZ are proportional to the virial masses with $>$10$^4$~M$_\odot$.
\citet{Uehara2019} suggested that the molecular gas compression by the CCC efficiently formed the massive bound cores or massive cold cores with high masses of 2500--3000~M$_\odot$ or more.
Our data points of the HMCCs  at the bottom-left corner of Figure~\ref{Fig15} are actually smoothly connected  to the data points  of the most massive cold cores  presented by \citet{Uehara2019}.

\begin{figure}
\includegraphics*[bb=20 190 550 650, scale=0.40]{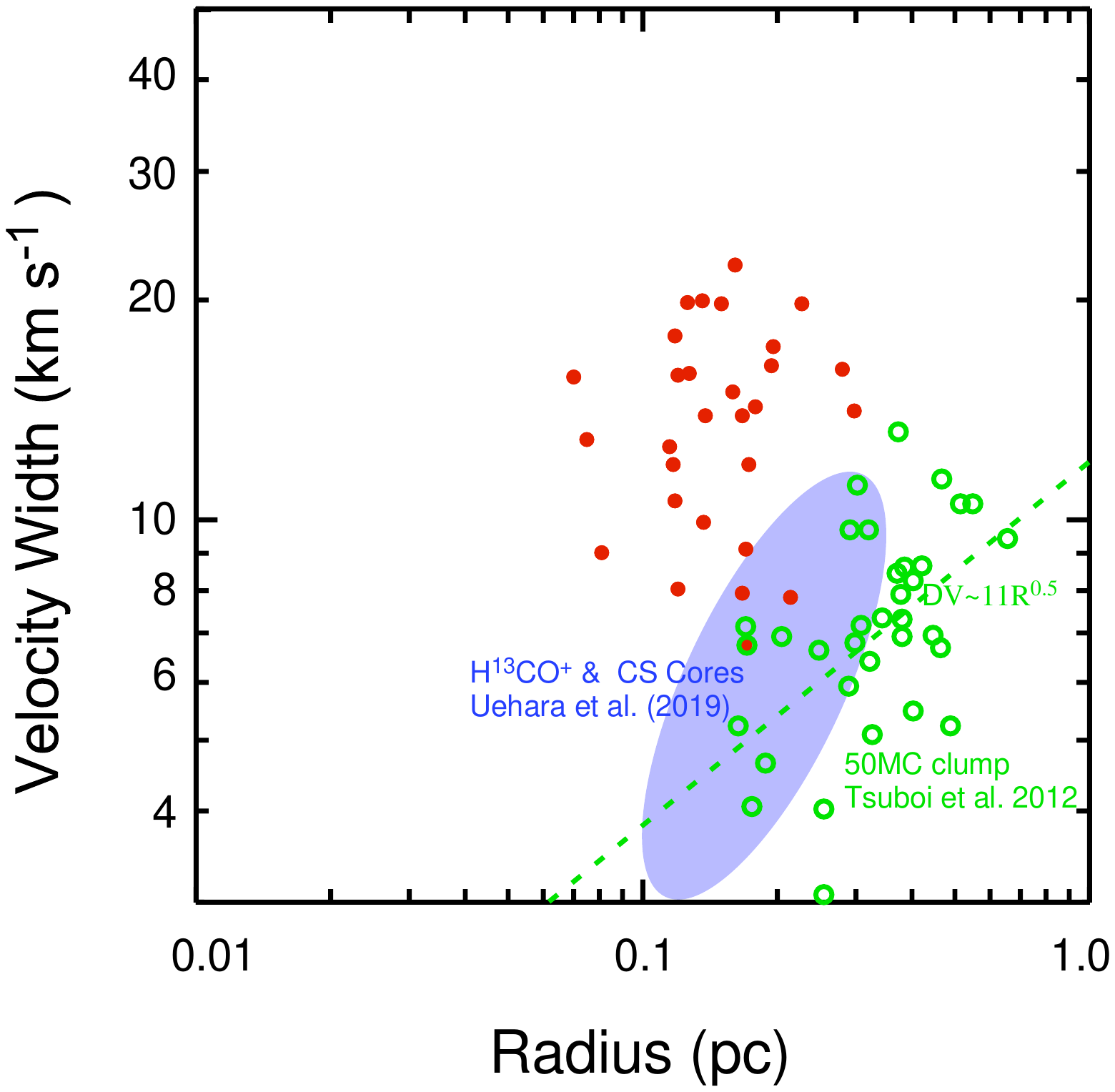}
\caption{
Scattered plot of the velocity width vs. radius  of the HMCCs of our sample. 
Red filled circles indicate the velocity widths of the HMCCs. 
Green circles indicate  those of the molecular clumps in the 50~MC \citep{Tsuboi2012}.
Purple oval  near the bottom  center shows the range of the velocity width-radius relation of cold cores derived from the H$^{13}$CO$^+$ and CS emissions \citep{Uehara2019}.
\label{Fig16}}
\end{figure}

The  scattered plot between the size and  velocity dispersion of the HMCCs in the 50~MC is shown in Figure~\ref{Fig16}. 
 A high degree of scatter is apparent in the plot, as quantitatively supported with  the small correlation coefficient, indicating that the size and linewidth of the HMCCs are not correlated.
 However, Larson's first law \citep{Larson1981} shows power-law dependence of the velocity dispersion on the region size.
We find that the data points of several HMCCs overlap with those of  the cold cores \citep{Uehara2019} and that the others have larger velocity widths 
 than those of the cold cores.
Therefore, the turbulence in the HMCs  must be more active than that  in the cold cores.

Although the outflow components are known to be significant at higher frequencies \citep{Tak2003}, these components are not significant in our observations.
Therefore, the velocity width of the SO emission line means the turbulent velocity of the HMC itself.
The results of turbulent velocities show  a hierarchical structure  for the molecular clump,  cold core, and  HMC (see, e.g., HMC05 and HMC27).
If we adopt the hypothesis that  cold cores  coalesce in CCC  and form a large-mass core,  which is the parent body of a HMC, 
Figure~\ref{Fig15} and Figure~\ref{Fig16}  give indication about  the coalescence process in which the radius does not change much  while the velocity width and density increase.
If a single large-mass core is formed,  a HMPO should be formed due to the contraction of the core.

\begin{figure}
\includegraphics*[bb=20 190 550 650, scale=0.40]{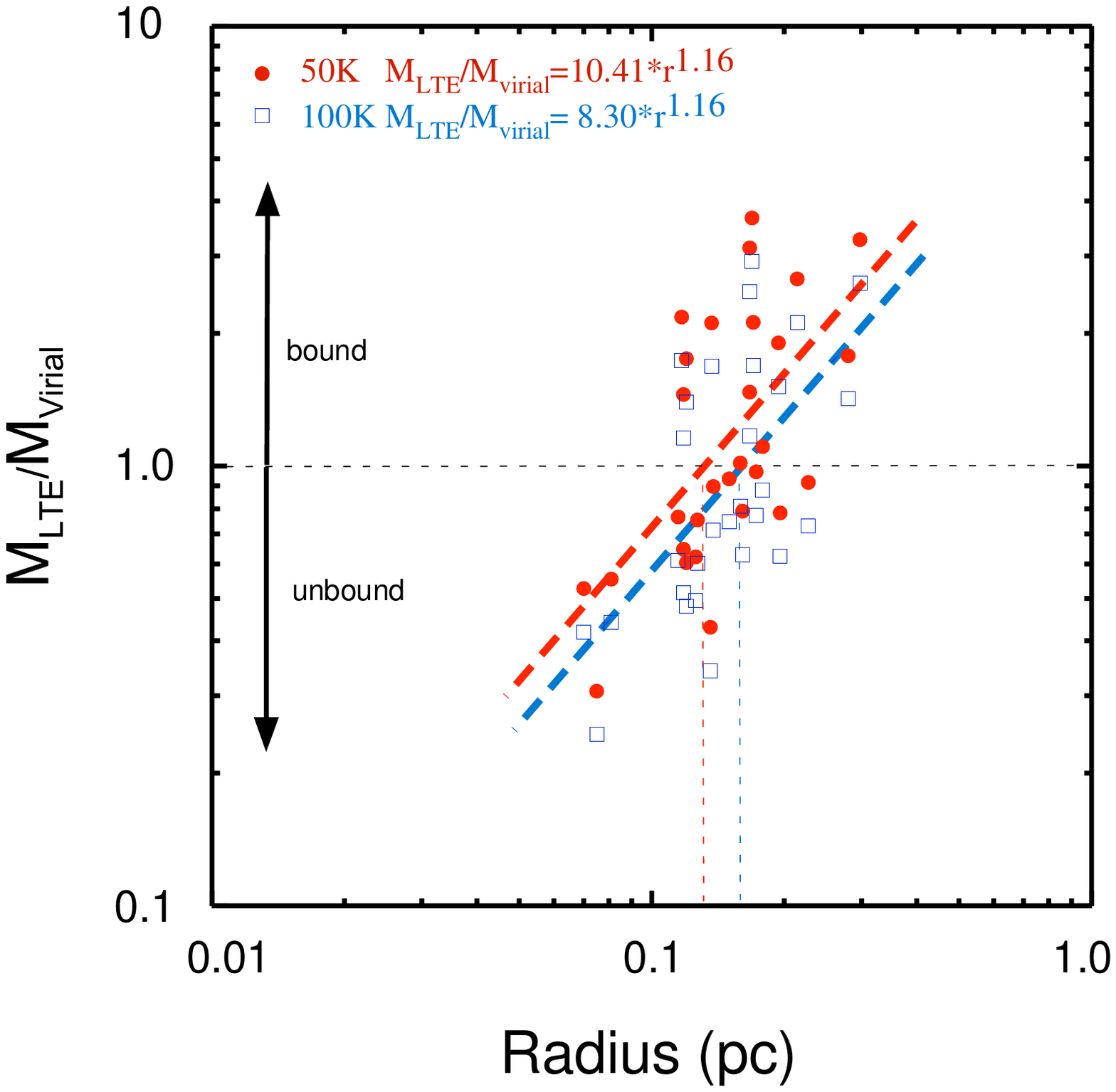}
\caption{
Relation between the ratio of the virial  to  LTE masses  and radius of the HMCCs of our sample. 
Red filled circles and blue open squares indicate the LTE masses calculated  for {\it T}$_{\mathrm{ex}}$ = 50~K and 100~K, respectively. 

\label{Fig17}}
\end{figure}

Figure~\ref{Fig17} shows the relation between the ratio of the virial  to  LTE masses  and radius  of the HMCCs of our sample:
M$_{\mathrm{LTE}}$/M$_{\mathrm{virial}}$ = 10.41 $\times$ (r/(pc))$^{1.16}$  and 
 8.30 $\times$ (r/(pc))$^{1.16}$ for  {\it T}$_{\mathrm{ex}}$ = 50~K and 100~K, respectively.
The ratio M$_{\mathrm{LTE}}$/M$_{\mathrm{virial}}$  is 1.0 at  radii of 0.13~pc  and 0.16~pc for  {\it T}$_{\mathrm{ex}}$ = 50~K and 100~K, respectively.
Here, M$_{\mathrm{LTE}}$/M$_{\mathrm{virial}}$ $>$ 1  means gravitationally bound and $<$1, gravitationally unbound.
In this case, the HMCCs with r $<$ 0.1~pc are gravitationally unbound,  whereas those with  r $>$ 0.1~pc are bound. 

\begin{figure}
\includegraphics*[bb=20 190 550 650, scale=0.40]{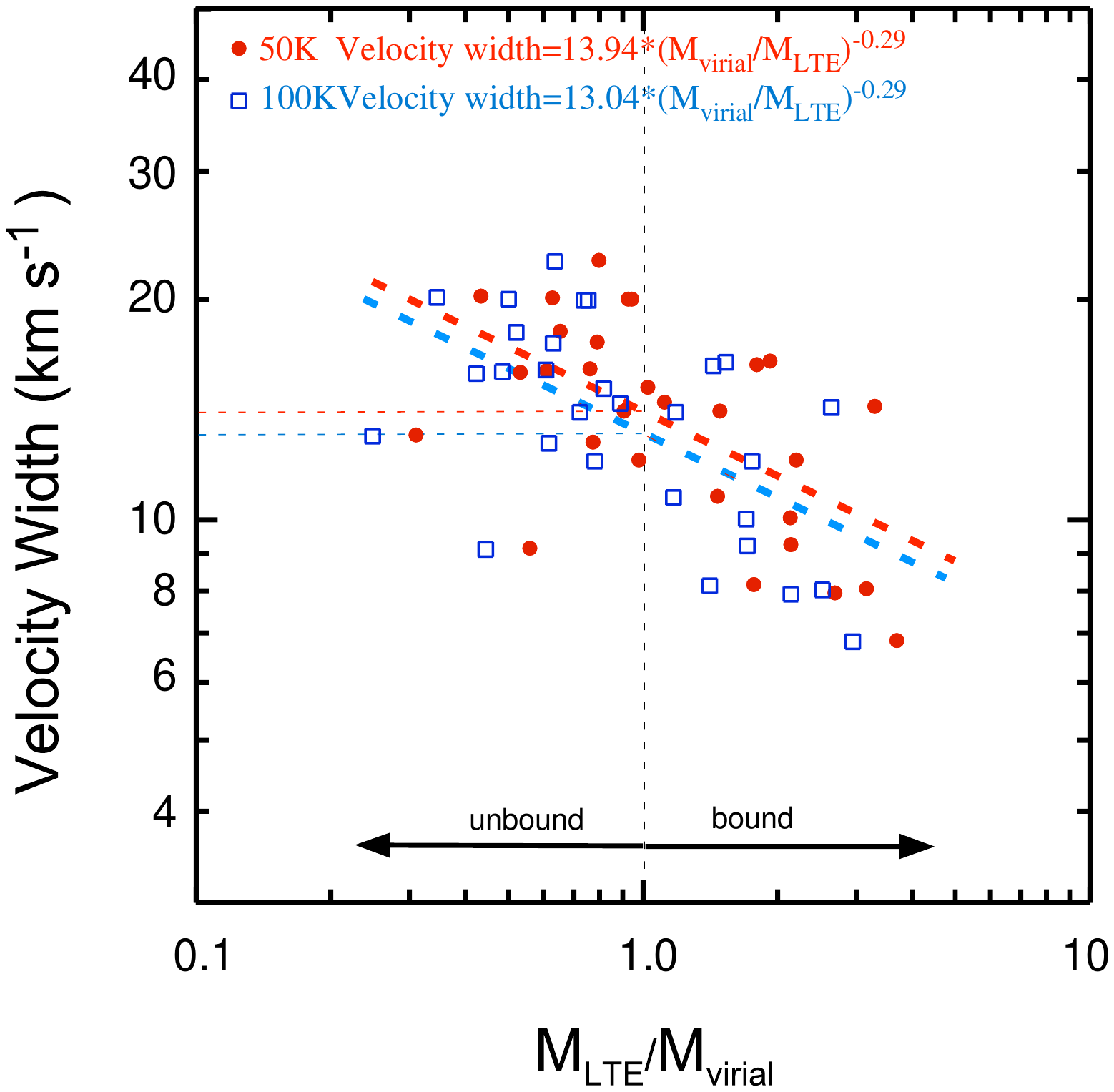}
\caption{
Relation between the velocity width  and the ratio of the LTE to virial mass  of HMCCs of our sample. 
Red filled circles and blue open squares indicate the LTE masses calculated  for {\it T}$_{\mathrm{ex}}$ = 50~K and 100~K, respectively. 
\label{Fig18}}
\end{figure}

Figure~\ref{Fig18} shows the relation between the velocity width  and the ratio of the LTE  to  virial masses  of HMCCs of our sample: 
$\Delta$V$_{\mathrm{FWHM}}$ = 13.94 $\times$ (M$_{\mathrm{LTE}}$/M$_{\mathrm{virial}}$)$^{0.29}$  and 
 13.04 $\times$ (M$_{\mathrm{LTE}}$/M$_{\mathrm{virial}}$)$^{0.29}$ for  {\it T}$_{\mathrm{ex}}$ = 50~K and 100~K, respectively.
When M$_{\mathrm{LTE}}$/M$_{\mathrm{virial}}$ = 1.0, $\Delta$V$_{\mathrm{FWHM}}$ are 13.94 \,km\,s$^{-1}$  and 13.04 \,km\,s$^{-1}$,  respectively.
These  imply that the HMCCs with $\Delta$V$_{\mathrm{FWHM}}$ $<$ 13 \,km\,s$^{-1}$  are gravitationally bound,  whereas those with  $\Delta$V$_{\mathrm{FWHM}}$ $>$ 14 \,km\,s$^{-1}$  are gravitationally unbound.
In the case of the velocity range, 13 \,km\,s$^{-1}$ $<$ $\Delta$V$_{\mathrm{FWHM}}$ $<$ 14 \,km\,s$^{-1}$, 
it depends on the temperature of HMCC,
 and it is unclear whether HMCC is gravitationally bound or unbound. 

It is unclear  whether the unbound regions in Figures~\ref{Fig17} and \ref{Fig18} indicate  the evolution of the HMC or  characteristics of a small HMC with a small LTE mass.
On the basis of star formation theories, however,
if the surrounding material is accreting to the central HMSCs, dissipation by outflows will let the HMCs gravitationally unbound and their central density profiles will be flattened \citep[e.g.][]{Shu1987}. 
Outflows with high mass-loss rates as well as radiation pressure and strong winds from the HMPOs or HCHIIs may accelerate the dissipation process.
Therefore, some of the HMCCs with a large velocity width in the 50~MC may evolve into the HMCs with HMPOs but  without HCHIIs, yet.

\subsection{Formation and Evolution of Massive Star in the 50~MC}
In \S4.1, 
 we have concluded that most of the HMCCs in the 50~MC are in the early stages. 
In \S4.2, 
 some correlations have been presented between the HMCC radius,  mass, density, mass accretion rate, etc. 
The overall inactive star formation in these CMZ clouds (see \S1) may be a result of the strong turbulence in this region and/or clouds being still in the very early evolutionary stage where the collapse has only recently started \citep{Lu2019a}.

 Formation of massive stars requires that 
molecular clumps with a moderate density ($\sim10^{4-5}$~cm$^{-3}$)  be compressed  with some mechanism, such as CCC, supernova explosion,  etc. 
In the 50~MC, CCC \citep[e.g.][]{Inoue2013, Fukui2020} is required efficiently to form massive and unstable cores as some simulation works show \citep[e.g.][]{Takahira2014, Takahira2018}.
It  would be reasonable to assume that a large amount of gas is compressed by not only gravity but some other factors triggered in r $\sim$ 0.1~pc  before a HMC is formed, which is unstable and eventually collapses to form massive stars with a high mass-accretion rate.
Indeed, observational indications of CCC having triggered  gas compression have been  reported   with regard to massive star formation in the 50~MC  \citep{Tsuboi2011, Tsuboi2015a, Tsuboi2015b, Tsuboi2019}.

The timescale of the evolution of the HMC   should depend on the mass of the HMC. 
Since the duration of the CCC is longer than the lifetime of the HMC as described  below, 
 we conjecture that the HMCs are formed and evolve  in an order of 10$^6$~yr.
Larger the mass is, faster the HMC evolves.   
 When  HMPOs are formed in the center of the HMC, 
 the periphery of the HMC is left behind and eventually dissipates.

\citet{Uehara2017} identified 27 molecular-cloud filaments in the 50~MC.
In addition,  \citet{Uehara2019, Uehara2021}  examined whether the CS cores were influenced by CCC or not and argued that CCC efficiently formed massive bound cores even  when the slope of the core mass function (often abbreviated as the CMF)   was not greatly changed  by CCC.
Some massive CS cores produced by CCC may have evolved to  the HMCCs that we observed. If so,  the ages of the CS cores are probably (1--2) $\times$ 10$^4$~yr, which  is comparable  with those of HMCs01--28, HIIs-A, B, C, and D.
In the 50~MC, the events of CCC are still ongoing  in the 27 filaments \citep{Uehara2017, Uehara2021} with a duration timescale of $\sim$10$^6$~yr. 
Then, we conjecture that  the above-described three evolutionary stages of the HMC, i.e., dense core, hot core, and HHMC,  take a combined timescale of 10$^{4-5}$~yr.

A burst of star formation or a mini starburst may be taking place in the 50~MC region  in a similar way as in Sgr~B2 in the CMZ \citep[e.g.][]{Hasegawa1994, Sato2000}, and W49A \citep[e.g.][]{Miyawaki2009} and W51A \citep[e.g.][]{Okumura2001} in the Galactic disk. In this case, it may be  induced by CCC.


\section{CONCLUSIONS}
We present the results of ~2\farcs5-resolution observations made with ALMA at 86 GHz in the continuum and SO $(N_J=2_2-1_1)$ emissions  of the region of the Galactic Center Molecular Cloud G$-$0.02$-$0.07 (the 50~MC) in the CMZ.

\begin{enumerate}
\item The 86-GHz continuum emission, which mainly traces HII regions, found  four HII regions of HII-A to D in the central part of the 50~MC. 
\item No new UCHIIs,  HCHIIs,  or  Class-II CH$_3$OH masers were detected.
\item  Ten dust cores  were identified in which five dust cores are positionally associated with  HMCCs.
\item We identified 28 HMCCs around the HII regions, using ``clumpfind'' and visually inspecting the channel  and  integrated-intensity maps.
\item The masses of the identified dust cores and HMCCs  were estimated under the LTE condition and  they were found to be almost proportional to their virial masses.
 No correlation between the size and velocity width of the HMCCs, as expected in the Larson's first law, was found.
\item The relation between the ratio of the LTE to virial masses  and radius  shows that the HMCCs with radii r $<$ 0.1~pc are bound,  whereas those with  r $>$ 0.1~pc are unbound.
We conclude that  a contraction of the HMC ceases by the time the size has decreased to 0.1~pc, after which  the size of the HMC  remains almost stable. 
\item The HMCCs  were likely to be formed through compression of molecular clumps  triggered by external force, such as CCC  and supernova explosion.
\item The HMCCs in the 50~MC are in  an evolutionary stage of pre-HMC.

\end{enumerate}

\begin{ack} 
We thank the anonymous referee for their  comments that helped us improve the manuscript considerably. 
We thank Dr. Yoshimi Kitamura for helpful discussions.
This work is supported in part by the Grant-in-Aid from the Ministry of Education, Sports, Science
and Technology (MEXT) of Japan, No.16K05308 and No.19K03939. 
 The work presented in this paper makes use of the following ALMA data: ADS/JAO.ALMA \#2012.1.00080.S. 
ALMA is a partnership of ESO (representing its member states), NSF (USA) and NINS (Japan), together with NRC (Canada), NSC and ASIAA (Taiwan), and KASI (Republic of Korea), in cooperation with the Republic of Chile. 
The Joint ALMA Observatory is operated by ESO, AUI/NRAO, and NAOJ.
Data analysis was in part carried out on the open-use data analysis computer system at the Astronomy Data Center, ADC, of the NAOJ.
\end{ack}


\vspace{20mm}

\begin{thebibliography}{}

\bibitem[Akeson \& Carlstrom (1996)]{Akeson1996}
Akeson, R. L. \& Carlstrom, J. E. 1996, ApJ, 470, 528

\bibitem[Amo-Baladr\'{o}n et al. (2011)]{Amo2011}
Amo-Baladr\'{o}n, M. A., Martin-Pintado, J., \& Martin, S. 2011, A\&A, 526, A54



 \bibitem[Andr\'{e} et al.(2010)]{Andre2010}
Andr\'{e} P., Men'shchikov A., Bontemps S. et al. 2010, A\&A, 518, L102
 
\bibitem[Bally et al.(2014)]{Bally2014}
Bally, J., Rathborne, J. M., Longmore, S. N., Jackson, J. M., Alves, J. F., Bressert, E., Contreras, Y., Foster, J. B., Garay, G., Ginsburg, A., Johnston, K. G., Kruijssen, J. M. D., Testi, L., \& Walsh, A. J.. 2014, ApJ, 795, 28

\bibitem[Battersby et al.(2017)]{Battersby2017}
Battersby, C., Bally, J., \& Svoboda, B. 2017, ApJ, 835, 263

\bibitem[Barnes et al.(2019)]{Barnes2019}
Barnes, A. T.; Longmore, S. N.; Avison, A., Contreras, Y., Ginsburg, A., Henshaw, J. D., Rathborne, J. M., Walker, D. L., Alves, J., Bally, J., Battersby, C., Beltr\'{a}n, M. T.,  Beuther, H., Garay, G., Gomez, L., Jackson, J., Kainulainen, J., Kruijssen, J. M. D., Lu, X., Mills, E. A. C., Ott, J., \& Peters, T.  2019, MNRAS, 486, 283.

\bibitem[Beltr\'{a}n \& de Wit(2016)]{Beltran2016}
Beltr\'{a}n, M. T. \& de Wit, W. J. 2016, A\&Arv, 24, 6

\bibitem[Berry(2015)]{Berry2015}
Berry, D. S. 2015, Astronomy \& Computing, 10, 22

\bibitem[Beuther et al.(2002)]{Beuther2002}
Beuther, H., Walsh, A., Schilke, P., Sridharan, T. K., Menten, K. M., \& Wyrowski, F. 2002 A\&A, 390, 289

\bibitem[Beuther et al.(2007)]{Beuther2007}
Beuther, H., Churchwell, E., McKee, C., \& Tan, J. 2007, in Protostars and Planets V, ed. B. Reipurth, D. Jewitt, \& K. Keil (Tucson: University of Arizona Press), 165 

\bibitem[Boonman et al.(2001)]{Boonman2001}
Boonman, A. M. S., Stark, R., van der Tak, F. F. S., van Dishoeck, E. F., van der Wal, P. B., Sch\"{a}fer, F., de Lange, G., \& Laauwen, W. M. 2001, ApJ, 553, L63

\bibitem[Cotton \& Yusef-Zadeh (2016)]{Cotton2016}
Cotton, W. D., \& Yusef-Zadeh, F.  2016, ApJS, 227, 10

\bibitem[Churchwell et al.(1990)]{Churchwell1990}
Churchwell, E., Walmsley, C.M. \& Cesaroni, R. 1990, A\&AS 83, 119

\bibitem[Churchwell(2002)]{Churchwell2002}
Churchwell, E. 2002, ARA\&A, 27

\bibitem[Coil \& Ho (2000)]{Coil2000}
Coil, Alison L.; Ho, Paul T. P., 2000, ApJ, 533, 245

\bibitem[Cragg et al.(2005)]{Cragg2005}
Cragg, D. M., Sobolev, A. M., \& Godfrey, P. D. 2005, MNRAS, 360, 533

\bibitem[de la Fuente et al.(2018)]{Fuente2018}
de la Fuente, E., Trinidad, M. A., Porras, A., Rodr\'{i}guez-Rico, C., Araya, E. D., Kurtz, S., Hofner, P., \& Nigoche-Netro, A. 2018, RMxAA, 54, 129

\bibitem[De~Pree et al.(2000)]{DePree2000}
 De Pree, C. G., Wilner, D. J., Goss, W. M., Welch, W. J., \& McGrath, E.  2000, ApJ, 540, 308


\bibitem[De~Pree et al.(2004)]{DePree2004}
De Pree, C. G., Wilner, D. J., Mercer, A. J., Davis, L. E., Goss, W. M., \& Kurtz, S.  2004, ApJ, 600, 286

\bibitem[Ekers et al.(1983)]{Ekers1983}
Ekers, R. D., van Gorkom, J. H., Schwarz, U. J.,  \& Goss, W. M. 1983, A\&A,  122, 143

\bibitem[Frail et al.(1996)]{Frail1996}
Frail, D. A., Goss, W. M., Reynoso, E. M., et al. 1996, AJ, 111, 1651

\bibitem[Fazal et al.(2008)]{Fazal2008}
Fazal, F. M., Sridharan, T. K., Qiu, K.,  Robitaille, T., Whitney, B., \& Zhang, Q. 2008, ApJ, 688, L41

\bibitem[Fontani et al.(2010)]{Fontani2010}
Fontani, F., Cesaroni, R., \& Furuya, R. S., 2010, A\&A, 517, 56

\bibitem[Fukui et al.(2020)]{Fukui2020}
Fukui, Y.,Inoue, T., Hayakawa, T., \& Torii K.. 2020, PASJ, in press

\bibitem[Furuya et al.(2005)]{Furuya2005}
Furuya, R.S., Cesaroni, R., Takahashi, S., Momose, M., Testi, L., Shinnaga, H. \& Codella, C. 2005, ApJ, 624, 827

\bibitem[Furuya et al.(2011)]{Furuya2011}
Furuya, R.S., Cesaroni, R., \& Shinnaga, H. 2011, A\&A, 525, 72

\bibitem[Garay \& Lizano(1999) ]{Garay1999}
Garay, G., \& Lizano, S. 1999, PASP, 111, 1049

\bibitem[Goddi et al.(2011)]{Goddi2011}
Goddi, C.; Greenhill, L. J.; Humphreys, E. M. L.; Chandler, C. J.; Matthews, L. D. 2011, ApJ, 739, 13

\bibitem[(Goldsmith \& Langer(1999)]{Goldsmith1999}
Goldsmith, P. F., \& Langer, W. D. 1999, ApJ, 517, 209

\bibitem[Goss et al.(1985)]{Goss1985}
Goss, W. M., Schwarz, U. J.;, van Gorkom, J. H.\& Ekers, R. D., 1985, MNRAS, 215, 69


\bibitem[Hasegawa et al.(1994)]{Hasegawa1994}
Hasegawa, T., Sato, F., Whiteoak, J. B.,  \& Miyawaki, R. 1994, ApJ,  429, L77

\bibitem[Herbst \& van Dishoeck(2009)]{Herbst2009}		
Herbst, E. \& van Dishoeck, E. F., 2009, ARA\&A, 47, 427

\bibitem[Hildebrand (1983)]{Hildebrand1983}
Hildebrand, R. H. 1983, Quart, J. R. A. S., 24, 267

\bibitem[Ho \& Towns(1983)]{Ho1983}		
Ho, P. T. P., \& Townes, C. H. 1983, ARA\&A, 21, 239

\bibitem[Ikeda et al.(2002)]{Ikeda2002}		
Ikeda, M., Hirota, T., \& Yamamoto, S., 2002, ApJ, 575, 250

\bibitem[Inoue \& Fukui(2013)]{Inoue2013}		
Inoue,T., \& Fukui, Y., 2013, ApJ, 774, L31


\bibitem[Jim\'{e}nez-Serra et al.(2012)]{Serra2012}		
Jim\'{e}nez-Serra, I., Zhang, Q., Viti, S., Martin-Pintado, J., \& de Wit, W.-J. 2012, ApJ., 753, 34

\bibitem[Kauffman et al.(1998)]{Kauffman1998}		
Kauffman, M. J., Hollenbach, D. J., \& Tielens, A. G. G. M. 1998, ApJ., 497, 276

\bibitem[Kauffmann et al.(2010a)]{Kauffmann2010a}		
Kauffmann, J., Pillai, T., Shetty, R., Myers, P. C., \& Goodman, A. A. 2010a, ApJ., 712, 1137

\bibitem[Kauffmann et al.(2010b)]{Kauffmann2010b}		
Kauffmann, J., Pillai, T., Shetty, R., Myers, P. C., \& Goodman, A. A. 2010b, ApJ., 716, 433


\bibitem[Kauffmann et al.(2017a)]{Kauffmann2017a}	
 Kauffmann, J,.Pillai, T., Zhang, Q., Menten, K. M., Goldsmith, P. F., Lu, X., \& Guzm\'{a}n, A. E. 2017a, A\&A, 603, A89; 

\bibitem[Kauffmann et al.(2017b)]{Kauffmann2017b}	
Kauffmann, J,.Pillai, T., Zhang, Q., Menten, K. M., Goldsmith, P. F., Lu, X., \& Guzm\'{a}n, A. E. 2017b, A\&A, 603, A90; 

\bibitem[Kratter \& Matzner(2006)]{Kratter2006}		
Kratter, K. M., \& Matzner, C. D. 2006, MNRAS, 373, 1563

\bibitem[Kurtz et al.(2000)]{Kurtz2000}
Kurtz, S., Cesaroni, R., Churchwell, E., Hofner, P., \& Walmsley, C. M. 2000, in Protostars and Planets IV, eds Mannings, V., Boss, A.P., Russell, S. S. (Tucson: University of Arizona Press), 299

\bibitem[Larson (1981)]{Larson1981}
Larson, R. B. 1981, MNRAS, 194, 809

\bibitem[Li et al.(2015)]{Li2015}
Li, J., Wang, J., Zhu, Q., Zhang, J., \& Li, D. 2015, ApJ, 802, 40

\bibitem[Li et al.(2020)]{Li2020}
Li, C., Wang, H., Wu, Y., Ma, Y., \& Lin, Li. 2020, RAA, 20, 31

\bibitem[Lu et al.(2019a)]{Lu2019a}	
Lu, X., Zhang, Q. Kauffmann, J.,Pillai, T., Ginsburg, A., Mills, E. A. C., Kruijssen, J. M. D., Longmore, S. N., Battersby, C., Liu, H. B., \& Gu, Q., 2019a, ApJ, 872, 171

\bibitem[Lu et al.(2019b)]{Lu2019b}
Lu, X., Mills, E. A. C., Ginsburg, A., Walker, D. L., Barnes, A. T., Butterfield, N. Henshaw, J. D., Battersby, C., Kruijssen, J. M. D., Longmore, S. N., Zhang, Q.,Bally, .J., Kauffmann, J. Ott, J., Rickert, M., \& Wang, K. 2019b, ApJ Suppl, 244, 35

\bibitem[McEwen et al.(2016)]{McEwen2016}	
McEwen, B. C., Sjouwerman, L. O.\& Pihlstr\``{o}m, Y. M. 2016, ApJ, 832, 129

\bibitem[Mac\ Low et al.(2007)]{MacLow2007}	
Mac Low, M-M, Toraskar, J., Oishi, J. S., \& Abel, T. 2007, ApJ, 688, 980

\bibitem[McMullin et al.(2007)]{McMullin2007}
McMullin, J. P., Waters, B., Schiebel, D., Young, W., \& Golap, K. 2007, Astronomical Data Analysis Software and Systems XVI (ASP Conf. Ser. 376), ed. R. A. Shaw, F. Hill, \& D. J. Bell (San Francisco, CA: ASP), 127 

\bibitem[Mills et al.(2011)]{Mills2011}
Mills, E. J., Goss, W. M., \& De~Pree, C. G. 2011 ApJ, 735, 84 

\bibitem[Minier et al. (2005)]{Minier2005}
Minier, V., Burton, M. G., Hill, T., Pestalozzi, M. R., Purcell, C., Garay, G., Walsh, A., \& Longmore, S., 2005, A\&A, 429, 945

\bibitem[Miyawaki et al.(2002)]{Miyawaki2002}
Miyawaki, R., Hasegawa, T., \& Hayashi, M.\  2002, The Proceedings of the IAU 8th Asian-Pacific Regional Meeting (the Astronomical Society of Japan, Tokyo) 171
\bibitem[Miyawaki et al.(2009)]{Miyawaki2009}
Miyawaki, R., Hayashi, M., \& Hasegawa, T., 2009, PASJ, 61, 39

\bibitem[Miyazaki \& Tsuboi(2000)]{Miyazaki2000}
Miyazaki, A. \& Tsuboi, M. 2000, ApJ, 536, 357

\bibitem[Motte et al.(2018)]{Motte2018}
Motte, F.,  Bontemps, S., \& Louvet, F. 2018, ARA \& A, 56, 41

\bibitem[Morris \& Serabyn(1996)]{Morris1996}
Morris, M.,  \& Serabyn, E., 1996, ARA \& A, 34, 645

\bibitem[Nomura \& Millar(2004)]{Nomura2004}
Nomura, H. \& Millar, T. J.. 2004, A\&A, 414, 409

\bibitem[Okumura et al.(2001)]{Okumura2001}
Okumura, S., Miyawaki, R., Sorai, K., Yamashita, T., \& Hasegawa, T.  2001, PASJ, 53, 793

\bibitem[Pestalozzi et al.(2002)]{Pestalozzi2002}
Pestalozzi, M., Humphreys, E. M. L., \& Booth, R. S., 2002, A\&A, 384, L15

\bibitem[Phillips et al.(1998)]{Phillips1998}
Phillips, C. J., Norris, R. P., Ellingsen, S. P., \& McCulloch, P. M. 1998, MNRAS, 300, 1131

\bibitem[Pihlstr\"{o}m et al.(2011)]{Pihlstrom2011}
Pihlstrr\"{o}m, Y. M., Sjouwerman, L. O., \& Mesler, R. A. 2011, ApJ, 740, 66

\bibitem[Pineda et al.(2009)]{Pineda2009}
Pineda, J. E., Rosolowsky, Erik W., \& Goodman, A. A. 2009, ApJ, 699, L134

\bibitem[Pillai et al.(2015)]{Pillai2015}
Pillai, T., Kauffmann, J., Tan, J. C., et al. 2015, ApJ, 799, 74

\bibitem[Plambeck et al.(1982)]{Plambeck1982}
Plambeck, R. L., Wright, M. C. H., Welch, W. J., Bieging, J. H., Baud, B., Ho, P. T. P., \& Vogel, S. N.  1982, ApJ, 259, 617

\bibitem[Rathborne et al.(2014)]{Rathborne2014}
Rathborne, J. M., Longmore, S. N., Jackson, J. M., Kruijssen, J. M. D., Alves, J. F., Bally, J., Bastian, N., Contreras, Y., Foster, J. B., Garay, G., Testi, L., \& Walsh, A. J 2014, ApJ, 795, L25

\bibitem[Rathborne et al.(2015)]{Rathborne2015}
Rathborne, J. M., Longmore, S. N., Jackson, J. M., Alves, J. F., Bally, J., Bastian, N., Contreras, Y., Foster, J. B., Garay, G., Kruijssen, J. M. D., Testi, L., \& Walsh, A. J. 2015, ApJ, 802, 125

\bibitem[Rolffs et al.(2011)]{Rolffs2011}
Rolffs, R., Schilke, P., Zhang, Q., \& Zapata, L. 2011, A\&A, 536, 33

\bibitem[Rosolowsky et al.(2008)]{Rosolowsky2008}
Rosolowsky, E. W., Pineda, J. E., Kauffmann, J., \& Goodman, A. A. 2008, ApJ, 679, 1338

\bibitem[Sanna et al.(2014)]{Sanna2014}
Sanna, A., Cesaroni, R., Moscadelli, L., Zhang, Q., Menten, K. M., Molinari, S., Caratti, A., o Garatti, A., \& De~Buizer, J. M. 2014, A\&A, 565, 34

\bibitem[Sanna et al.(2019)]{Sanna2019}
Sanna, A., K\"{o}lligan, A., Moscadelli, L, Kuiper, R., Cesaroni, R.,  Pillai, T., Menten, K. M., Zhang, Q., Caratti~o~Garatti,  A.,  Goddi, A., Leurini, S., \& Carrasco-Gonz\'{a}lez, C. 2019, A\&A, 623, 77


\bibitem[Sato et al.(2000)]{Sato2000}
Sato, F., Hasegawa, T., Whiteoak, J. B., \& Miyawaki, R.  2000, ApJ, 535, 857

\bibitem[Sewilo et al.(2004)]{Sewilo2004}
Sewilo, M., Churchwell, E., Kurtz, S., Goss, W. M. \& Hofner, P. 2004, ApJ, 609, 285

\bibitem[Sewilo et al.(2011)]{Sewilo2011}
Sewilo, M., Churchwell, E., Kurtz, S., Goss, W. M. \& Hofner, P. 2011, ApJS, 194, 44

\bibitem[Schilke et al.(2001)]{Schilke2001}
Schilke, P., Benford, D. J., Hunter, T. R., Lis, D. C., \& Phillips, T. G. 2001, ApJS, 132, 281

\bibitem[Shu et al.(1987)]{Shu1987}
Shu, F. H., Adams, F. C., \& Lizano, S. 1987, ARA\&A, 25, 23

\bibitem[Silva et al.(2017)]{Silva2017}
Silva, A., , Zhang, Q., Sanhueza, P., Lu, X., Beltr\'{a}n, M. T., Fallscheer, C., Beuther, H., Sridharan, T. K., \& Cesaroni, R., 2017, ApJ, 847, 87

\bibitem[Sjouwerman \& Pihlstr\"{o}m (2008)]{Sjouwerman2008}
Sjouwerman, L. O., \& Pihlstrr\"{o}m, Y. M.,  2008, ApJ, 681, 1287

\bibitem[Sridharan et al.(2002)]{Sridharan2002}
Sridharan, T. K.,  Beuther, H., Schilke, P., Menten, K. M., \& Wyrowski, F. 2002, ApJ, .566, 931

\bibitem[St\'{e}phan et al.(2018)]{Stephan2018}
St\'{e}phan, G., S., Schilke, P., Le Bourlot, J., Schmiedeke, A., Choudhury, R., Godard, B.,  \& S\'{a}nchez-Monge, \'{A}.  2018, A\&A, 617, 60

\bibitem[Stutzki \& Guesten(1990) ]{Stutzki1990}
Stutzki, J., \& Guseten, R. 1990, ApJ, 356, 513

\bibitem[Takahira et al.(2014)]{Takahira2014}
Takahira, K.,  Tasker, E. J., \& Habe, A., 2014, ApJ, 792, 63

\bibitem[Takahira et al.(2018)]{Takahira2018}
Takahira, K., Shima, K, Habe, A., \& Tasker, E. J. 2018, PASJ, 70, 1

\bibitem[Takekawa et al.(2017a)]{Takekawa2017a}
Takekawa, S., Oka, T., \& Tanaka, K. 2017a, ApJ, 834, 121

\bibitem[Takekawa et al.(2017b)]{Takekawa2017b}
Takekawa, S., Oka, T., Iwata, Y., Tokuyama, S., \& Nomura, M. 2017b, ApJ, 843, L11

\bibitem[Tanaka et al.(2016)]{Tanaka2016}
Tanaka, K. E. I. , Tan, J. C., Zhang, Y. 2016, ApJ, 818, 52

\bibitem[Tsuboi et al.(2009)]{Tsuboi2009}
Tsuboi, M., Miyazaki, A., \& Okumura, S., 2009, PASJ, 61, 29

\bibitem[Tsuboi et al.(2011)]{Tsuboi2011}
Tsuboi, M., Tadaki, K., Miyazaki, A., \& Handa, T. 2011, PASJ, 63, 763

\bibitem[Tsuboi et al.(2012)]{Tsuboi2012}
Tsuboi, M., \&  Miyazaki, A.,  2012, PASJ, 64, 111


\bibitem[Tsuboi et al.(2015a)]{Tsuboi2015a}
Tsuboi, M., Miyazaki, A., \& Uehara, K. 2015a, PASJ, 67, 90

\bibitem[Tsuboi et al.(2015b)]{Tsuboi2015b}
Tsuboi, M., Miyazaki, A., \& Uehara, K. 2015b, PASJ, 67, 109

\bibitem[Tsuboi et al.(2019)]{Tsuboi2019}
Tsuboi, M., Kitamura,, Y., Uehara, K., Miyazaki, A., Miyawaki, R., Tsutsumi, T.,  \& Miyoshi, M. 2019, PASJ, 71, 128

\bibitem[Tsuboi et al.(2021)]{Tsuboi2021}
Tsuboi, M., Kitamura,, Y., Uehara, K., Miyawaki, R., Tsutsumi, T.,\& Miyazaki, A. 2021, PASJ, 73, S91 

\bibitem[Uehara et al.(2017)]{Uehara2017}
Uehara, K., Tsuboi, M., Kitamura, Y., Miyawaki, R., \& Miyazaki, A. 2017, in The Multi-Messenger Astrophysics of the Galactic Centre, Proceedings of the International Astronomical Union, IAU Symposium, 322, pp.162

\bibitem[Uehara et al.(2019)]{Uehara2019}
Uehara, K., Tsuboi, M., Kitamura, Y., Miyawaki, R., \& Miyazaki, A. 2019,  ApJ, 872, 121

\bibitem[Uehara et al.(2021)]{Uehara2021}
Uehara, K., Tsuboi, M., Kitamura, Y., Miyawaki, R., \& Miyazaki, A. 2021 to be submitted

\bibitem[van~der~Tak et al.(2003)]{Tak2003}
van der Tak, F. F. S., Boonman, A. M. S., Braakman, R., \& van Dishoeck, E. F. 2003, A\&A, 412, 133. 

\bibitem[van~der~Tak et al.(2007)]{Tak2007}
 van der Tak, F.F.S., Black, J.H., Sch\"{o}ier, F.L., Jansen, D.J., van Dishoeck, E.F. 2007, A\&A, 468, 627

\bibitem[Walker et al.(2018)]{Walker2018}
Walker, D. L., Longmore, S. N., Zhang, Q., Battersby, C.; Keto, E., Kruijssen, J. M. D., Ginsburg, A., Lu, X., Henshaw, J. D., Kauffmann, J., Pillai, T., Mills, E. A. C., Walsh, A. J., Bally, J., Ho, L. C., Immer, K., \& Johnston, K. G. 2018, MNRAS, 474, 2373

\bibitem[Walsh et al.(1998)]{Walsh1998}
Walsh, A.J., Burton, M.G., Hyland, A.R. \& Robinson, G. 1998, MNRAS, 301, 640

\bibitem[Walsh et al.(2003)]{Walsh2003}
Walsh, A. J., Macdonald, G. H., Alvey, N. D. S., Burton, M. G., \& Lee, J.-K., 2003, A\&A, 410, 597


\bibitem[Williams et al.(1994)]{Williams1994}
Williams, J. P., de Geus, E. J., \& Blitz, L. 1994, ApJ, 428, 693

\bibitem[Wilner et al.(2001)]{Wilner2001}	
Wilner, D. J., De Pree, C. G., Welch, W. J., \& Goss, W. M. 2001, ApJ, 550, L81

\bibitem[Wright et al.(1996)]{Wright1996}	
Wright, M. C. H., Plambeck, R. L., \& Wilner, D. J. 1996, ApJ, 469, 216

\bibitem[Wyrowski et al.(2012)]{Wyrowski2012}	
Wyrowski, F., G\"{u}sten, R.,  Menten, K. M.,  Wiesemeyer, H. \&Klein, B. 2012 A\&A, 542, 2012

\bibitem[Yusef-Zadeh et al.(2010)]{Yusef-Zadeh2010}	
Yusef-Zadeh, F., Lacy, J. H., Wardle, M., Whitney, B., Bushouse, H., Roberts, D. A., \& Arendt, R. G.  2010, ApJ, 725, 1429

\bibitem[Yusef-Zadeh et al.(2013)]{Yusef-Zadeh2013}	
Yusef-Zadeh, F., Cotton, W., Viti, S., Wandle, M.,  \& Royster, M. J.  2013, ApJL, 452, L19

\bibitem[Zinchenko \& Henkel(2018)]{Zinchenko2018}	
Zinchenko, I., \& Henkel, C. 2017, in Proceedings of the IAU Symposium No. 332 , Astrochemistry VII - Through the Cosmos from Galaxies to Planets, 274


\end{thebibliography}
\end{document}